ORGANISATION EUROPÉENNE POUR LA RECHERCHE NUCLÉAIRE

**CERN** EUROPEAN ORGANIZATION FOR NUCLEAR RESEARCH

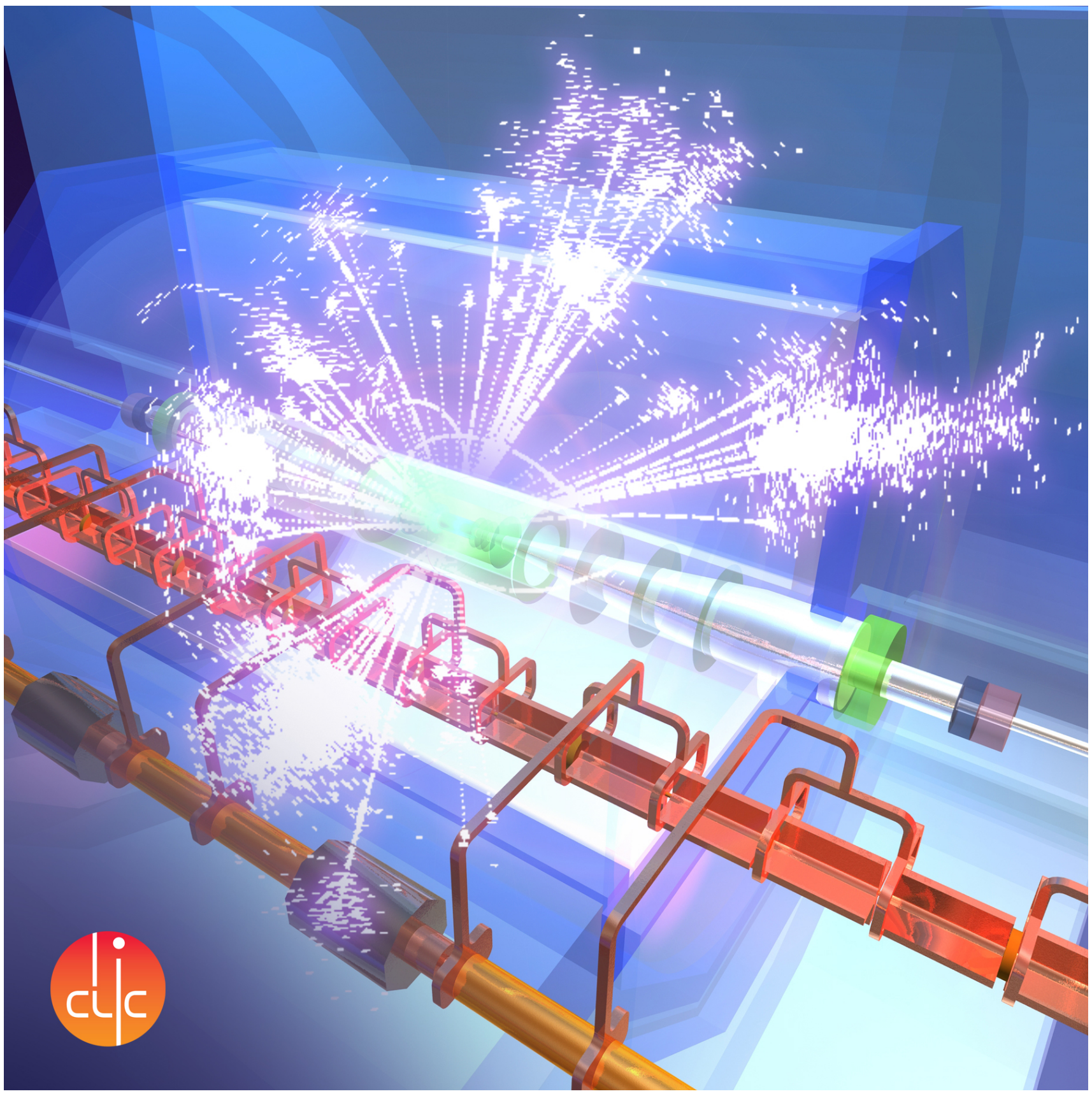

# THE CLIC PROGRAMME: TOWARDS A STAGED $e^+e^-$ LINEAR COLLIDER EXPLORING THE TERASCALE

CLIC CONCEPTUAL DESIGN REPORT

**Abstract**

This report describes the exploration of fundamental questions in particle physics at the energy frontier with a future TeV-scale $e^+e^-$ linear collider based on the Compact Linear Collider (CLIC) two-beam acceleration technology. A high-luminosity high-energy $e^+e^-$ collider allows for the exploration of Standard Model physics, such as precise measurements of the Higgs, top and gauge sectors, as well as for a multitude of searches for New Physics, either through direct discovery or indirectly, via high-precision observables. Given the current state of knowledge, following the observation of a $\sim$125 GeV Higgs-like particle at the LHC, and pending further LHC results at 8 TeV and 14 TeV, a linear $e^+e^-$ collider built and operated in centre-of-mass energy stages from a few-hundred GeV up to a few TeV will be an ideal physics exploration tool, complementing the LHC. In this document, an overview of the physics potential of CLIC is given. Two example scenarios are presented for a CLIC accelerator built in three main stages of 500 GeV, 1.4 (1.5) TeV, and 3 TeV, together with operating schemes that will make full use of the machine capacity to explore the physics. The accelerator design, construction, and performance are presented, as well as the layout and performance of the experiments. The proposed staging example is accompanied by cost estimates of the accelerator and detectors and by estimates of operating parameters, such as power consumption. The resulting physics potential and measurement precisions are illustrated through detector simulations under realistic beam conditions.

This report constitutes the strategic summary document of the CLIC Conceptual Design Report (CDR), putting emphasis on the construction and operation of CLIC in successive energy stages. The CLIC CDR comprises two other reports, providing detailed descriptions of the CLIC accelerator and of the CLIC physics and detectors. These two reports have a common list of more than 1300 signatories[1].

CORRESPONDING EDITORS

Philippe Lebrun, Lucie Linssen, Angela Lucaci-Timoce, Daniel Schulte, Frank Simon, Steinar Stapnes, Nobukazu Toge, Harry Weerts, James Wells

[1] https://edms.cern.ch/document/1183227/

# Contents

# Chapter 1

# Introduction

This report is part of the Conceptual Design Report (CDR) of the Compact Linear Collider (CLIC). CLIC is a high-energy linear $e^+e^-$ collider with the potential to operate at centre-of-mass energies ranging from a few hundred GeV up to 3 TeV and with luminosities of a few $10^{34}$ cm$^{-2}$s$^{-1}$. The CLIC accelerator complex and the CLIC physics and detector studies are described in separate CDR reports [1, 2]. The CLIC accelerator CDR [1] provides detailed descriptions of the accelerator layout, its components and the expected performance of CLIC. In particular, it describes technical solutions to the key feasibility issues, thus proving the validity of the CLIC concept. Prototypes of many of the technical subsystems have been successfully tested at the CLIC test facility at CERN and at other facilities around the world. The test results are reported in detail in the CDR.

The CLIC physics and detector CDR [2] gives an overview of the extensive CLIC physics potential. The physics aims together with the challenging beam-induced background conditions are driving the two detector designs CLIC_ILD and CLIC_SiD. These detector concepts are based on the ILD and SiD concepts, initially designed for the International Linear Collider (ILC). Detailed detector benchmark studies, using key physics processes as examples, demonstrate that physics measurements can be performed to high precision, despite the beam-induced background.

The focus of the physics and detector CDR and the accelerator CDR was on the maximum CLIC centre-of-mass energy of 3 TeV. This energy corresponds to the most challenging situation for both the accelerator and the detector technologies, while simultaneously providing an outlook on the ultimate physics reach. Exploring the full physics potential of an $e^+e^-$ collider under optimal conditions, however, requires the availability of a broad range of centre-of-mass energies. For example, Standard Model physics parameters such as several Higgs decay branching ratios, the Higgs mass and the top sector, can be measured precisely at centre-of-mass energies below 500 GeV. The measurement of other Higgs properties, such as the top-Yukawa coupling, the Higgs potential and rare Higgs decay modes will profit from centre-of-mass energies at or above 1 TeV. For the investigation of New Physics phenomena beyond the Standard Model, positive observations from LHC are eagerly awaited. Given the results from LHC searches so far, it seems likely that high $e^+e^-$ centre-of-mass energies will be required for these studies.

In addition, for technical reasons the staged construction of CLIC is highly desirable. While a 3 TeV collider could be operated at much lower centre-of-mass energies, this would only be possible at significantly reduced luminosities. However, many of the important physics processes have small cross-sections, and investigating such channels requires the highest possible luminosity. Building CLIC in a few main energy stages, with some flexibility to operate below the nominal energy at each stage, is the optimum approach.

It is currently too early to predict which CLIC staging scenario will combine the best physics opportunities with an optimal accelerator implementation. More insight is expected through additional results from LHC at 8 TeV and 14 TeV. In the meantime, this document explores an example of a staged construction and operation scenario for CLIC. The staging scenario is based on current knowledge of Standard Model physics, including the $\sim$125 GeV Higgs-like particle observed at the LHC. To illustrate the CLIC capabilities for precision measurements of phenomena beyond the Standard Model, a supersymmetric model that is compatible with current LHC results (summer 2012) is used as a basis for this case study. This example allows a detailed insight into the physics potential of such a staged machine, together with realistic scenarios for the construction and operation of the accelerator and the detectors along with estimates of the deployment of resources and of the energy consumption.

Chapter 2 of this report provides a physics overview, with emphasis on potential $e^+e^-$ collider observables and on complementarity to the LHC discovery potential. Chapter 3 summarises the CLIC accelerator design, technologies and performance. It also presents a scenario for a staged construction of CLIC with energy stages at nominal centre-of-mass energies of 500 GeV, 1.4 (1.5) TeV and 3 TeV, and with possible lower energy operation at every stage. In Chapter 4 the CLIC_ILD and CLIC_SiD detector concepts are presented. An overview of the experimental conditions at CLIC and the methodology to suppress the impact of beam-induced backgrounds on the data are described. Chapter 5 describes the implementation of the staged approach, including schedules, power consumption and cost estimates. The physics potential of the staged scenario is illustrated in Chapter 6 on the basis of realistic physics benchmark simulations using the CLIC_ILD and CLIC_SiD concepts. Chapter 7 describes the strategy and work objectives for the upcoming CLIC project phases.

# Chapter 2

# Physics Overview

## 2.1 Introduction

The recent observation at the LHC of a Higgs-like boson with mass 125 GeV [1, 2] has significantly bolstered the physics case for a linear collider. In the coming years many new developments and discoveries in particle physics are expected. Therefore, the physics landscape may require re-visiting the CLIC physics case taking into account future input. In the meantime, however, this report addresses the physics potential of the future collider in the context of the current understanding of particle physics. In this chapter a brief physics overview is presented, focusing on issues relevant for the physics potential of CLIC. In Chapter 6 the physics potential that springs from this overview will be presented, with examples of detailed simulations within the CLIC environment.

In the following sections the current state of knowledge for tests of the Standard Model (SM) and the status of various well-motivated theories of physics beyond the Standard Model are outlined. The focus is on the high-energy frontier, and less discussion is given of the valuable frontier experiments at lower energies, such as neutrino physics, $B$ physics, etc. Although there can be important connections of these phenomena with high-energy $e^+e^-$ collider observables, the focus here will remain on higher energy phenomena.

## 2.2 Standard Model Physics

### 2.2.1 Electroweak Precision Analysis

Precision tests of the SM started in earnest with LEP and SLC over twenty years ago, where measurements of many electroweak observables were combined in a global analysis. These observables included the $Z$ and $W$ boson masses, their widths and branching fractions in various final states. Some low-energy observables are taken from elsewhere, such as $\alpha_{QED}$ and the lifetime of the $\tau$ lepton. In addition, over time, the top mass measurement, $\alpha_s$ measurements, $W$ mass measurements, and even the Higgs boson mass constraints have greatly enhanced the capacity of precision measurements to test the SM at unprecedented levels.

One of the best measures of compatibility with the SM is the total $\chi^2$ analysis [3], which quantifies the compatibility of theory to the data. A multitude of observables have been measured and are included in the $\chi^2$ analysis. The Higgs boson is an important parameter in the analysis since it appears in the quantum loop of precision electroweak observables, such as $\sin^2\theta_W$ and $\Gamma(H \to \ell^+\ell^-)$. When $m_H = 125$ GeV, the value of $\chi^2_{tot}$ is well within the accepted statistical $\chi^2_{tot}$/ndf to declare that the theory is consistent with the data. This is a very meaningful test of the SM since if $m_H < 50$ GeV or $m_H > 170$ GeV were measured, the $\chi^2_{tot}$/ndf would have far exceeded what is tolerable for a theory to be consistent with the data.

In summary, the precision electroweak tests, which have combined the efforts of many experiments and many theory calculations, have shown no inconsistencies with the SM at this time, and an impressive concurrence with the direct search measurements and limits has been established.

On the other hand there is one observable that is traditionally not included in this $\chi^2$ analysis, but which bears mentioning in this report. It is the measurement of the $a_\mu \equiv (g-2)/2$ anomalous magnetic moment of the muon. The current measurement minus the SM expected value is [4]

$$a_\mu^{new} = a_\mu^{exp} - a_\mu^{SM} = (287 \pm 80) \times 10^{-11} \tag{2.1}$$

where the errors are added in quadrature. The deviation of the measurement from the SM is more than three times the $1\sigma$ error. This has lead to much speculation on whether the subtle SM computation has been done properly and the uncertainty understood well, or whether New Physics is required to account for this discrepancy. One of the most often studied New Physics explanations is supersymmetry, which has the feature that large contributions can be obtained from large Yukawa couplings, which are made possible by the two Higgs doublet nature of the model. A simple parametrisation of supersymmetry effects is [4]

$$a_\mu^{SUSY} = \pm 130 \times 10^{-11} \cdot \left( \frac{100\ \text{GeV}}{M_{SUSY}} \right)^2 \tan\beta \tag{2.2}$$

where $M_{SUSY}$ is the universal superpartner mass scale in this simplified expression. Nevertheless, it does give the correct implication that at least some superpartners should be sub-TeV in order for the $g-2$ observable to shift by more than $200 \times 10^{-11}$ within supersymmetry. This is one of the key data-driven motivations for the consideration of light supersymmetry accessible to a TeV collider.

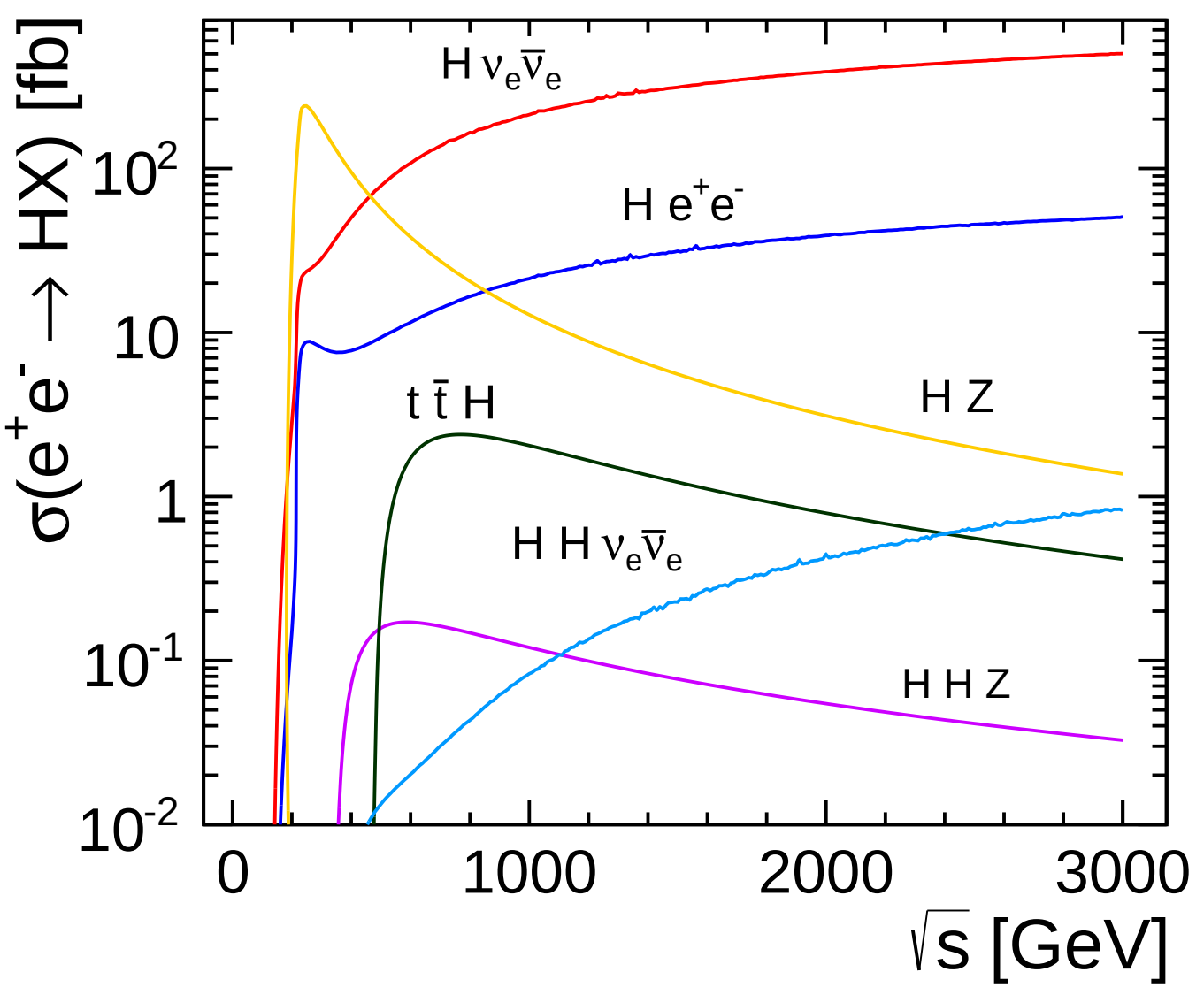

Fig. 2.1: Cross-sections for different production mechanisms for a 125 GeV Higgs boson as a function of the $e^+e^-$ centre-of-mass energy.

### 2.2.2 Higgs Boson

At different energy stages of CLIC, precision measurements of various observables of the SM Higgs boson can be carried out. There are several different modes for producing the Higgs boson in $e^+e^-$ interactions, which are exemplified in Figure 2.1 for a 125 GeV Higgs boson mass. The cross-sections are presented as a function of the centre-of-mass energy of $e^+e^-$ collisions. In Chapter 6 below, these production cross-sections and the various final states are discussed in the context of benchmark studies based on full detector simulations. It is there that details are given regarding how well certain properties of the Higgs boson can be measured for a given luminosity.

As can be seen from Figure 2.1, the vector-boson fusion contributions exceed the Higgsstrahlung process for higher energies, and indeed play an integral role in the Higgs boson study strategies at these higher energies (for more details see Chapters 1 and 12 of the CLIC CDR [5]). While several Higgs couplings are generally already well measured at centre-of-mass energies below 1 TeV, one can profit from higher luminosities and production cross-sections at larger energies for rare processes, like the Higgs decay to muons. The figure shows that the top-Yukawa coupling is best measured at centre-of-mass energies around 1 TeV. The extraction of the Higgs self-coupling using the $HH\nu_e\bar{\nu}_e$ final state requires the highest possible centre-of-mass energy.

### 2.2.3 Top Quark Physics

The top quark is the most massive particle in the SM, and is expected to have the strongest interaction with the Higgs boson. Because of this large interaction it is anticipated in some frameworks that its phenomenology, cross-section and decays, may be most vulnerable to beyond the SM influences. This is especially true since modern theory generally does not accept that the SM Higgs boson can alone be a stable solution to the dynamics of electroweak symmetry breaking, and anything that couples strongly to it may be sensitive to new effects.

Many measurements of the top quark sector have been accomplished at the Tevatron and LHC experiments. These include the total production cross-sections at different collider energies for both one and two top quarks, invariant mass distribution of the production, and searches for unexpected decay modes of the top quark. So far there is no indication that New Physics is at play in top quark production. The only possible exception to this conclusion is the persistent anomaly in the forward-backward asymmetry of the top quark production.

Although top quark production is mainly through QCD, which couples to fermions without chiral distinctions, there can still be at $\mathcal{O}(\alpha_s^3)$ an asymmetry in the production cross-section due to interference between box diagrams and tree diagrams, or interference between initial state and final state QCD radiation [6]. The resulting asymmetry amounts to approximately $(5\pm1)\%$. However, both CDF and D0 have recorded central values of the measurements of approximately $(20\pm7)\%$ [7], well in excess of the SM value of 5%. Statistically this result is not conclusive of New Physics, but it is tantalising. The current LHC top-quark asymmetry measurements are not inconsistent with the SM, but they are not decisive at this time.

There have been many New Physics ideas that have tried to explain the asymmetry while not disrupting other observables, most notably other top quark observables. A summary of these efforts and the implications for other observables [8], including at the LHC, shows what a challenging task this is. Nevertheless, refined analysis of Tevatron data and incoming LHC data should be able to re-establish the SM preeminence for this observable or find evidence of New Physics through a persistent asymmetry and non-SM contributions to other observables.

## 2.3 Beyond the Standard Model Physics

There is general agreement that the SM is not the complete description of particle physics. Reasons for this viewpoint are data driven and theory driven. On the data-driven side, there is strong evidence for dark matter in the universe that cannot be composed of SM particles. Furthermore, the baryon asymmetry of the universe implies new dynamics, including first-order phase transition(s) and extra sources of CP violation beyond the SM. On the more theory-driven side, nature has often agreed with more unified descriptions, and indeed there are attractive ideas on how to unify the gauge couplings of the three gauge forces into a simpler theory. Furthermore, the quantum instabilities of a fundamental scalar Higgs boson suggest a supporting cast of other particles and interactions that stabilise it. This can be accomplished through several scenarios: a symmetry (e.g. supersymmetry), banishing fundamental scalars (e.g. technicolor), or understanding high scales as mirages (e.g. extra dimensional theories).

The plethora of theories of physics beyond the SM makes it impossible here to summarise well the status of all of them. However, it can be said that most of these beyond the SM ideas have an overall mass scale associated with them that is unknown. For example, supersymmetry has a supersymmetry breaking scale that sets the mass of the superpartners; conformal field theory has the scale of conformal breaking; technicolor has the scale of the fermion condensates; Higgs compositeness theories have the scale of compositeness; and, extra dimensional theories have the scale of compactification. In each of these theories there are usually particles with masses tied to these scales, such as KK excitations, techni-$\rho$, superpartners, etc. As long as they couple to the SM particles, which they generally must do if they are to solve the hierarchy problem in particular, good sensitivity can be achieved at a high-energy

Table 2.1: A partial representative list of the masses probed at the LHC for various exotic physics scenarios beyond the SM. These are extracted from the summary plots of the ATLAS collaboration [9].

| New Physics scenario | Search strategy | Mass limit (TeV) |
|---|---|---|
| Large extra dimensions ($\delta = 2$) | jets + $E_{T,miss}$ | 3.2 |
| Universal extra dimensions | $\gamma\gamma + E_{T,miss}$ | 3.0 |
| Randall-Sundrum KK graviton | dilepton, $m_{\ell\ell}$ | 2.1 |
| Sequential SM $Z'$ | $ee/\mu\mu$ resonance | 2.2 |
| Scalar lepto-quarks | $\mu\mu + 2\,$jets, $\mu\nu + 2\,$jets | 0.68 |
| Fourth generation quark | $Q\bar{Q} \to WqWq$ | 0.35 |
| Excited quark | $q^* \to \gamma +$jet | 2.4 |
| Excited electron | $e^* \to e\gamma$ | 2.0 |
| Techni-hadrons | $ee/\mu\mu$ resonance | 0.47 |
| $W_R$ from $SU(2)_R$ | $\ell\ell +$jets | 2.4 |
| Color octet scalar | di-jet resonance | 1.9 |
| SUSY gluino ($\tilde{q} = \tilde{g}$) | jets + $E_{T,miss}$ | 1.4 |
| SUSY gluino ($\tilde{q} \gg \tilde{g}$) | jets + $E_{T,miss}$ | 0.85 |
| Chargino ($\chi_1^\pm \chi_2^0$) | $3\ell + E_{T,miss}$ | 0.25 |
| Chargino (stable) | charged tracks | 0.12 |
| Sneutrino (R-parity violation) | $e\mu$ resonance | 1.3 |

collider with sufficient luminosity, all tied to some mass scale.

With this understanding, we show in Table 2.1 a list of several representative mass limits of New Physics scenarios beyond the SM from the ATLAS experiment, obtained at up to 8 TeV. One can see from the table that the New Physics mass reach sensitivities vary widely in scale, from a fraction of a TeV to several TeV, depending on what observables can access them and what the underlying cross-section strength is. Leptons are much more favourable than jets, and invariant mass peaks are much more favourable than wide-spread distributions. This high variability in mass-scale sensitivity is typical of the hadron collider environment. On the other hand, as we will see in the later benchmarks discussion, the variability in mass limit reach is tied much more closely to the $e^+e^-$ collision energy than the specifics of the observable. Thus, states produced in pairs are generally discoverable at CLIC up to $\sim \sqrt{s}/2$. However, their effects on observables can extend to mass scales much higher than that, despite not being directly produced, such as a $Z'$ boson influencing $e^+e^- \to \mu^+\mu^-$ with mass even an order of magnitude beyond $\sqrt{s}$. An important point to emphasise here is that among the many New Physics scenarios possible there is nearly always a role for high-energy CLIC to augment our understanding of nature beyond what the LHC can do, either through discovery or complementary precision measurements.

What is noticeably absent from the current LHC limits are meaningful limits on purely electroweak interacting particles. The limits from these particles are currently not better than the $\sim 100$ GeV limits from LEP II, unless the particle decays very spectacularly, like $H^{++} \to \mu^+\mu^+$ producing nearly background free same-sign dimuon resonances (limit is 355 GeV now) or is charged and stable on detector time scales, such as a stable stau with its ionising tracks all the way out to the muon chamber (limit is 136 GeV now). Despite the spectacular nature of their unusual phenomenology, the limits and sensitivity achievable at the LHC do not compare to CLIC, where one expects to discover such states up to a mass of $\sim \sqrt{s}/2$.

Another case where the sensitivity of CLIC can greatly outstrip that of the LHC is in the detection of effects from higher order operators such as $qq\ell\ell$, $\ell'\ell'\ell\ell$, etc. An example is $Z'$ physics, where the LHC can find a new neutral gauge boson resonance with electroweak strength couplings to the SM states up to $\sim 5$ TeV, but cannot see effects of the $Z'$ even indirectly on $q\bar{q} \to \ell^+\ell^-$ at mass scales above that [10]. On the other hand, the linear collider cannot compete for the direct resonance bump searches due to its centre-of-mass energy being limited; yet, the sensitivity to the $Z'$ through non-resonance $e^+e^- \to \mu^+\mu^-$ signals, for example, can far exceed the limits of the LHC.

Given the current sensitivities and limits of the LHC to strongly coupled physics and to new resonance states, we focus our discussion in Chapter 6 on supersymmetry and $Z'$ analyses. Supersymmetry is an excellent example theory to explore CLIC's capacity to measure new electroweak states. It is a highly motivated theory of physics beyond the SM for a wide range of reasons, and it is a complete, calculable and rich theory of electroweak-scale physics by which to formulate benchmark processes in order to investigate the collider's capabilities in a well-understood way. The results will be largely transferable to any other electroweak theory nature may choose.

To make the study concrete, and to provide an illustration of the physics potential of a staged CLIC machine, we have produced an example supersymmetric model (*model III*) [11]. This scenario is consistent with all known constraints on supersymmetry, and furthermore has a lightest supersymmetric particle with relic abundance computed to be what is needed to be the cold dark matter of the universe. The mass spectrum of this model in units of GeV is

$$
\begin{aligned}
\text{Neutralinos } (\tilde{\chi}^0_{1,2,3,4}): &\quad 357, 487, 904, 911 \\
\text{Charginos } (\tilde{\chi}^\pm_{1,2}): &\quad 487, 911 \\
\text{Sleptons } (\tilde{e}_R,\ \tilde{e}_L,\ \tilde{\nu}_e): &\quad 559, 650, 644 \\
(\tilde{\tau}_1,\ \tilde{\tau}_2,\ \tilde{\nu}_\tau): &\quad 517, 642, 630 \\
\text{Gluino } (\tilde{g}): &\quad 1114 \\
\text{Squarks } (\tilde{t}_1,\ \tilde{t}_2,\ \tilde{b}_1,\ \tilde{b}_2): &\quad 844, 1120, 1078, 1191 \\
(\tilde{d}_R,\ \tilde{u}_R,\ \tilde{d}_L,\ \tilde{u}_L): &\quad 2167, 2181, 2197, 2196 \\
\text{Higgs Bosons } (h^0,\ A^0,\ H^0,\ H^\pm): &\quad 118, 765, 765, 769
\end{aligned}
$$

This spectrum is close in spirit to the simplified mSUGRA models but with non-uniform squark masses. This spectrum was chosen for our benchmark studies before the 125 GeV boson discovery was made. With small alterations a spectrum with the same qualitative features can be achieved. Here, second generation sfermions have the same mass as the first generation sfermions. The value of $\tan\beta$ is 10.

In order to have an overview of the various pair-production cross-sections in the theory, we show in Figure 2.2 the cross-sections as a function of $e^+e^-$ centre-of-mass energy. These cross-sections generally range from a fraction of a fb to tens of fb. Therefore, in order to achieve the $< 1\%$ accuracies necessary to discover and study supersymmetry well at CLIC, we need at least hundreds of fb$^{-1}$ of integrated luminosity for the full panoply of possible measurements.

## 2.4 Energy Staging

As summarised in Chapter 3, the CLIC technology has been shown to be flexible and robust enough to offer the possibility of building towards its final energy in several stages. There is flexibility in the energy choices for these stages, but one possibility discussed in this report is to have a first energy stage at $\sqrt{s} = 500$ GeV. This stage, which includes the possibility to tune to lower centre-of-mass energies, enables a guaranteed physics case of top quark physics and precision measurements of the Higgs boson sector. The Higgs boson mass and several coupling measurements can be made there. In addition, it will provide the ability to perform scans around the $t\bar{t}$ production threshold, devoting about 100 fb$^{-1}$ to measure the top mass with $< 100$ MeV precision.

The physics potential of the remaining CLIC energy stages is more speculative. Generically, a theory of physics beyond the SM will have multiple mass scales by which the exotic states are organised, and which are tied by $\mathcal{O}(1)$ couplings to an overall mass scale of the theory. One hopes that the results from the LHC can shed light on the various appropriate CLIC energy stages. Prior to the data, one approach to considering staging is to assume that all New Physics particles not charged under the strong interaction will have a different mass than particles that are. This is due to $g_s$ being the largest known coupling constant, and quantum renormalisation effects generally will separate strongly interacting particle masses from electroweak interacting masses. This is a qualitatively true statement for the SM fermions, and it is well-known to be the case in other frameworks beyond the SM. Thus, we can state that the two other stages are an "electroweak stage" and a "strong stage". There are other ways in which nature may order the mass thresholds, for example a "chiral stage" (new chirally interacting fermions) and "non-chiral stage" (vector-like fermions). There are numerous other possibilities. Whatever is the underlying principle that gives mass gaps in the New Physics sector, it is not a priori known what precise energies are needed in order to separate out the study of these scales.

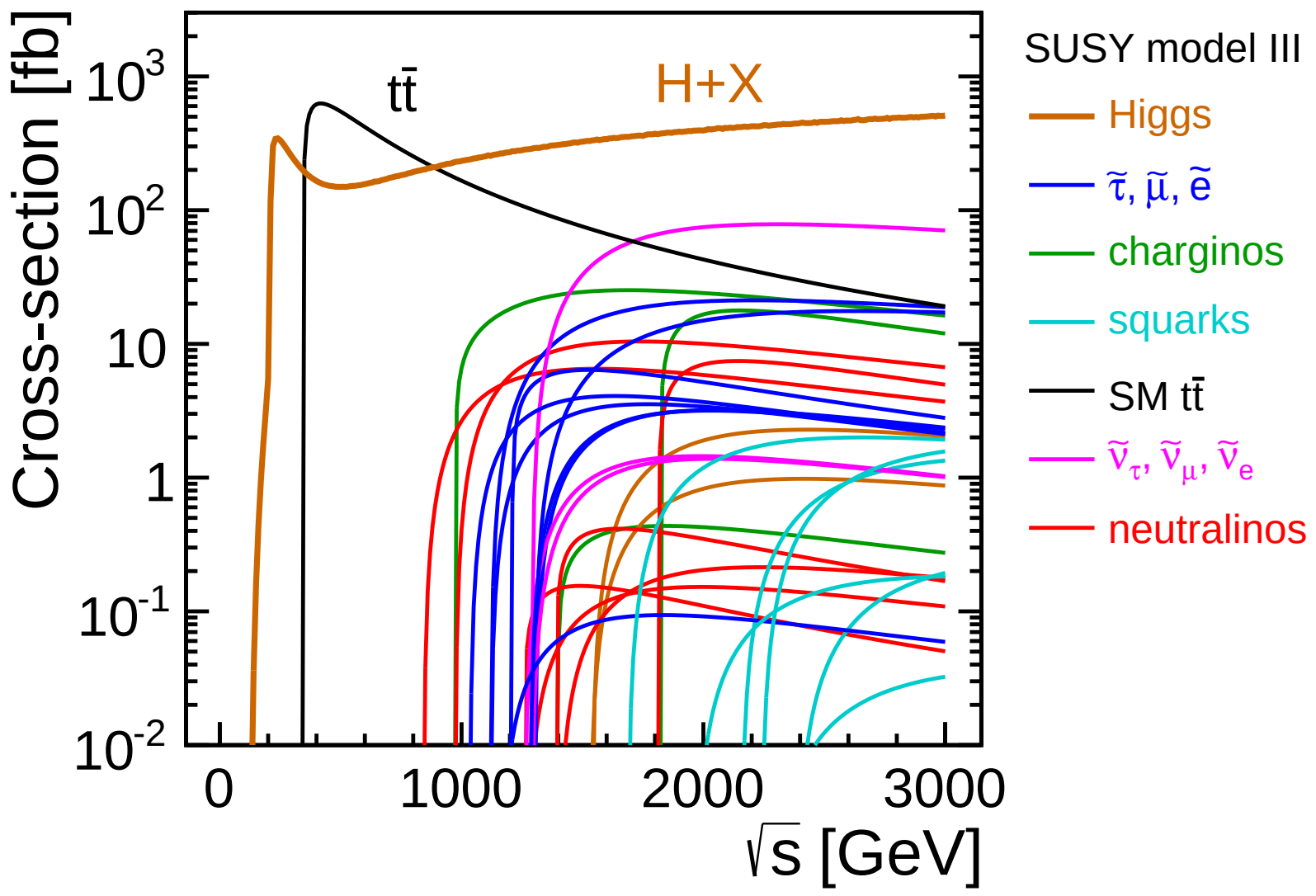

Fig. 2.2: Supersymmetry cross-sections for pairs of superpartners in *model III* as a function of $e^+e^-$ centre-of-mass energy. The $H+X$ cross-section is for a SM Higgs boson with mass 125 GeV. The lightest Higgs boson of supersymmetry is expected to have couplings very close to those of the SM Higgs boson.

Keeping these challenges in mind, let us nevertheless for illustration discuss the case of an "electroweak stage" and a "strong stage" in the context of our example supersymmetry *model III*, discussed in the previous section. For example, in stage 1 with energy 500 GeV, adjustable down to e.g. $\sim 2 \cdot m_t$, one can study the Higgs boson and top quark properties with high precision. This forms the core of the physics case for a low-energy linear collider. In stage 2 at 1.4 (1.5) TeV we can repeat many of the measurements. This repetition is useful because at higher energies cross-section ratios are different and the extraction of mass parameters, mixings and widths in a different physics background environment is an excellent cross-check on the results obtained from the lower energy. In addition, new processes open up. In the Higgs analysis the $t\bar{t}H$ study is now possible with reasonable event rates and the Higgs self-coupling can be measured. Furthermore, in the example supersymmetry model, there are many electroweak states that become kinematically accessible at this higher energy. These particles and cross-sections can then be measured with less of the supersymmetry background than a higher energy collider would have to face when many more new particles are kinematically accessible. Ultimately, in stage 3 at 3 TeV one can repeat all the measurements of stage 2, which again will cross-check results in a different background environment with different observables at higher energy. The triple Higgs coupling can now be measured with higher precision. In addition, more states become kinematically accessible. In particu-lar, we have the higgsinos and the heavy Higgs bosons and also third generation squarks consistent with a "strong phase". Although stage 3 is chosen here to be the maximum deliverable energy anticipated by a CLIC machine, it is important to note that the current limits at the LHC, and the discovery sensitivity anticipated for the future, retain for CLIC the prospect of many new particles and direct discoveries.

## 2.5 Conclusions

In summary, the overview of the physics situation is that the SM is holding up well to the recent high-energy data. Recently a Higgs-like boson has been discovered with mass 125 GeV. In Chapter 6 we show why CLIC would be an exceptional machine to test the precise nature of this Higgs boson, to determine if it is the SM Higgs or whether it has non-standard properties that could signify new dynamics.

Despite the fact that the LHC did not yet see any indicators for physics beyond the SM, there are nevertheless compelling reasons to believe that the SM is not complete and New Physics will have to show up at some scale. Planning to reach as far as possible into the energy frontier to search for new phenomena, at mass scales that are not already ruled out by the LHC, in addition to securing the physics case through accurate measurements of top quark and Higgs boson properties, are key elements to the physics potential of a linear collider. CLIC's capacity to do this will be demonstrated in Chapter 6.

# Chapter 3

# CLIC Accelerator Technology

The aim of the CLIC study has been to develop a technology that can be used to build a multi-TeV linear electron-positron collider. The study therefore concentrated on a 3 TeV design and demonstrated the feasibility of the technology, as documented in [1]. A design for 500 GeV has also been developed, although in less detail. After a short summary of the design, the following sections describe two examples of a staged approach to CLIC, the status of the feasibility studies and the energy flexibility of each energy stage of CLIC after construction.

## 3.1 The CLIC Design at 3 TeV

The conceptual layout of CLIC is shown in Figure 3.1 and the fundamental parameters are given in Table 3.1. These parameters are the result of a full cost optimisation, see Chapter 2.1 in [1].

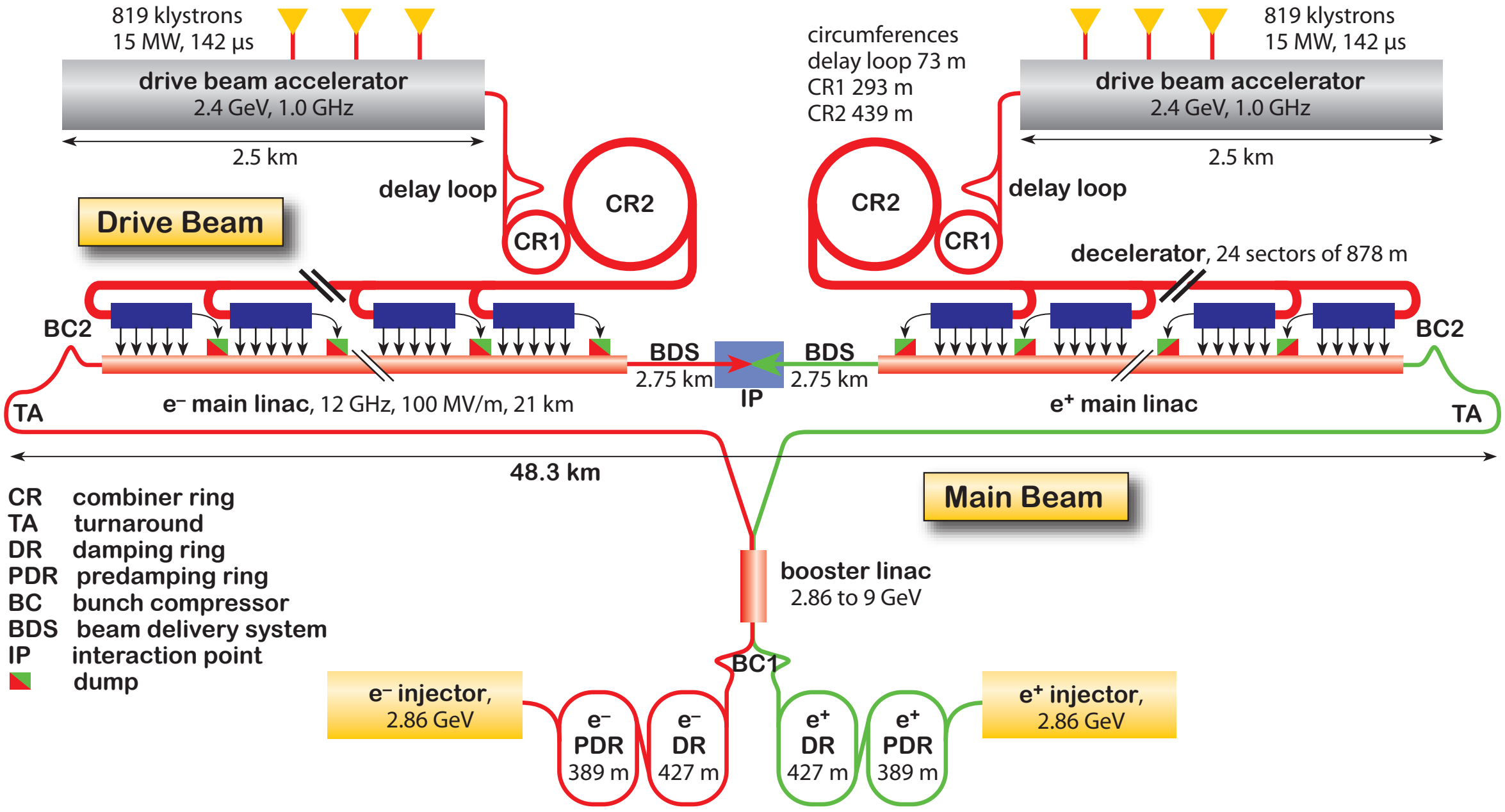

Fig. 3.1: Overview of the CLIC layout at $\sqrt{s} = 3$ TeV.

The main (colliding) beams are produced in conventional electron and positron sources and accelerated to 2.86 GeV. The beam emittances are reduced in a pre-damping ring followed by a damping ring. In the ring-to-main-linac transport system the beams are compressed longitudinally and accelerated to 9 GeV. The main linac uses 100 MV/m 12 GHz accelerating structures to achieve the final beam energy. In the Beam Delivery System (BDS) the beam is cleaned by collimation and compressed to the very small size at collision. The main challenge for the CLIC main beam is to achieve the main linac accelerating gradient (i.e. the high beam energy) and the good beam quality (i.e. the high luminosity). An additional challenge arises from the strong beam-beam interaction and is discussed in Chapter 4. During the collision particles will emit beamstrahlung, which reduces their energy and leads to the development of a luminosity spectrum. We therefore quote the total luminosity as well as the luminosity above 99% of the nominal centre-of-mass energy. The design foresees 80% polarisation of the electrons at collision, and it is compatible with the addition of a polarised positron source.

The necessary RF power for the main linac is extracted from a high-current, low-energy drive beam that runs parallel to the colliding beam through a sequence of power extraction and transfer structures (PETS). In these structures the drive beam generates RF power that is transferred via waveguides into the main linac accelerating structures.

Table 3.1: Key parameters of the 3 TeV and 500 GeV designs.

| Parameter | Symbol | Unit | 500 GeV | 3 TeV |
|---|---|---|---|---|
| Centre-of-mass energy | $\sqrt{s}$ | TeV | 0.5 | 3.0 |
| Repetition frequency | $f_{rep}$ | Hz | 50 | 50 |
| Number of bunches per train | $n_b$ | | 354 | 312 |
| Bunch separation | $\Delta t$ | ns | 0.5 | 0.5 |
| Accelerating gradient | $G$ | MV/m | 80 | 100 |
| Total luminosity | $\mathscr{L}_{total}$ | $10^{34}$cm$^{-2}$s$^{-1}$ | 2.3 | 5.9 |
| Luminosity above 99% of $\sqrt{s}$ | $\mathscr{L}_{0.01}$ | $10^{34}$cm$^{-2}$s$^{-1}$ | 1.4 | 2.0 |
| Number of photons per electron/positron | $n_\gamma$ | | 1.3 | 2.1 |
| Average energy loss due to beamstrahlung | $\Delta E/E$ | | 0.07 | 0.28 |
| Number of coherent pairs per bunch crossing | $N_{coh}$ | | $2\times10^{-2}$ | $6.8\times10^{8}$ |
| Energy of coherent pairs per bunch crossing | $E_{coh}$ | TeV | 15 | $2.1\times10^{8}$ |
| Number of incoherent pairs per bunch crossing | $n_{incoh}$ | $10^6$ | 0.08 | 0.3 |
| Energy of incoherent pairs per bunch crossing | $E_{incoh}$ | $10^6$ GeV | 0.36 | 23 |
| Hadronic events per bunch crossing | $n_{had}$ | | 0.3 | 3.2 |

The drive beam is generated in a central complex with a fundamental RF frequency of 1 GHz. The injector produces a 140 µs-long electron beam pulse. Every second RF bucket is filled, i.e. the bunch spacing is 60 cm. Every 244 ns the injector switches from filling odd to filling even buckets and vice versa, creating 244 ns-long sub-pulses. The drive beam accelerator (DBA) accelerates this beam to about 2.4 GeV with an RF to beam efficiency of 97%. An 0.5 GHz RF deflector then sends the sub-pulses filling even buckets into a delay loop, so that the bunches of each can be interleaved with those of the next sub-pulse, which fills odd buckets. This produces a sequence of 244 ns-long sub-pulses spaced by 244 ns-long gaps and with 30 cm bunch spacing. In a similar fashion, three of these sub-pulses are merged in a first combiner ring and four of the new sub-pulses in a second ring. Thus each final sub-pulse has 24-times the initial current and only 2.5 cm bunch spacing. Each of the generated 24 sub-pulses will feed one drive beam decelerator in the main linac. This drive beam scheme allows a total effective compression of the drive beam power by a factor 576, i.e. from 140 µs to 244 ns.

## 3.2 The CLIC Design at 500 GeV

The design parameters for a 500 GeV CLIC machine have been investigated and are based on the 3 TeV design, see Table 3.1. The design, which is shown in Figure 3.2, respects a number of basic boundary conditions in order to facilitate the upgrade from 500 GeV to 3 TeV but no full integration has been done. The main linac components at 500 GeV are the same as at 3 TeV and can be re-used, except the accelerating structures, which have the same length and almost the same input power but a larger aperture and lower gradient than the 3 TeV structures. This allows a larger bunch charge and slightly more bunches per pulse, which increases the luminosity for the 500 GeV machine. The main beam generation complex at 500 GeV is the same as for 3 TeV, but more installed klystrons are required due to the larger beam current. In contrast to the 3 TeV case only one drive beam complex is required, which produces the drive beam pulses for both linacs. The design of this complex is exactly the same as for one of the two required for 3 TeV, but a slightly higher RF power and drive beam current is needed.

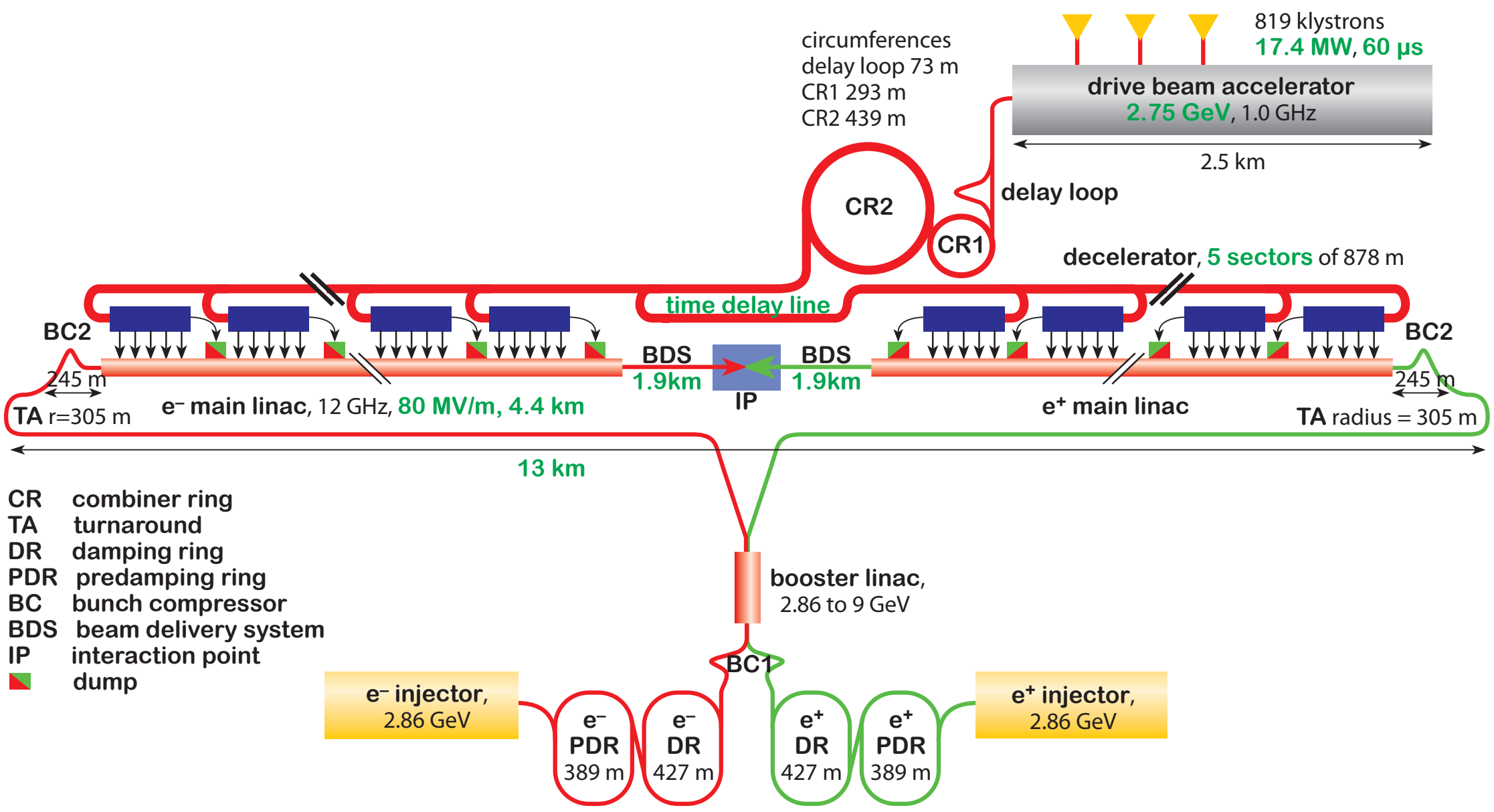

Fig. 3.2: Overview of the CLIC layout at $\sqrt{s}$ =500 GeV.

At energies of 500 GeV or less one can also consider to base CLIC on the use of klystrons to power the main linacs. A first exploration of the resulting parameters has been carried out [2] in preparation of a more detailed analysis.

## 3.3 CLIC Feasibility

The following four areas have been identified as the most critical for the CLIC machine design:

- The ability to achieve the high main linac gradient of 100 MV/m;
- The generation of the drive beam, the production of RF power from the drive beam and the stable deceleration of the drive beam to extract efficiently the energy, as well as the use of the power to accelerate the main beam;
- The generation of the ultra-low emittances of the main beam in the damping rings and the preservation of the emittance during the beam transport and acceleration in the main linac, which requires pre-alignment with unprecedented accuracy and active magnet stabilisation against mechanical vibrations;
- The ability to protect the machine against damage while still providing a high availability.

The experimental programme that addressed these issues has demonstrated the feasibility of CLIC, as detailed below.

### 3.3.1 Accelerating Structure and Main Linac Gradient

Each main linac contains about 70000 accelerating structures with a length of 23 cm. The total ratio of active length to linac length is almost 80%. The structure design has been carefully optimised using empirical constraints to achieve a gradient of 100 MV/m. The main limitation arises from so-called breakdowns, i.e. sparks that can occur in the structure during the RF pulse, which can give transverse kicks to the beam. The breakdown probability depends strongly on the gradient and also on the pulse length as well as on the structure design and material. We conservatively assume that a single breakdown in any main linac structure renders the beam pulse useless for luminosity. This should happen only in 1% of the beam pulses at the target gradient of 100 MV/m, which results in a target rate of $3 \times 10^{-7}$ events/(m $\times$ pulse).

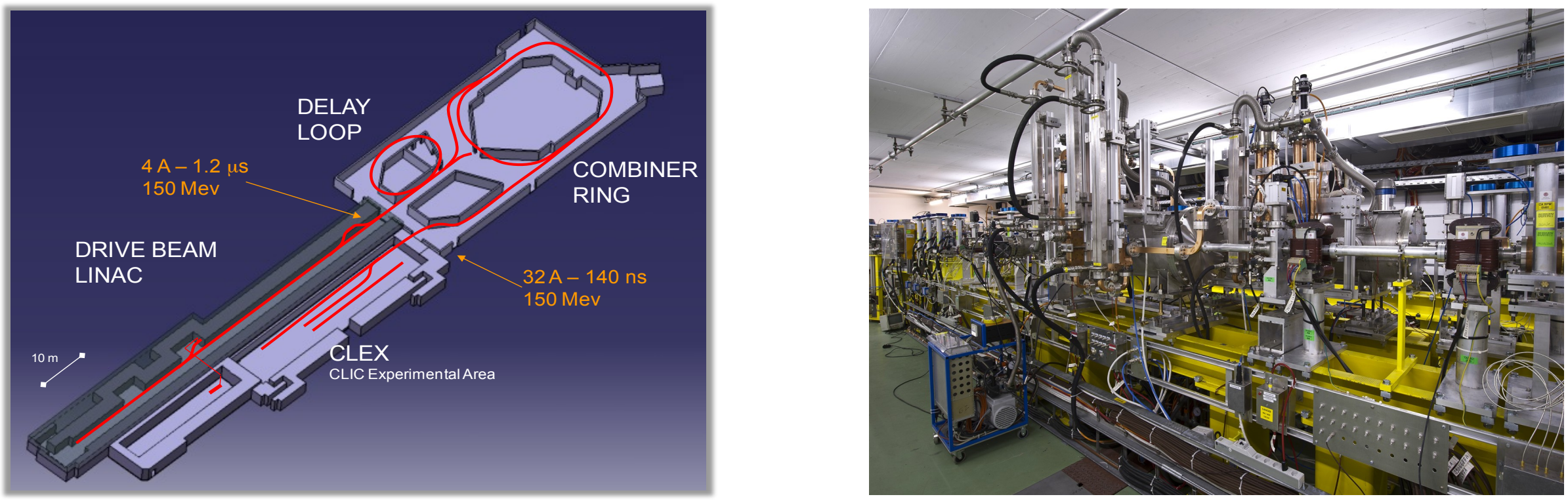

Fig. 3.3: Left: Layout of CTF3. Right: The two-beam test stand in CTF3.

Table 3.2: Typical design parameters of the CLIC and CTF3 drive beam.

| Parameter | Unit | CLIC | CTF3 |
|---|---|---|---|
| Accelerated current | A | 4.2 | 3.5 |
| Combined current | A | 101 | 28 |
| Accelerated pulse length | µs | 140 | 1.6 |
| Final pulse length | ns | 240 | 140 |
| Acceleration frequency | GHz | 1 | 3 |
| Final bunch frequency | GHz | 12 | 12 |

To verify the gradient, different structures have been tested at SLAC and KEK using klystrons. Prototypes of the CLIC structure achieved an unloaded gradient of 100 MV/m for pulse lengths longer than in CLIC.

### 3.3.2 The Two-Beam Scheme

The main challenges of the two-beam scheme are:

- The stable production of the drive beam;
- The RF power production. The design of the power extraction and transfer structures (PETS) in which the drive beam generates the RF power. The stable deceleration of the drive beam to extract a maximum of the beam power into the RF;
- The RF power production of the RF beam and the acceleration of the main beam with this power.

#### *3.3.2.1 Generation of the Drive Beam*

To demonstrate the two-beam scheme, CTF3 (CLIC Test Facility 3) has been constructed and commissioned at CERN; the layout is shown in Figure 3.3 (left) and the fundamental parameters in Table 3.2. CTF3 consists of a drive beam source, the drive beam accelerator operating at 3 GHz, the delay loop and one combiner ring. This allows to increase the initial beam current by a factor eight. The produced drive beam can be used in the two-beam test stand (TBTS), shown in Figure 3.3 (right), which also includes a probe beam that simulates the CLIC main beam. Alternatively it can be sent into the test beam line (TBL) which is a small decelerator.

The drive beam accelerator of CTF3 accelerates routinely a current of above 3.5 A. It has shown full beam loading, in which case 95% of the RF that is coupled into the accelerating structure is trans-mitted to the beam. Using the delay loop and the combiner ring the beam combination by a factor eight has been demonstrated, yielding a current of up to 29 A, slightly in excess of the nominal value.

In addition, theoretical tolerance studies have been performed for the CLIC drive beam source and the drive beam accelerator, which show that the required performance can be reached.

#### 3.3.2.2 RF Power Production

*The Power Extraction and Transfer Structures (PETS)*

The PETS design has been based on the same scaling laws as the main linac accelerating structures and provides the RF power for two accelerating structures. A mechanism has been developed to control the output power of the PETS. This allows to reduce the power into a bad accelerating structure without compromising the other structures that are powered by the same drive beam. A klystron-driven test of the PETS has established the feasibility by showing operation at an output power of 147 MW for a pulse length of 266 ns and a breakdown rate less than $3\times10^{-7}$ events/(m × pulse).

*Drive Beam Deceleration*

The CLIC decelerator will decelerate the beam from 2.4 GeV to 0.24 GeV. It is mandatory to achieve very small losses and to avoid any instability, which is challenging due to the 100 A beam current and the large energy spread. Simulations of the decelerator have been performed to study the drive beam stability and the impact of static and dynamic imperfections. They show that the beam remains stable even if the wake field damping is less efficient than expected and that the alignment tolerances are less stringent than for the main linac.

The concept has been verified in the TBL at CTF3; this line currently contains nine PETS and more will be installed in 2012. A 21 A beam has been decelerated by 26% (corresponding to the expectation) with no measurable losses. After installation of all PETS deceleration to about 50% will be possible.

#### 3.3.2.3 The Two-Beam Acceleration

The two-beam acceleration has been demonstrated in the TBTS, which currently consists mainly of one PETS, one accelerating structure and the necessary instrumentation. A full two-beam module will be installed later, followed by a string of modules. The CTF3 drive beam generates power in the PETS and a test beam can be sent through the accelerating structure. Since the drive beam current is lower in CTF3 than in CLIC recirculation is used. A part of the output power of the PETS is injected at the PETS entrance, which seeds the produced RF and increases the output power at the cost of a reduced pulse length at full power.

Gradients up to 145 MV/m have been achieved in the TBTS. The deceleration of the drive beam, the RF power measured and the probe beam acceleration are all consistent, also with the theoretical predictions.

### 3.3.3 Generation and Preservation of Ultra-low Emittances

The main effects that impact on the CLIC ultra-low emittance generation and preservation and therefore on the achievement of high luminosities have been addressed:

- The damping ring to achieve the emittances;
- The design of the beam transport system from the damping ring to the interaction point, which is required to achieve the very small spot size at collision;
- The emittance degradation due to static imperfections in the main linac and beam delivery system, where the main issue is the accuracy of the pre-alignment of the beam line components;
- The luminosity loss due to dynamic imperfections in the main linac and beam delivery system. Important sources of imperfections are ground motion, which is mitigated using active stabilisation of the magnets, and fluctuations of the drive beam intensity and phase.

#### *3.3.3.1 Damping Ring Design and Low Emittance Generation*

A conceptual design of the CLIC damping ring exists that, according to extensive simulation studies, meets the CLIC specifications. Existing third generation light sources and damping ring test facilities have normalised emittances that are not too far from the CLIC damping ring goal. In particular, the Swiss Light Source (SLS) has achieved a normalised vertical emittance slightly better than the CLIC target [3] and the Accelerator Test Facility (ATF) at KEK reached a value that is only 2.5 times larger than the CLIC goal [4]. Both have single bunch charges comparable to CLIC. The horizontal emittances in these machines are larger than the CLIC goal (e.g. 26000 nm at SLS and 3800 nm at ATF vs. 500 nm for CLIC). This limitation is overcome in the CLIC design by using strong wigglers, which achieve a very fast damping and allow small equilibrium emittance also in the horizontal plane. Two planned light sources, MAX-IV [5] in Lund and PEP-X [6] at SLAC, aim for horizontal emittances of 1500 nm and 100 nm, which are close to or beyond the CLIC design.

#### *3.3.3.2 Transport Lattice Design and Nanometre Beam Sizes*

Designs exist for the different beam lines from the damping ring to the interaction point; a particular challenge has been the optics for the beam delivery system. Simulations show that these designs would allow to reach 250% of the CLIC luminosity goal in the absence of imperfections.

#### *3.3.3.3 Static Imperfections and Pre-Alignment*

The most important static imperfections are due to misalignments of the beam position monitors and the accelerating structures. The beam position monitors will be used to define the beam trajectory. Misalignments will result in unwanted residual dispersion, leading to an increase of the beam emittance. Similarly, offsets of the accelerating structures will lead to the generation of parasitic transverse fields, so-called wake fields, which kick the tails of the bunches, thus generating emittance growth.

To ensure excellent alignment of the main linac components, they are mounted on girders equipped with movers, which can be remotely controlled, and sensors, which measure their position with respect to a reference system of overlapping wires, similar to that used in the LHC insertions. A reference system has been developed and built for CLIC, the first prototype showing an RMS accuracy of 14 µm, which is already very close to the target of 10 µm and would lead to very little reduction of the luminosity. The other system components have been developed for the test module and will be tested soon in CTF3. This technology will be used to pre-align the components after installation. Then orbit measurements at different beam energies will be used to measure and improve the effective alignment of the beam position monitors and magnets. Finally, the generation of wake fields in the accelerating structures is minimised by measuring the beam offset in the structures using novel wake monitors. Simulations show that with these procedures the emittance growths stay within specifications.

In the beam delivery a similar pre-alignment procedure is foreseen, with an accuracy of 10 µm for all components. The beam-based alignment and tuning is somewhat more complex due to the sophisticated nature of the system design. However, simulations show that one can achieve the target luminosity. In 65% of the simulated cases this can be achieved without another access to adjust the pre-alignment. With further improvements of the algorithms it is expected to further increase this probability. The beam-based tuning is currently under test in ATF2.

#### *3.3.3.4 Dynamic Imperfections and Stabilisation*

The main beam is very sensitive to magnet motions, in particular in the main linac and BDS. An important source is ground motion, which is site specific. Technical systems can also induce vibrations, but those can be mitigated by the design of the technical component itself. Hence they will be addressed in the technical design phase. As a conservative benchmark, we use a ground motion model that is based on measurements on the floor of the CMS experimental hall, which is significantly more noisy than the LEP tunnel.

The main linac and BDS magnets are equipped with active stabilisation systems, which use motion sensors and piezo-electric actuators controlled by a local feedback/feed-forward system. A prototype system has been developed and the transfer of the ground motion to the magnet has been measured and compared to simulations with reasonable agreement. Based on the results of the first simulation studies, an improved system concept has also been developed. The final quadrupoles, which are most sensitive to motion, are mounted on a pre-isolator, consisting of a large concrete block that is supported by air springs.

The expected impact on luminosity is calculated using a simulation code that models the ground motion, the transfer through the stabilisation system and the beam-based feedback. The nominal luminosity includes a 20% safety margin for dynamic imperfections in the main linac and BDS. With the existing prototype system 13% luminosity would be lost, using most of this margin. With the new system concept only 3% would be lost.

Fluctuations of the drive beam phase or current change the CLIC main beam acceleration and lead to luminosity loss since the energy bandwidth of the beam delivery system is limited. This places tight tolerances on the drive beam current stability ($7 \times 10^{-4}$) and on the phase and power stability of the drive beam accelerator klystrons (0.05° and 0.2%) in order to limit the luminosity loss to 1%. Measurements in the CTF3 drive beam accelerator show that the beam current has a stability of $5 \times 10^{-4}$ and a good klystron has a phase stability of 0.07° and a power stability of 0.21%, very close to the CLIC requirements.

### 3.3.4 Machine Protection

The machine protection for CLIC has to cope with a wide variety of failures. Based on the LHC experience, a strategy has been developed to ensure protection:

- Slow errors and drifts that develop over several pulses will be detected by a post-pulse analysis;
- Faster failures that can develop between pulses are addressed by
  - an interlock system that detects the equipment failure directly;
  - failures that occur immediately before the beam pulse and cannot be caught by the interlock system are avoided by a "safe by design" design of the components. In such a design the inertia of the component is large and slows down the failure.
- Against even faster failures the machine is protected by masks and other passive protection.

Also a fast intensity ramp has been developed for the drive and main beam.

## 3.4 Energy Flexibility of a 3 TeV and 500 GeV Machine after Construction

It will be necessary to operate CLIC at lower than nominal energies to study resonances or thresholds. Typically these studies would require running at a few energies that are close together but well below the nominal centre-of-mass energy. This requires the capability to significantly change the operation energy once in a while and to change more frequently the energy in small steps close to the new energy.

The baseline option to reduce the collision energy significantly consists in a reduction of the drive beam current. This will reduce the main linac gradient proportionally, which in turn reduces collision energy. In order to ensure the main beam stability, the main beam bunch charge will have to be reduced proportionally to the gradient for larger changes. This will reduce the luminosity per collision significantly. This luminosity loss can be partially mitigated by an intelligent use of the drive beam generation complex. Using an appropriate switching pattern in the drive beam, one can reduce the current of the drive beam trains in the decelerators by reducing the number of bunches per unit time. This method allows to increase the drive beam train length in proportion – the total number of bunches per train is constant. This allows to increase the main beam pulse length, which increases the luminosity. The CLIC main beam injection system is compatible with this mode of operation. In addition one has to adjust the magnet strengths in the main linac and the beam delivery system proportionally to the new beam energy, which does not pose a problem since all magnets have the required flexibility. It may be beneficial to replace the final quadrupoles with quadrupoles that have a larger aperture, which increases the collimation aperture and may simplify the operation, but this is not mandatory. A retuning of the machine is also required to optimise the luminosity. Figure 3.4 shows the luminosity for the 3 TeV design at different energies. The energies at which the pulse length can be increased are clearly visible. The peak luminosity is approximately proportional to the centre-of-mass energy.

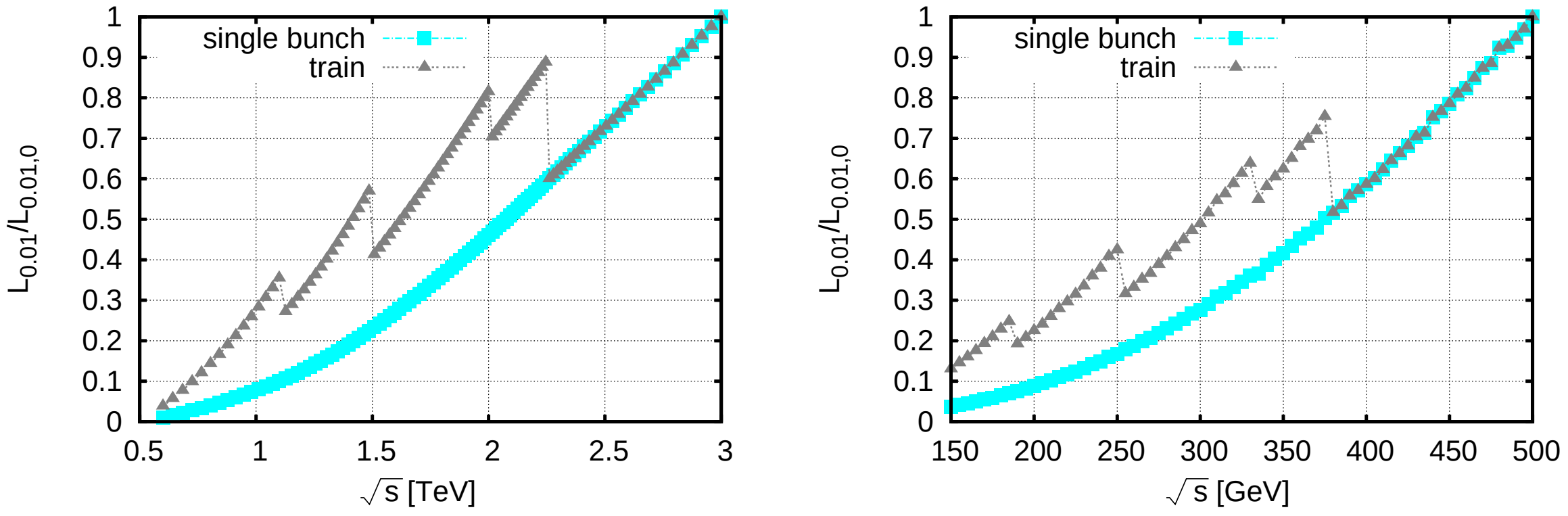

Fig. 3.4: Luminosity in the peak at different centre-of-mass energies, for the 3 TeV (left) and 500 GeV (right) designs. The luminosity is normalised to the value at full energy.

At a given operation energy, small changes of the energy can be achieved by reducing the drive beam bunch charge. It is also possible to vary the relative phases of the drive beam sectors to reduce the effective main beam acceleration. In both cases the beam delivery system magnets have to be reduced in strength proportionally with the beam energy, which poses no problem, and some tuning is required to optimise the luminosity. In this fashion a reduction of energy up to a few percent can be implemented rapidly and the luminosity would be proportional to the energy. These steps require much smaller modification of the operation compared to the large energy change, and could be carried out rather rapidly, smoothly transiting from one energy to the next.

It should be noted that the unloaded gradient of the CLIC accelerating structures is 20% higher than the loaded gradient. It is therefore possible to increase the energy reach by up to 20% by compromising the luminosity.

## 3.5 Energy Stages

In order to satisfy the physics demands to provide luminosity at very different energies, CLIC can be built in stages. The choice of stages will depend on the results of the LHC. However, in the following we show some specific scenarios to illustrate the capabilities of the CLIC technology.

Parameters for the different energy stages can be based on the existing CLIC designs at 3 TeV and 500 GeV by shortening the main linac to change the final beam energy. All other beam parameters are kept constant. This allows to use the same main beam generation complex before the main linac as for the original energy, only the beam delivery system has to be re-designed to accommodate the lower energy beam. Down to about 1 TeV, the beam delivery system length will be the same as for 3 TeV. We use the same lattice design, only the magnets will be scaled in strength, since this allows most easily for an upgrade. In this scheme, the turn-arounds of the drive beam sectors of the lower energy machine are at the same location as those of the 3 TeV machine. At 500 GeV, the beam delivery system is about 900 m shorter. The different dynamic and static tolerances remain either constant or can be relaxed, depending on their nature. Also the drive beam generation complex can remain almost unchanged, since the RF pulse length and power in the main linac accelerating structures is the same as in the original design. This means that the drive beam current and combination factor in the production complex remain unchanged and the design unmodified. However, the total drive beam pulse length at production can be reduced proportionally to the number of decelerators that needs to be fed in the main linac, i.e. proportionally to the length (and energy) of the main linac.

Table 3.3: Parameters for the CLIC energy stages of scenario A.

| Parameter | Symbol | Unit | | | |
|---|---|---|---|---|---|
| Centre-of-mass energy | $\sqrt{s}$ | GeV | 500 | 1400 | 3000 |
| Repetition frequency | $f_{rep}$ | Hz | 50 | 50 | 50 |
| Number of bunches per train | $n_b$ | | 354 | 312 | 312 |
| Bunch separation | $\Delta_t$ | ns | 0.5 | 0.5 | 0.5 |
| Accelerating gradient | $G$ | MV/m | 80 | 80/100 | 100 |
| Total luminosity | $\mathscr{L}$ | $10^{34}$ cm$^{-2}$s$^{-1}$ | 2.3 | 3.2 | 5.9 |
| Luminosity above 99% of $\sqrt{s}$ | $\mathscr{L}_{0.01}$ | $10^{34}$ cm$^{-2}$s$^{-1}$ | 1.4 | 1.3 | 2 |
| Main tunnel length | | km | 13.2 | 27.2 | 48.3 |
| Charge per bunch | $N$ | $10^9$ | 6.8 | 3.7 | 3.7 |
| Bunch length | $\sigma_z$ | µm | 72 | 44 | 44 |
| IP beam size | $\sigma_x/\sigma_y$ | nm | 200/2.6 | $\approx$ 60/1.5 | $\approx$ 40/1 |
| Normalised emittance (end of linac) | $\varepsilon_x/\varepsilon_y$ | nm | 2350/20 | 660/20 | 660/20 |
| Normalised emittance (IP) | $\varepsilon_x/\varepsilon_y$ | nm | 2400/25 | — | — |
| Estimated power consumption | $P_{wall}$ | MW | 272 | 364 | 589 |

We consider two staging scenarios, A and B. Scenario B is uniquely based on the 3 TeV design, while scenario A is based on a combination of those at 3 TeV and 500 GeV. Each scenario consists of three stages, the first at 500 GeV is based on the physics demands and the third at 3 TeV corresponds to the design in [1]. The second stage is chosen to have exactly half the main linac length than the final stage, resulting in 1.4 and 1.5 TeV. This choice allows to feed both main linacs from a single drive beam generation complex that does not exceed the specifications in [1].

In both scenarios, first the tunnel for the first 500 GeV stage is built and the machine installed. During the installation and operation of this stage the tunnel construction is continued to full length required for the second stage. In order to install the second stage, all modules of the first stage are moved from their initial positions to the outer end of the new main linac tunnel and the new modules are installed toward the interaction point. This is necessary since the quadrupole type and spacing at the beginning and end of the main linac are different.

### 3.5.1 Energy Staging Scenario A

The parameters of scenario A can be found in Table 3.3 and a simplified scheme of the implementation in Figure 3.5. The three stages are designed in the following way:

- The first stage is the 500 GeV design described in [1]. Its main linac consists of five drive beam decelerator sectors and has a gradient of 80 MV/m. Only one drive beam complex will be built.
- For the second stage, an additional tunnel length corresponding to eight sectors is being constructed. The tunnel of the last drive beam sector at 500 GeV is added to the beam delivery system. The existing modules are moved to the beginning of the new tunnel and seven sectors are added to the main linac, using the 100 MV/m structures from the 3 TeV design. This requires that the main beam bunch charge and number of bunches per pulse is reduced to the same level as for the 3 TeV design. Together with the re-used modules from the first stage, which will provide a gradient slightly above 80 MV/m, the final centre-of-mass energy reached is 1.4 TeV. The beam delivery system will have to be re-designed and the damping rings have to be upgraded to provide smaller horizontal emittances.

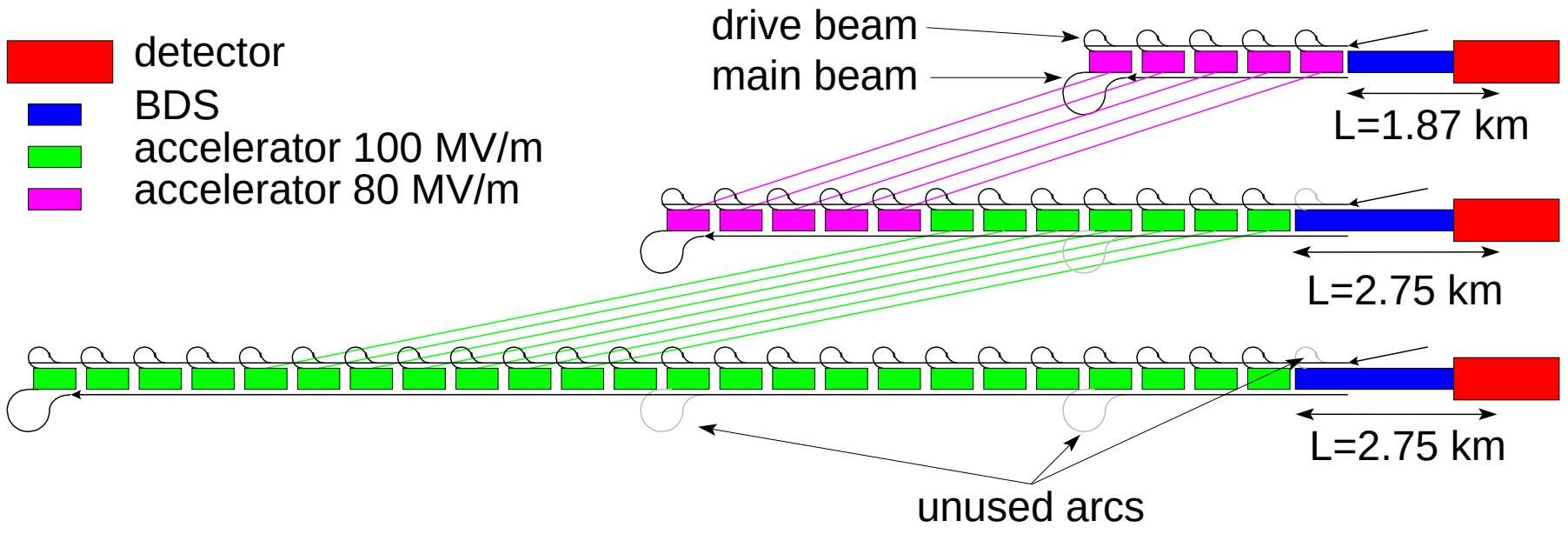

Fig. 3.5: Simplified upgrade scheme for CLIC staging scenario A. The coloured lines indicate the required movement of the modules from one stage to the next.

– In the third stage the tunnel is lengthened by twelve sectors. The modules with 80 MV/m are replaced with 100 MV/m and the existing modules are moved to a new location in the tunnel. This allows to achieve a centre-of-mass energy of 3 TeV. Alternatively one could avoid to replace the 80 MV/m structures. This reduces the cost but allows reaching only 2.9 TeV. In both cases a second drive beam generation complex has to be built to feed the additional sectors and the beam delivery system needs to be adjusted.

### 3.5.2 Energy Staging Scenario B

The parameters of scenario B can be found in Table 3.4 and a simplified scheme of the implementation in Figure 3.6. It consists of the following stages:

– The first stage uses already the structures of the 3 TeV design with a gradient of 100 MV/m. Therefore only four drive beam sectors are required for a centre-of-mass energy of 500 GeV. The damping ring will have to deliver the smaller horizontal emittance already for this stage. This stage would achieve roughly half the luminosity of the same stage in scenario A. Potential options to recover the luminosity are discussed below. Only one drive beam complex will be built.
– In the second stage the tunnel is lengthened by nine sectors. The existing modules are moved to the beginning of the tunnel and eight additional sectors are installed. This allows to reach 1.5 TeV. The beam delivery system has to be re-designed and increases in length by one drive beam sector.
– Twelve more sectors are added to the main linac in a similar fashion as for scenario A and a second drive beam generation complex is built. The beam delivery system is adjusted to the energy.

This scenario provides a more consistent sequence of stages. No structures need to be replaced and the injection complex remains unmodified from the beginning. But the luminosity at 500 GeV is smaller than for scenario A.

The luminosity in the first stage could be increased by increasing the repetition rate of the whole complex by a factor two. This will have consequences for the different active components, which need to be studied. One can also increase the luminosity by generating a longer drive beam pulse. This allows to feed the main linac more than once. In this scheme the first eight drive beam trains would feed the four sectors of each linac. About 48 µs later, the next eight trains would feed the main linac accelerating structures again. This allows to accelerate a second main beam pulse. Ultimately one could even have three main beam pulses as one can produce 24 drive beam trains. However, this scenario would require changes in the main beam generation complex to be able to produce the larger number of pulses. Also the impact of RF pulses with short spacing on the main linac accelerating structures remains to be investigated.

Table 3.4: Parameters for the CLIC energy stages of scenario B.

| Parameter | Symbol | Unit | | | |
|---|---|---|---|---|---|
| Centre-of-mass energy | $\sqrt{s}$ | GeV | 500 | 1500 | 3000 |
| Repetition frequency | $f_{rep}$ | Hz | 50 | 50 | 50 |
| Number of bunches per train | $n_b$ | | 312 | 312 | 312 |
| Bunch separation | $\Delta_t$ | ns | 0.5 | 0.5 | 0.5 |
| Accelerating gradient | $G$ | MV/m | 100 | 100 | 100 |
| Total luminosity | $\mathscr{L}$ | $10^{34}$ cm$^{-2}$s$^{-1}$ | 1.3 | 3.7 | 5.9 |
| Luminosity above 99% of $\sqrt{s}$ | $\mathscr{L}_{0.01}$ | $10^{34}$ cm$^{-2}$s$^{-1}$ | 0.7 | 1.4 | 2 |
| Main tunnel length | | km | 11.4 | 27.2 | 48.3 |
| Charge per bunch | $N$ | $10^9$ | 3.7 | 3.7 | 3.7 |
| Bunch length | $\sigma_z$ | µm | 44 | 44 | 44 |
| IP beam size | $\sigma_x/\sigma_y$ | nm | 100/2.6 | $\approx$ 60/1.5 | $\approx$ 40/1 |
| Normalised emittance (end of linac) | $\varepsilon_x/\varepsilon_y$ | nm | — | 660/20 | 660/20 |
| Normalised emittance | $\varepsilon_x/\varepsilon_y$ | nm | 660/25 | — | — |
| Estimated power consumption | $P_{wall}$ | MW | 235 | 364 | 589 |

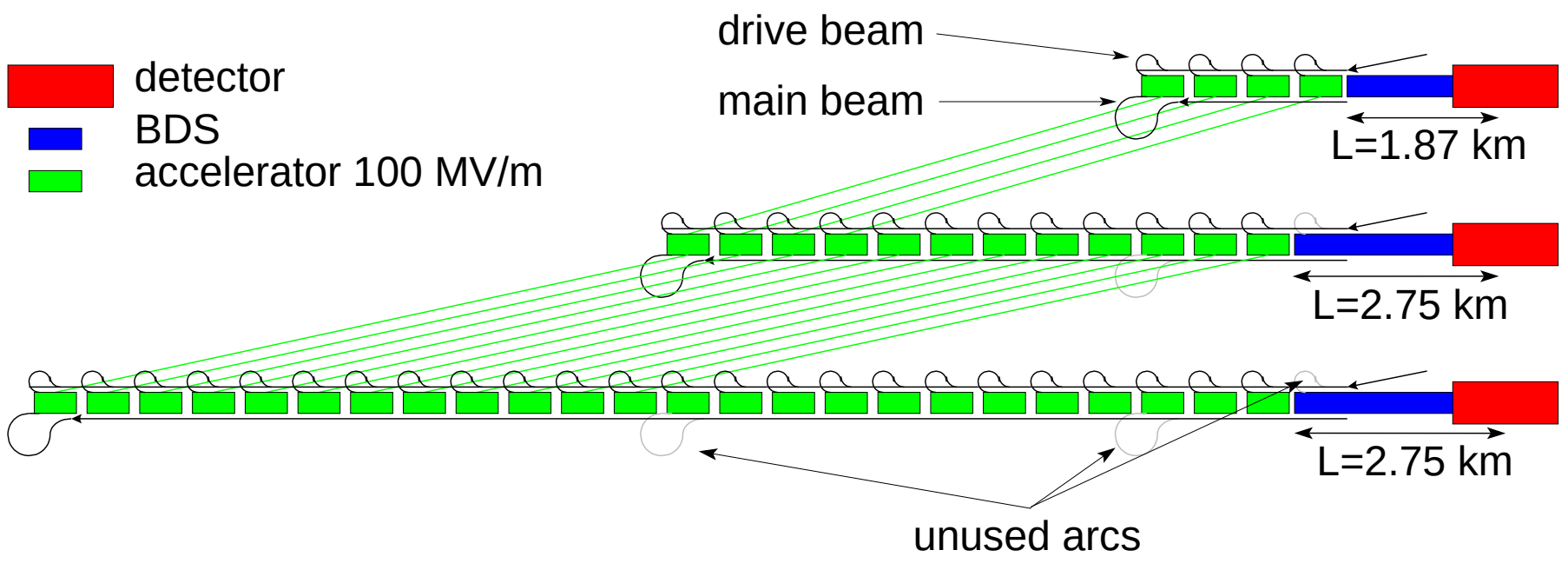

Fig. 3.6: Simplified upgrade scheme for CLIC staging scenario B.

## 3.6 Summary

Based on the conceptual design and the experimental demonstration the feasibility of a 3 TeV linear electron-positron collider has been established. The project can be implemented in energy stages, which allow to re-use most or even all of the components of one stage for the next.

# Chapter 4

# CLIC Detector Concepts

This chapter gives an overview of the experimental conditions at CLIC, followed by a description of possible detector technologies for a future experiment at CLIC. Two detector concepts are described, CLIC_ILD and CLIC_SiD. They combine high precision measurement capability with the ability to operate in the CLIC environment. They are based on the general purpose detector concepts ILD [1] and SiD [2], initially developed for the ILC. More detailed descriptions can be found in [3].

## 4.1 Detector Requirements for Physics Reconstruction

To set the detector requirements, a staged approach is assumed for CLIC, with a possible initial operation at centre-of-mass energies of a few-hundred GeV, followed by higher energy stages up to a centre-of-mass energy of 3 TeV. In view of the physics aims, described in Chapter 2, the minimum detector requirements are:

– Jet energy resolution of $\sigma_E/E \lesssim 3.5\%$ for jet energies from 100 GeV to 1 TeV ($\lesssim 5\%$ at 50 GeV);
– Track momentum resolution of $\sigma_{p_T}/p_T^2 \lesssim 2 \cdot 10^{-5}$ GeV$^{-1}$;
– Impact parameter resolution with $a \lesssim 5$ µm and $b \lesssim 15$ µm GeV, where the resolution is expressed as:

$$\sigma_{d_0}^2 = a^2 + \frac{b^2}{p^2 \sin^3\theta}, \tag{4.1}$$

– Lepton identification efficiency better than 95% over the full range of energies;
– Detector coverage for electrons down to very low angles.

As described below, the jet energy resolution requirement is the main driver of the detector concept designs. As a result both detector concepts are based on fine-grained calorimeters and particle-flow analysis techniques.

## 4.2 Experimental Environment

The time structure of the CLIC beam corresponds to 50 bunch trains per second, occurring at 20 ms time intervals. Each 156 ns long bunch train consists of 312 distinct bunch crossings separated by 0.5 ns.

One consequence of the small bunch sizes required to achieve high luminosities at CLIC is the phenomenon of the strong electromagnetic radiation (beamstrahlung) from the electron and positron bunches in the high field of the opposite beam. Beamstrahlung results in the creation of a large background of $e^+e^-$ pairs which are predominantly produced in forward directions and with low transverse momenta. Whereas numerous background particles from beamstrahlung will be created in every single bunch crossing, typically at most one hard $e^+e^-$ physics interaction will be produced per bunch train. The presence of the pair background mainly impacts the design of the inner vertex and very forward regions of the detector and results in potentially high detector occupancies in the inner layers of the vertex detector and in the forward tracking detectors.

The main source of background particles with higher transverse momenta are hadronic two-photon interactions (where the photons either can be virtual or originate from beamstrahlung). For a 500 GeV machine, there are on average 0.3 $\gamma\gamma \rightarrow$ hadrons interactions for each bunch crossing, while at 3 TeV, this number increases to 3.2 $\gamma\gamma \rightarrow$ hadrons interactions. At 3 TeV the pile-up of this background over the entire 156 ns bunch-train deposits approximately 20 TeV of energy in the calorimeters, of which about 90% occurs in the endcaps and 10% in the barrel regions of the calorimeters. The presence of the $\gamma\gamma \rightarrow$ hadrons background is a major consideration for the design of a CLIC detector and its readout.

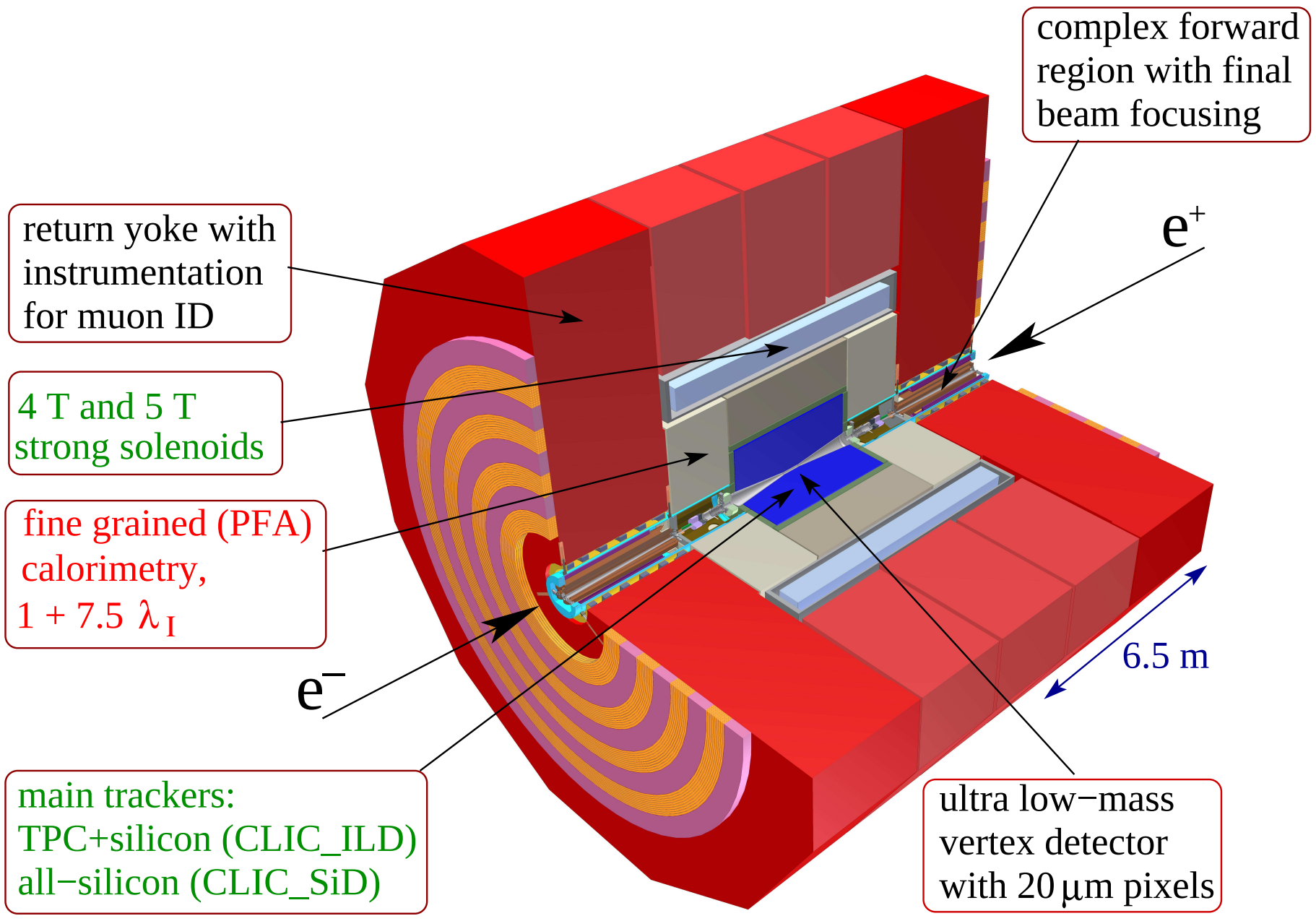

Fig. 4.1: Schematic overview of the CLIC detector concepts.

The beamstrahlung effects, which are largest at the highest centre-of-mass energy, also have a major impact on the effective luminosity spectrum, resulting in a peak at the nominal centre-of-mass energy and a long tail towards lower energies. For 3 TeV operation at a total luminosity of $5.9 \cdot 10^{34}$ cm$^{-2}$s$^{-1}$, the luminosity in the most energetic 1% fraction of the spectrum is $2.0 \cdot 10^{34}$ cm$^{-2}$s$^{-1}$. At 500 GeV the ratio between the most energetic 1% fraction of the spectrum and the total spectrum exceeds 50%. Most physics measurements at CLIC will be significantly above production threshold and will therefore profit from the major part of the total luminosity produced.

## 4.3 The Detector Concepts CLIC_ILD and CLIC_SiD

A schematic overview of a CLIC detector concept is presented in Figure 4.1, whereas longitudinal cross sections showing the major detector components of CLIC_ILD and CLIC_SiD are shown in Figure 4.2. Some key parameters of the two CLIC detector concepts are given in Table 4.1.

The **Vertex Detector** (VTX) consists of pixelated silicon detector layers integrated into the tracking system. In the case of CLIC_ILD, this detector has mainly a role of achieving optimal vertex reconstruction and flavour tagging. Therefore, as also proposed for the ILC, a geometry with three double layers was chosen, since with double layers one can reduce the material for the supports and exploit spatial correlations of hits to improve robustness against beamstrahlung background. In the CLIC_SiD case, the Vertex Detector provides additional space points for the track finding, hence the choice of 5 single vertex layers as part of the barrel tracking system with 10 layers in total. The inner radius of the beam pipes and vertex detectors is constrained by the rate of direct hits from $e^+e^-$ pair background to 31 mm for CLIC_ILD and to 27 mm for CLIC_SiD. For running at lower centre-of-mass energies, where background rates will be reduced, modified vertex-detector geometries are envisaged with smaller inner radii [4].

The CLIC vertex detector must have excellent spatial resolution, full geometrical coverage extending to low polar angles $\theta$, low occupancy, time-tagging at the 10 ns level, extremely low mass, and sufficient heat removal from sensors and readout. As none of the existing technologies is able to fulfil all these challenging goals, several options are being pursued. One approach is the so-called hybrid solution, composed of a thinned high-resistivity sensor bonded to an ultra compact and thinned readout ASIC in

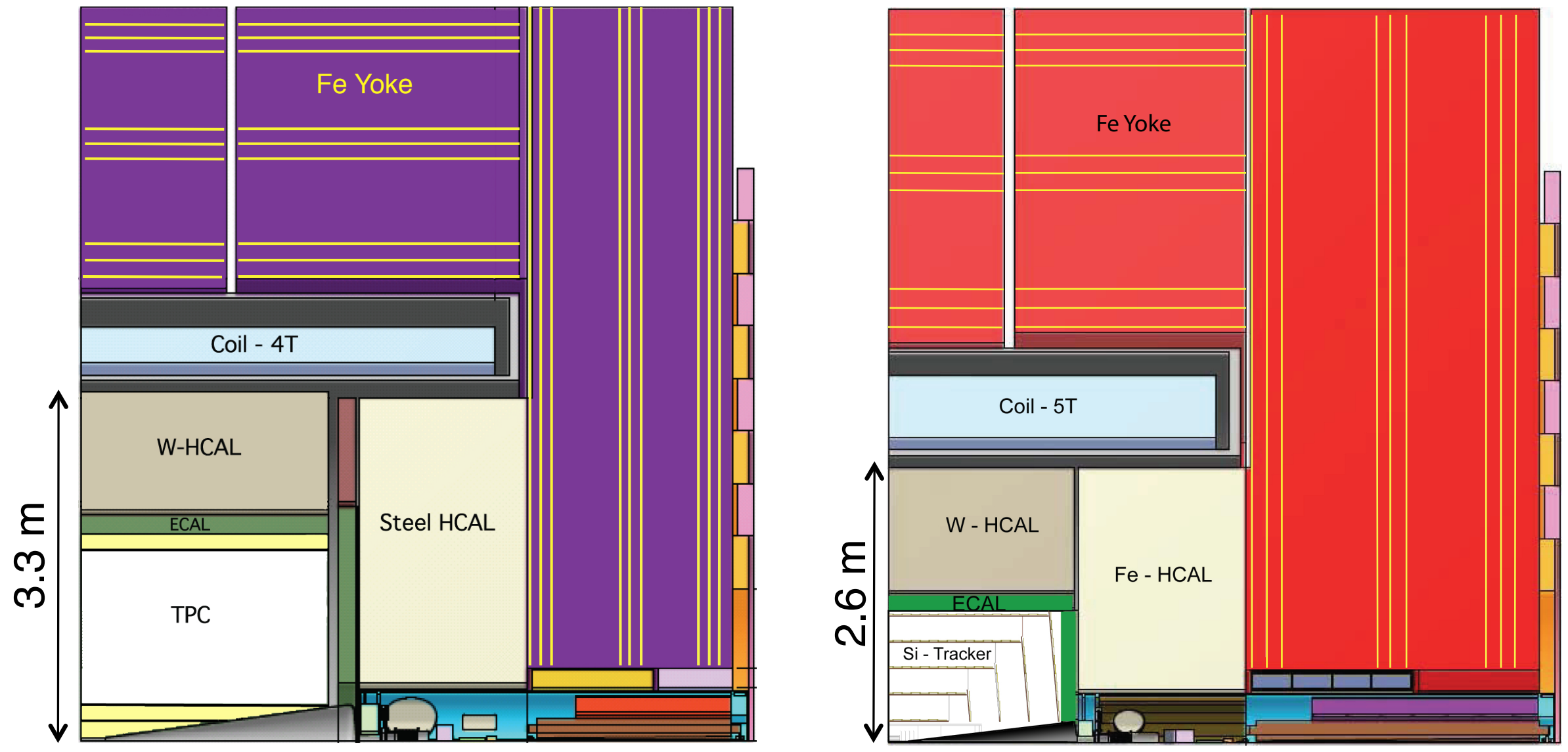

Fig. 4.2: Longitudinal cross section of the top quadrant of CLIC_ILD (left) and CLIC_SiD (right).

Table 4.1: Key parameters of the CLIC detector concepts. The inner radius of the electromagnetic calorimeter is given by the smallest distance of the calorimeter to the main detector axis.

| **Concept** | **CLIC_ILD** | | **CLIC_SiD** | |
|---|---|---|---|---|
| | Material or dimension | Technology options | Material or dimension | Technology options |
| VTX | Inner radius: 31 mm | Hybrid, integrated CMOS, SOI or 3D integrated silicon pixel technologies | Inner radius: 27 mm | Hybrid, integrated CMOS, SOI or 3D integrated silicon pixel technologies |
| Tracker | TPC/Silicon | Silicon micro-strips/pixels and TPC with MPGD readout | Silicon | Silicon micro-strips/pixels |
| ECAL | $r_{min} = 1.8$ m<br>$\Delta r = 172$ mm | Silicon, scintillator | $r_{min} = 1.3$ m<br>$\Delta r = 135$ mm | Silicon |
| HCAL | Absorber<br>barrel: W<br>endcap: Fe<br>7.5 $\lambda_I$ | Scintillator, glass RPC | Absorber<br>barrel: W<br>endcap: Fe<br>7.5 $\lambda_I$ | Glass RPC, scintillator, MPGD |
| Solenoid | Field: 4 T<br>Free bore: 3.4 m<br>Length: 8.3 m | | Field: 5 T<br>Free bore: 2.7 m<br>Length: 6.5 m | |
| Muon system | | Glass RPC, scintillator | | Glass RPC, scintillator |
| Overall height<br>Overall length | 14.0 m<br>12.8 m | | 14.0 m<br>12.8 m | |

Very Deep Sub-Micron (VDSM) technology. Other approaches build upon the experience with integrated Complementary Metal Oxide Semiconductor (CMOS) technologies developed for the ILC, or use new emerging technologies like the Silicon on Insulator (SOI) or the fully 3D integrated pixel solution.

For the **Outer Tracker**, there are two main technology options: a large Time Projection Chamber (TPC) complemented with a small inner silicon tracker and a silicon envelope surrounding the TPC for CLIC_ILD, and a compact full silicon tracker in a very high magnetic field for CLIC_SiD. The full silicon tracker version consists of 5 layers of thin silicon strips in the barrel section and 4 layers in the endcap section. The TPC design is based on a lightweight field cage, with a thin central high-voltage cathode, read out by Micro-Pattern Gas Detector (MPGD). The chamber is complemented with a silicon tracking system which extends the tracking acceptance down to small polar angles of 7°. In order to achieve the momentum resolution requirements, all vertex and tracking systems have to be ultra thin. The low accelerator duty cycle permits the use of power pulsing of the on-detector electronics. This will allow the detectors to be cooled by gas flow.

The design of the CLIC calorimeters is driven by the adoption of high granularity particle flow calorimetry. With this it is possible to achieve the CLIC objective of a $\sim$3.5% jet energy resolution for high-energy jets. As described below, high granularity particle flow calorimetry is also well suited to the relatively high levels of beam-induced background at 3 TeV CLIC; it has the potential to separate calorimetric energy deposits due to background particles from those of the hard interaction.

The **Electromagnetic Calorimeter** (ECAL) is copied without modifications from the original ILC concepts. Both the ILD and the SiD ECAL use silicon-tungsten sampling calorimeters optimised for particle flow, placing particular emphasis on the separation of adjacent electromagnetic showers. In addition, the use of small scintillator strips with Silicon Photomultiplier (SiPM) readout as active medium is being considered. The active layers of the CLIC_ILD ECAL consist of $5.1 \times 5.1$ mm$^2$ pads, while CLIC_SiD ECAL uses hexagonal pads of 13 mm$^2$. Both concepts use 30 longitudinal layers in total, where the rear ten absorber layers are twice as thick at the front twenty.

For the **Hadron Calorimeter** (HCAL) tungsten was chosen as absorber material for the barrel. It provides sufficient depth to contain high-energy showers, while limiting the diameter of the surrounding solenoid. Based on simulation studies, an optimal sampling structure with 1 cm tungsten absorber plates was chosen. The active layers have a thickness of 5 mm. For the HCAL endcaps, 2 cm steel plates are used as absorber, as depth restrictions are less stringent in this region.

The readout technologies considered for the HCAL are scintillator tiles coupled to SiPMs, and gaseous devices with different amplification structures, such as Resistive Plate Chambers (RPCs), Micro-MEsh Gaseous Structure (Micromegas) and Gas Electron Multiplier (GEM) foils with pad readout. The scintillator option implies analog readout electronics, thus providing energy deposition information, while the gaseous techniques are read out in digital or semi-digital mode (one or two threshold bits), the particle energy being inferred from the number of hit cells. The assumed HCAL cell sizes are $3 \times 3$ cm$^2$ for the analog readout option and $1 \times 1$ cm$^2$ for the digital readout options. Validation of these options with beam tests is ongoing.

To ensure a jet energy resolution at or below the required 3.5% for high-energy jets, an HCAL depth of 7.5 $\lambda_I$ is chosen. This results in a total thickness of 8.5 $\lambda_I$ for the calorimeter system, including the ECAL.

For the purpose of optimal particle flow reconstruction, the ECAL and HCAL are located within the inner bore of the **Solenoid** magnet. A high magnetic field (4 T for CLIC_ILD and 5 T for CLIC_SiD) is required to: confine low-$p_T$ particles, resulting from the beam-induced background, within the beam pipe; to achieve the desired momentum resolution; and also to separate tracks from nearby particles in high-energy jets. Superconducting solenoid coils based on reinforced conductor technology, as already successfully applied in the CMS and ATLAS experiments, are foreseen.

The magnetic flux is returned through an **iron yoke**. The **Muon System**, comprising nine track sensitive layers of either glass RPCs or scintillators, is located in the yoke. It enhances the muon identification capability of the detector. In addition, it serves as a tail catcher for showers developing late in the calorimeters, thereby slightly improving the energy measurement for high-energy hadrons.

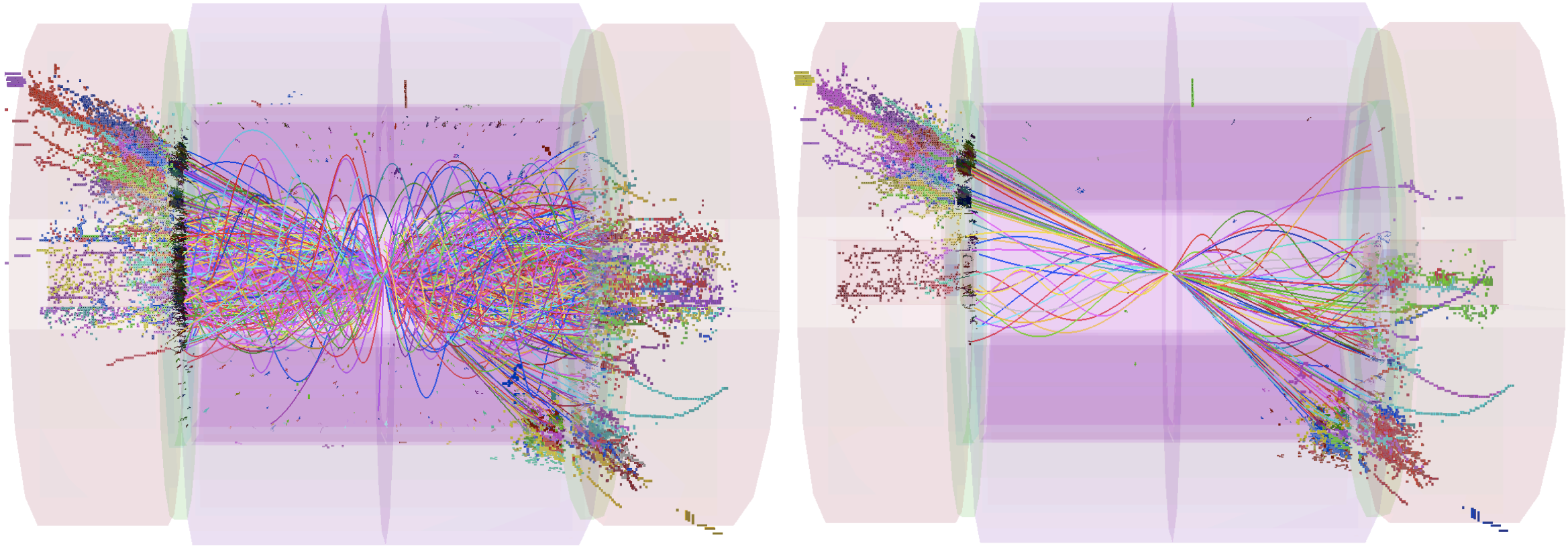

Fig. 4.3: Left: Reconstructed particles in a simulated $e^+e^- \to t\bar{t}$ event at 3 TeV in the CLIC_ILD detector concept with background from $\gamma\gamma \to$ hadrons overlaid. Right: The effect of applying tight timing cuts on the reconstructed cluster times.

At CLIC many Standard Model processes will produce particles at relatively low angles with respect to the incoming beams. It is therefore important to extend the detector coverage to small polar angles. The **very forward detectors** consist of two sandwich electromagnetic calorimeters. The Luminosity Calorimeter (LumiCal) is a silicon-tungsten detector, which covers the angular range from 40 mrad to 100 mrad and provides a precise measurement of the luminosity. The BeamCal is a radiation-hard semiconductor-tungsten calorimeter, which extends the angular coverage down to about 10 mrad for low-angle electron tagging and potentially also for fast beam feedback purposes.

The forward regions also house the final beam focusing quadrupoles. These are placed with sub-nm mechanical stability to achieve maximum luminosity performance with very small beams. This impacts on the hadron calorimeter coverage in the forward region.

## 4.4 Suppression of Beam-induced Background

The relatively high levels of beam-induced background can be suppressed using precise hit timing information. For this, a scheme was developed in the event reconstruction, which considers time-stamping capabilities of 10 ns for all silicon tracking hits, and of 1 ns time resolution for all calorimeter hits.

It is assumed that trigger-less data readout will be carried out at the end of each bunch train, making the information of the full 156 ns bunch train available for offline reconstruction. Within the bunch train, candidates for a hard interaction are identified, and the data in a window around the time of this interaction are passed to the event reconstruction. This reconstruction time window is defined to be 10 ns for all raw detector hits, except for the HCAL barrel, where it is increased to 100 ns, to account for the slower shower development in tungsten. In a next step, the particle flow reconstruction algorithm uses the information from the highly granular calorimeters to cluster together hits from a single particle. At this stage, the background can be further reduced by applying tighter timing cuts on the reconstructed calorimeter cluster time. These cuts are in the range of 1.0–2.5 ns, depending on the type of the reconstructed particle and its $p_T$, and are applied only to relatively low-$p_T$ particle flow objects.

As a result, the average background level can be reduced from approximately 20 TeV per bunch train to about 100 GeV per reconstructed physics event. This background rejection, which is exemplified in Figure 4.3, is achieved without significantly impacting the detector performance.

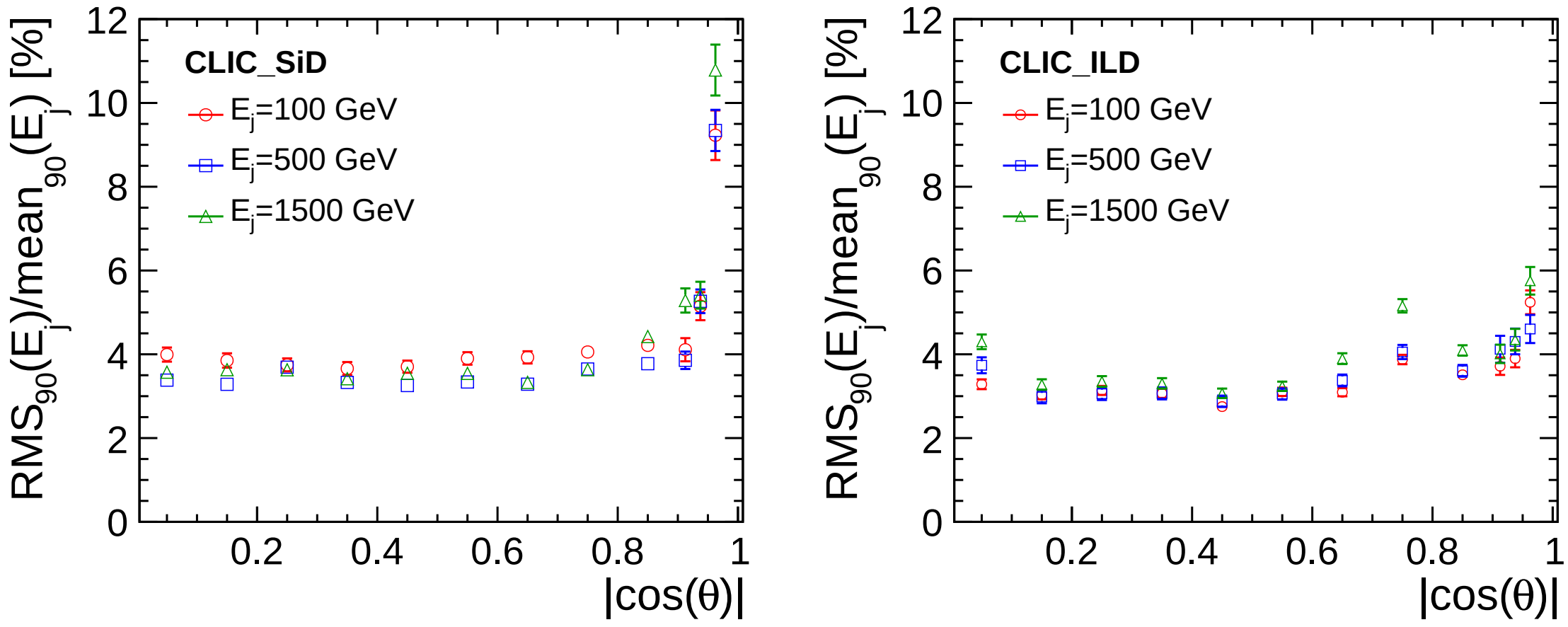

Fig. 4.4: Jet energy resolution dependence on event angle and jet energy for CLIC_SiD (left) and CLIC_ILD (right).

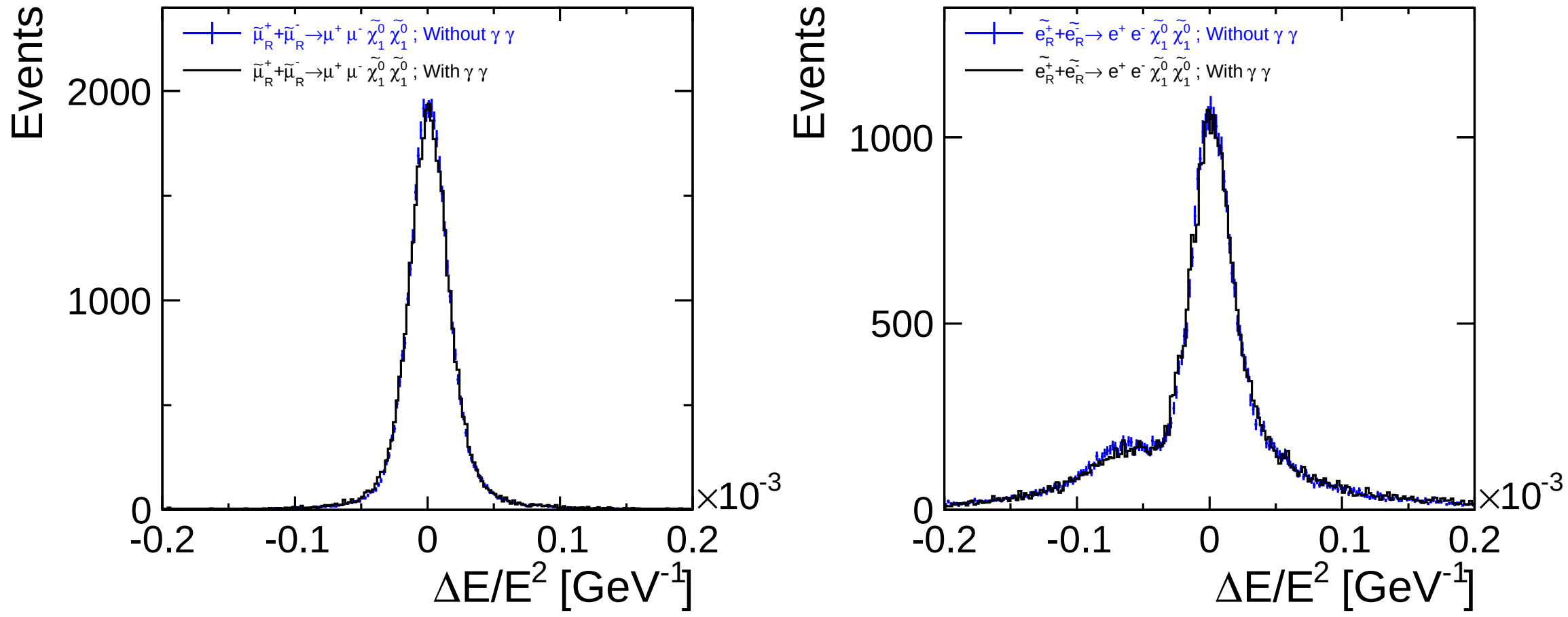

Fig. 4.5: Lepton energy resolution for processes with missing energy and two isolated leptons. Left: Muon energy resolution obtained from $e^+e^- \to \tilde{\mu}_R^+\tilde{\mu}_R^-$ events. Right: Electron energy resolution observed for $e^+e^- \to \tilde{e}_R^+\tilde{e}_R^-$ events. Both samples were reconstructed without time selection.

## 4.5 Detector Performance

In this section, two examples of detector performance are presented. For more examples, see [3].

The particle flow performance in terms of jet energy resolution is shown as a function of the angle of the jet in Figure 4.4 for both CLIC_SiD and CLIC_ILD. The study is based on single $Z$-like bosons decaying at rest into light quarks, hence producing two mono-energetic jets. The resolution of the jet energy $E_j$ is obtained by calculating the $\mathrm{RMS}_{90}(E_{jj})$ and the $\mathrm{mean}_{90}(E_{jj})$, and then applying a factor of $\sqrt{2}$. In the barrel region, both detectors show only small variations in performance according to the angle of the jet. As the tracker volume in CLIC_ILD is larger than in CLIC_SiD, its HCAL coverage extends to lower polar angles (see Figure 4.2). In the case of CLIC_ILD, the jet energy resolution worsens in the overlap region between the barrel and the end cap. This is due to the cabling gap between the ECAL barrel and the ECAL endcap, which is larger in CLIC_ILD than in CLIC_SiD.

To illustrate the detector performance for single particles at 3 TeV, the lepton energy resolution for SUSY smuon and selectron production processes with missing energy and two isolated leptons is shown in Figure 4.5 [5]. The distributions are shown without and with overlay of $\gamma\gamma \to$ hadrons events. The energy resolution is not affected by $\gamma\gamma \to$ hadrons interactions because a high $p_T$ cut of 4 GeV is applied in the analysis to reduce the Standard Model physics background. This removes almost all hadrons originating from this beam-induced background process. However, the inclusion of background in the data sample induces reconstruction inefficiencies of 1.0% and 4.6% for the smuon and selectron study respectively.

## 4.6 Conclusions

Detailed simulation studies, based on the CLIC_ILD and CLIC_SiD detector concepts, demonstrate that the CLIC detector performance goals are achievable (see also Chapter 12 in [3]). This holds up to the highest foreseen CLIC centre-of-mass energy of 3 TeV, where the background conditions are most challenging.

These detector concepts involve challenging detector technologies. Building on experience from a broad linear collider detector development programme, complemented with experience from the LHC, these technologies are considered feasible following a 5-year R&D programme.

# Chapter 5

# CLIC Project Implementation

## 5.1 Introduction

This chapter focuses on some important features of the project which condition its implementation, namely schedules for construction, commissioning and operation based on the staging scenarios defined above, electrical power and energy consumption, and cost. Other – equally important – aspects of the project implementation, such as site selection, environmental impact, industrial strategy, procurement rules and procedures, legal and institutional framework and more generally governance are not treated here, and will be addressed in the forthcoming project preparation phase described in Chapter 7.

## 5.2 Construction and Operation Schedules

This section addresses the two essential ingredients of the time development of the CLIC programme, namely the schedules for production of main components, installation and technical commissioning of the sequential stages ("construction" schedules), and the evolution of performance of machine and experiments leading to useful luminosity for physics at each stage ("operation" schedules). These two ingredients are not independent, as the operation schedules drive the duration of the time spans between construction of the successive stages, and conversely, the construction schedules impose minimum "dead times" for physics operation of the CLIC complex. This approach is substantially different from that described in Chapter 9.5 of [1], for which attaining the 3 TeV stage in the minimum amount of time was the prime driver and operation at intermediate collision energy occurred only in the shade of the 3 TeV construction schedule. It has consequences not only on time schedules proper, but also on related aspects such as power and energy consumption and cost (see next sections), as the CLIC programme up to 3 TeV now develops over a time span of some 25 years. This is well beyond the horizon of industrial procurement contracts, so that each stage is now considered as constituting a separate project *per se*, similar to the LEP1 and LEP2 projects at CERN, the construction of which spread over two decades. Production of similar components for the different stages will then likely be done by different industrial companies through separate contracts with the consequence that, *inter alia*, the benefits of learning curves on unit costs will be limited to the series quantities procured through each contract. Another consequence is that the lead times necessary to reach production "cruise rates" will need to be applied, at least partially, for each construction stage.

### 5.2.1 Operation for Physics

The growth of luminosity for physics per year at the different values of collision energy for the two staging scenarios is sketched in Figure 5.1, based on the expected development of instantaneous luminosity and assuming operation for 200 days per year at an overall net efficiency of 0.5, this number covering both accelerator complex and experiments. The duration of each stage is defined by the time needed to reach the targets of integrated luminosity set in Chapter 6, namely 500 $fb^{-1}$ at 500 GeV, 1.5 $ab^{-1}$ at 1.4 (1.5) TeV and 2 $ab^{-1}$ at 3 TeV collision energy.

Figure 5.2 shows the development of total integrated luminosity for the two staging scenarios considered. Scenario B ("low entry cost") requires operating two more years at 500 GeV collision energy to make up for the lower luminosity; this is partly regained at the next higher energy stage, as the higher instantaneous luminosity at 1.5 TeV enables to collect the required integrated luminosity faster than at the collision energy of 1.4 TeV achievable in scenario A ("luminosity optimised"). Overall, given the same integrated luminosity targets, the complete duration of the programme is comparable for both staging scenarios.

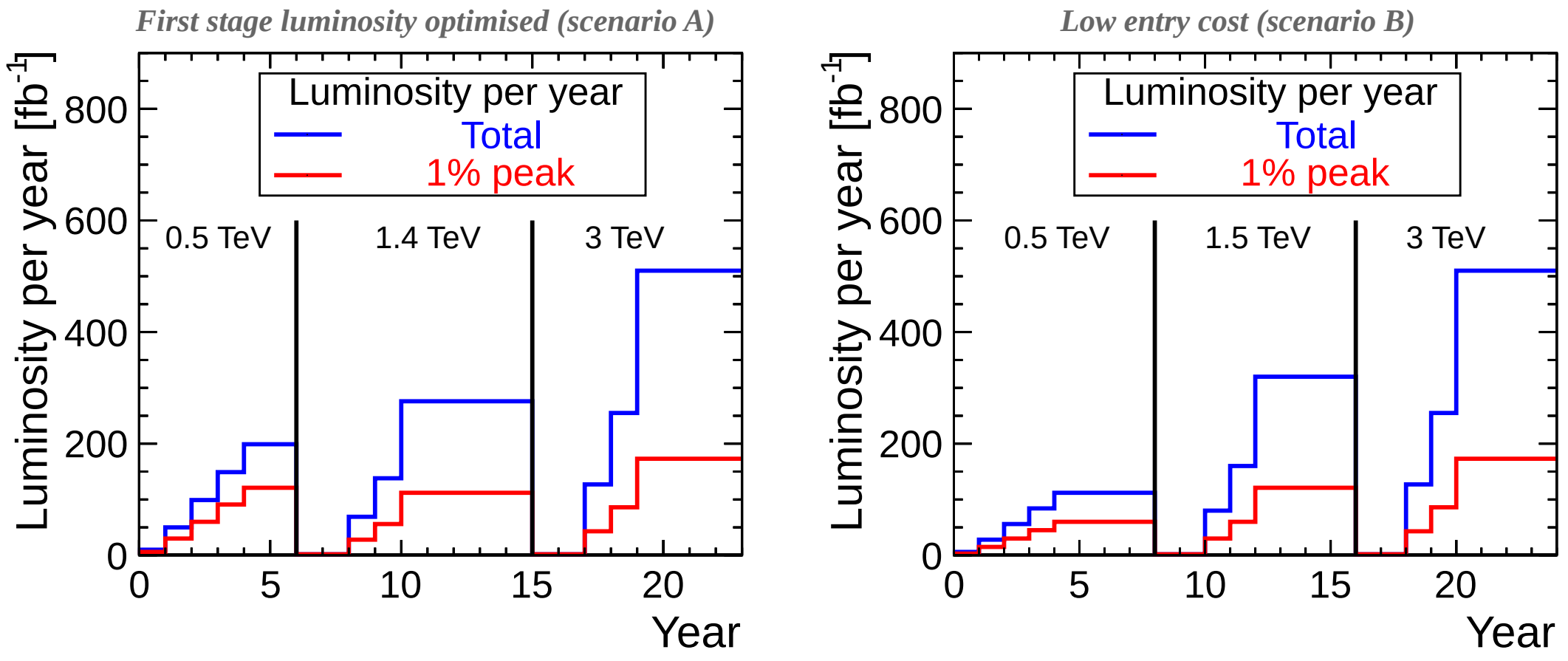

Fig. 5.1: Luminosity per year in the scenarios optimised for luminosity in the first energy stage (left) and optimised for entry costs (right).

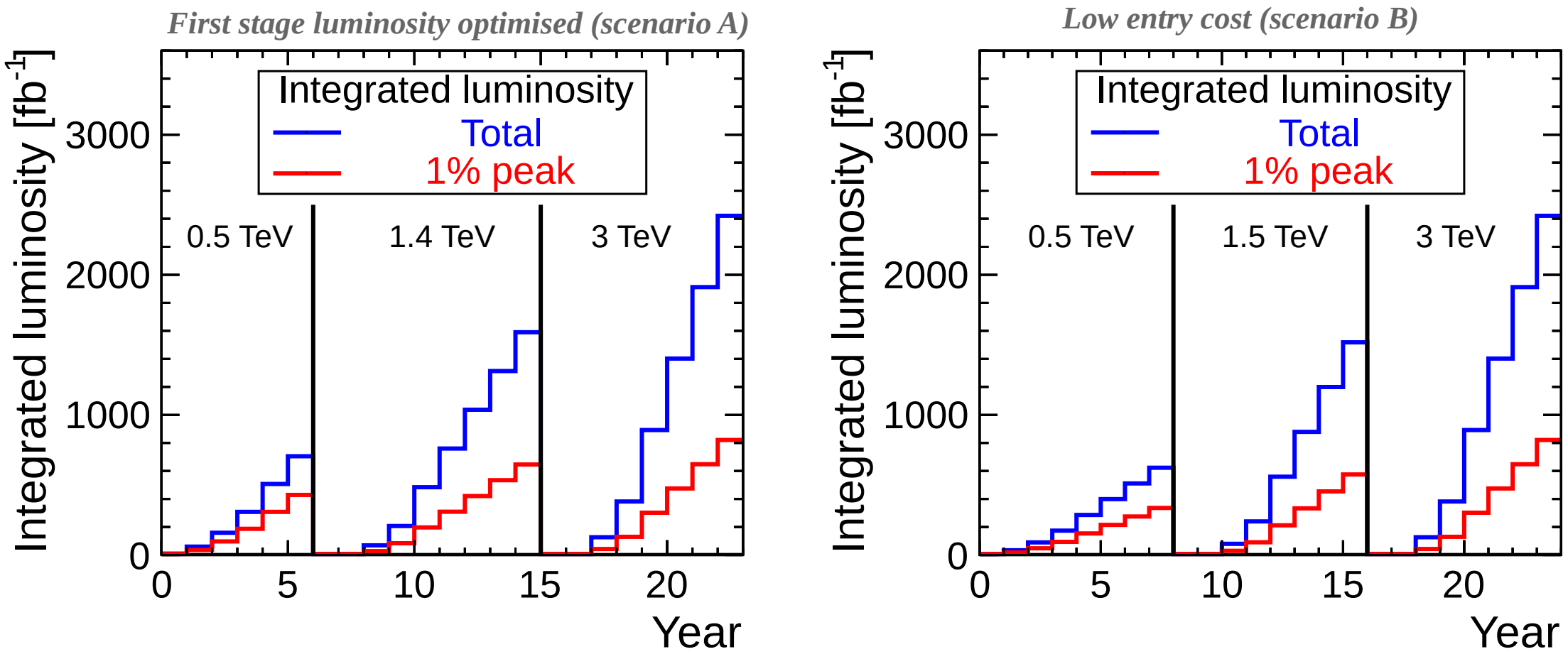

Fig. 5.2: Integrated luminosity in the scenarios optimised for luminosity in the first energy stage (left) and optimised for entry costs (right). Years are counted from the start of beam commissioning. These figures include luminosity ramp-up of four years (5%, 25%, 50%, 75%) in the first stage and two years (25%, 50%) in subsequent stages.

### 5.2.2 Construction

The tentative construction schedules presented in the following for scenarios A and B aim at reaching start of operation at each stage of collision energy as rapidly as permitted by the production and installation rates for the sequential activities, and within the interference constraints defined in Chapter 9.5 of [1], while accommodating the operation schedules defined above. In particular, production and reception of the main components for the first stage at 500 GeV collision energy must be such that they become available for installation as soon as preceding construction activities allow it. With the same assumption of three suppliers, the required delivery rates for the accelerating structures – the most numerous series components – are substantially lower than in [1]: after a ramp-up period of about one year, each supplier now has to deliver at the rate of 343 accelerating structures per month. These lower delivery rates favourably impact the fixed costs (manufacturing premises and specific investments).

At each stage of collision energy, construction will develop in four successive phases. Site preparation for the civil engineering phase will take four months, after which excavation of the main shafts will begin. Two tunnel-boring machines (TBM) will be assembled on site and start excavating the right and left tunnels. Handover by civil engineering will then allow the establishment of the tunnel geodetic network and the marking of fiducials on the floor. Installation of the general services, piping and cabling will immediately follow, each activity progressing along four fronts, in parallel in the left and right sectors. Once most of the cabling work is completed, installation of the ground supports will proceed. The two-beam modules will then be transported, pre-aligned and interconnected, working on two fronts in each sector. The rates of progress for civil engineering and general services, as well as for installation of the two-beam modules in the main linacs are identical to those in [1], derived from experience with installation of the LHC at CERN. After completion of installation, commissioning of the technical systems (without beam) will take about one year, followed by final alignment. In parallel with the main linac activities, construction, installation and gradual commissioning of the main-beam and drive-beam injector complexes will spread over some six years, while seven years will be needed to complete the interaction region and install and commission the detectors [1].

The main linac "railway" schedules resulting from this analysis [2] appear in Figures 5.3 and 5.4, for the first two stages of scenarios A and B, respectively. The overall schedules including construction of the injectors and experimental area, and installation of the detectors as described in Chapter 9.5 of [1] are shown in Figures. 5.5 and 5.6, for scenarios A and B, respectively. The time needed to get first beams at 500 GeV collision energy remains seven years from start of construction for both staging scenarios. For scenario A, the construction and installation of the main linacs just match those of the injectors and experimental area. For scenario B, they are in the shade of the injectors and experimental area, which drive the schedule.

Civil engineering and tunnel construction for the second stage must restart in years 10 and 11, respectively, for scenarios A and B, i.e. during operation for physics of the first stage. Possible interference due to ground vibration is mitigated by restarting excavation of the shafts and tunnel from the outer points, and ensuring that the TBMs remain at minimum one sector (about 900 m) away from the running linacs. At the TBM progress rate of 150 m/week, this safety distance can then be excavated during one of the yearly shutdown periods of the machine, thus allowing execution of subsequent tasks (finishing of tunnel and installation of general services and infrastructure) without interference with further operation of the accelerator. Similarly, series component production, assumed to proceed at the same rates as for the first stage, must restart in years 10 and 12 respectively for scenarios A and B, so as not to delay installation of the new sectors of the main linacs. The interruption of six to eight years in the series procurement, civil engineering and installation contracts warrants the decision to consider the first two stages as independent projects.

Installation of the new sectors for the second stage needs to proceed in parallel with dismounting of the two-beam modules of the first stage and their re-installation at the outer ends of the main linacs. This will require additional installation teams in order to fit within the two-year interruption of operation between stages.

## 5.3 Power and Energy Consumption

The electrical power and energy consumption of CLIC at 3 TeV collision energy is presented in Section 2.11 of [1], with detailed breakdown by domain and by technical system, and in Section 9.4 of [1] for the intermediate centre-of-mass energy stages at 500 GeV and 1.5 TeV. Reference is made to those chapters for discussion of the relative and absolute power and energy requirements, and of the potential paths for power and energy savings [3].

This section summarises and updates the information, introducing the two alternative staging scenarios and the revised operation schedule presented in the preceding section.

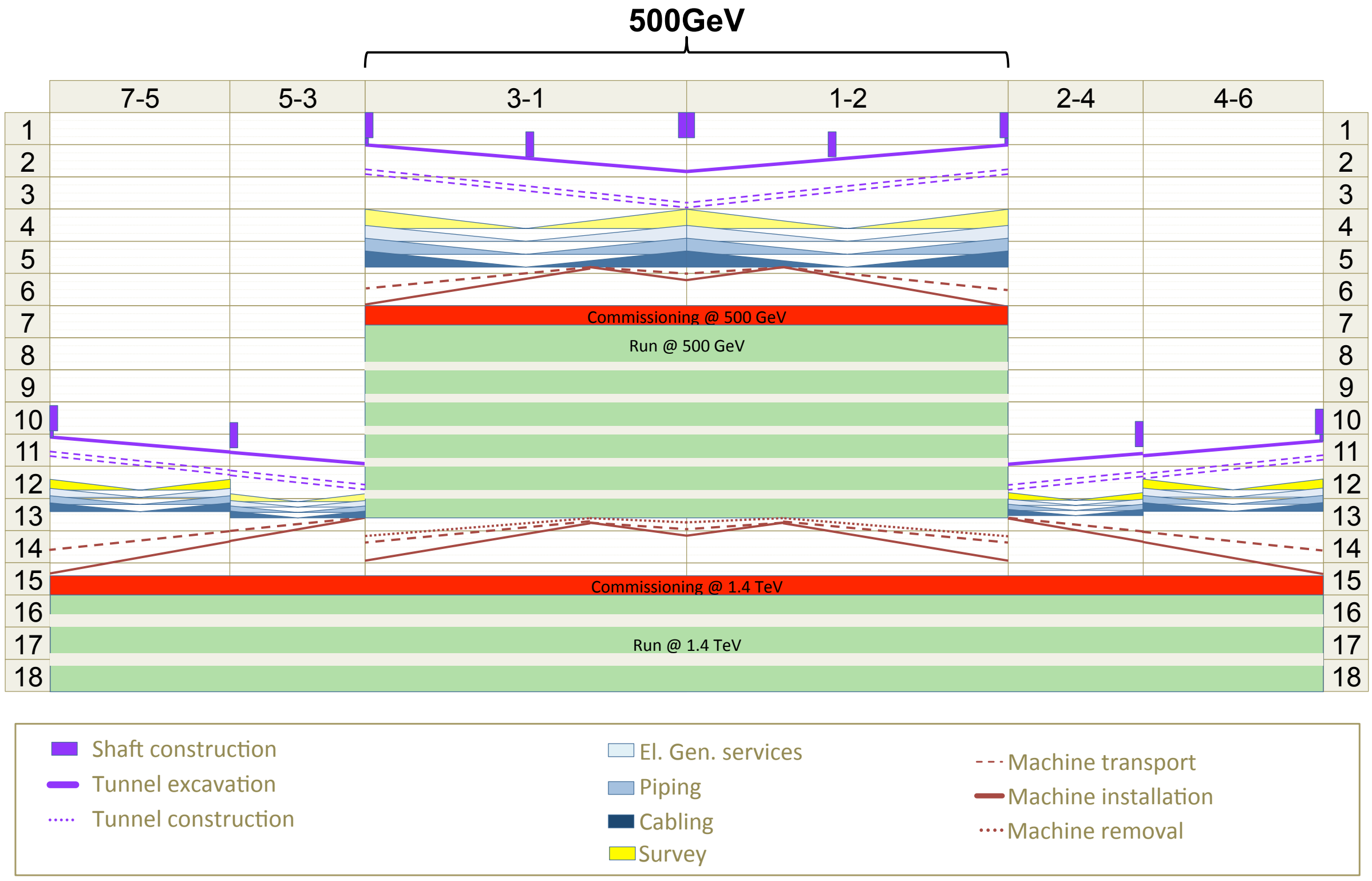

Fig. 5.3: Overall "railway" schedule for the first two stages of scenario A. The horizontal scale represents tunnel length, with the experimental area in the centre. The vertical scale shows years from the start of construction.

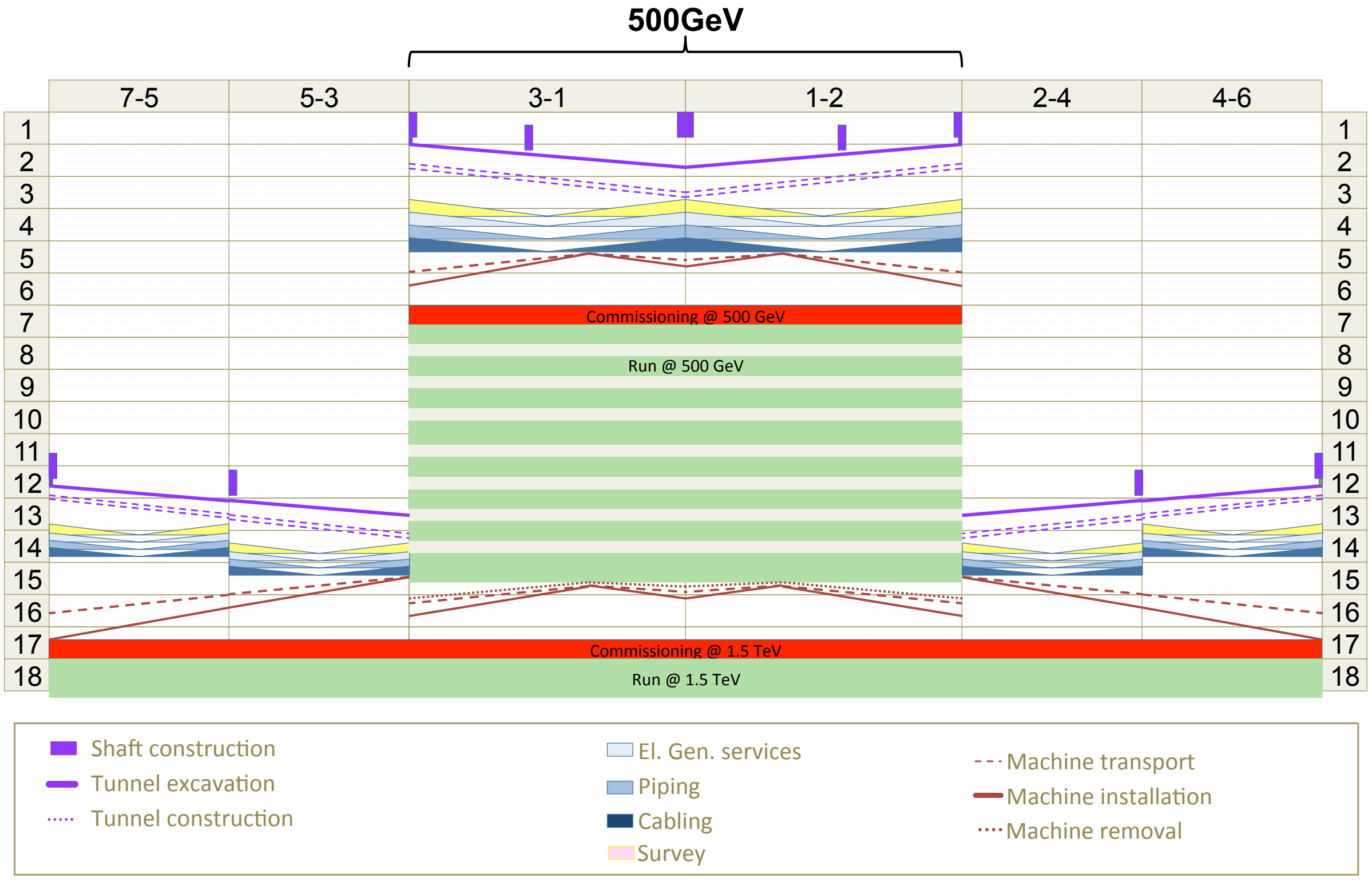

Fig. 5.4: Overall "railway" schedule for the first two stages of scenario B. The same conventions as in Figure 5.3 are used.

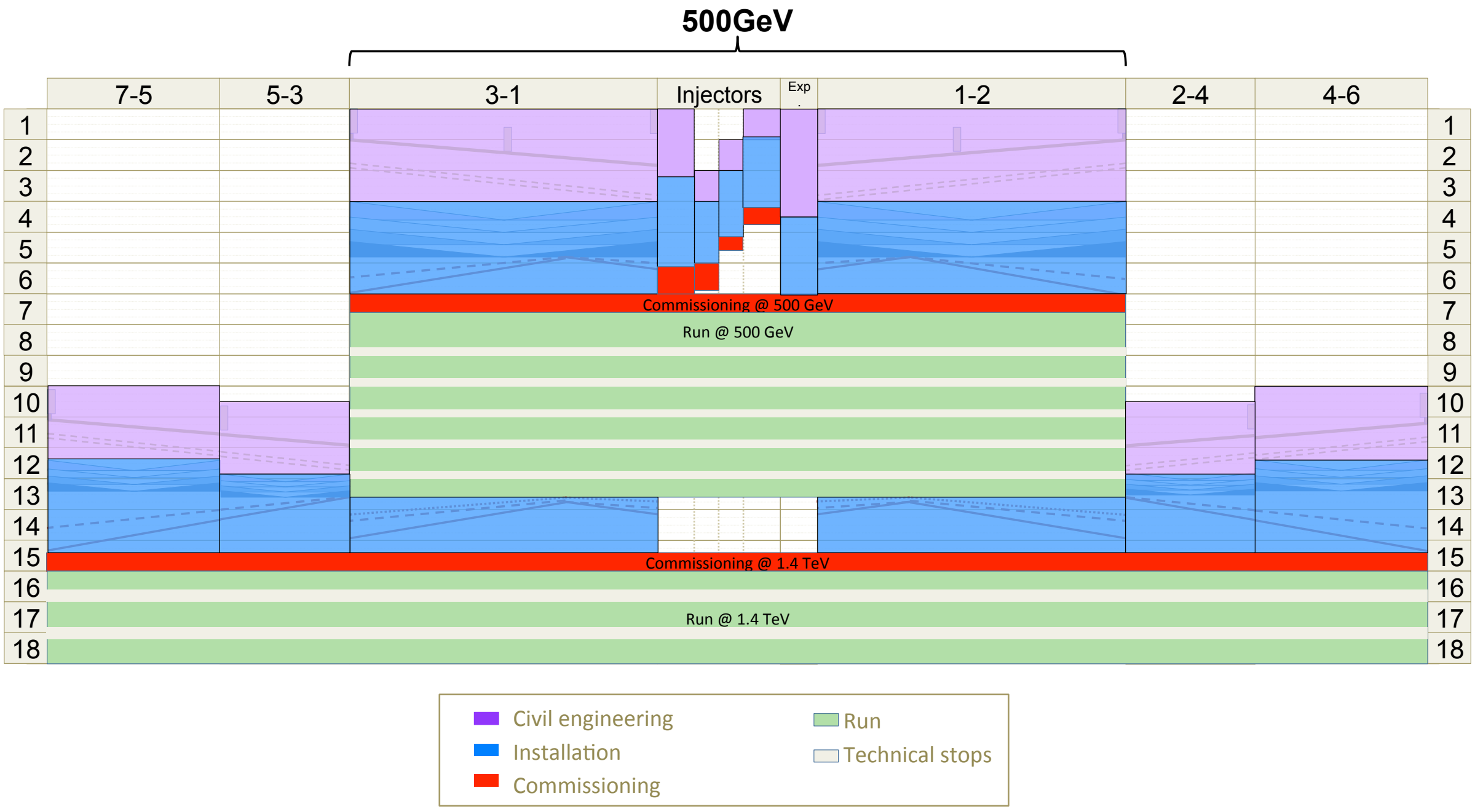

Fig. 5.5: Overall "railway" schedule for the first two stages of scenario A. The same conventions as in Figure 5.3 are used. Construction schedule for main-beam and drive-beam injectors, and for experimental area are shown in the centre.

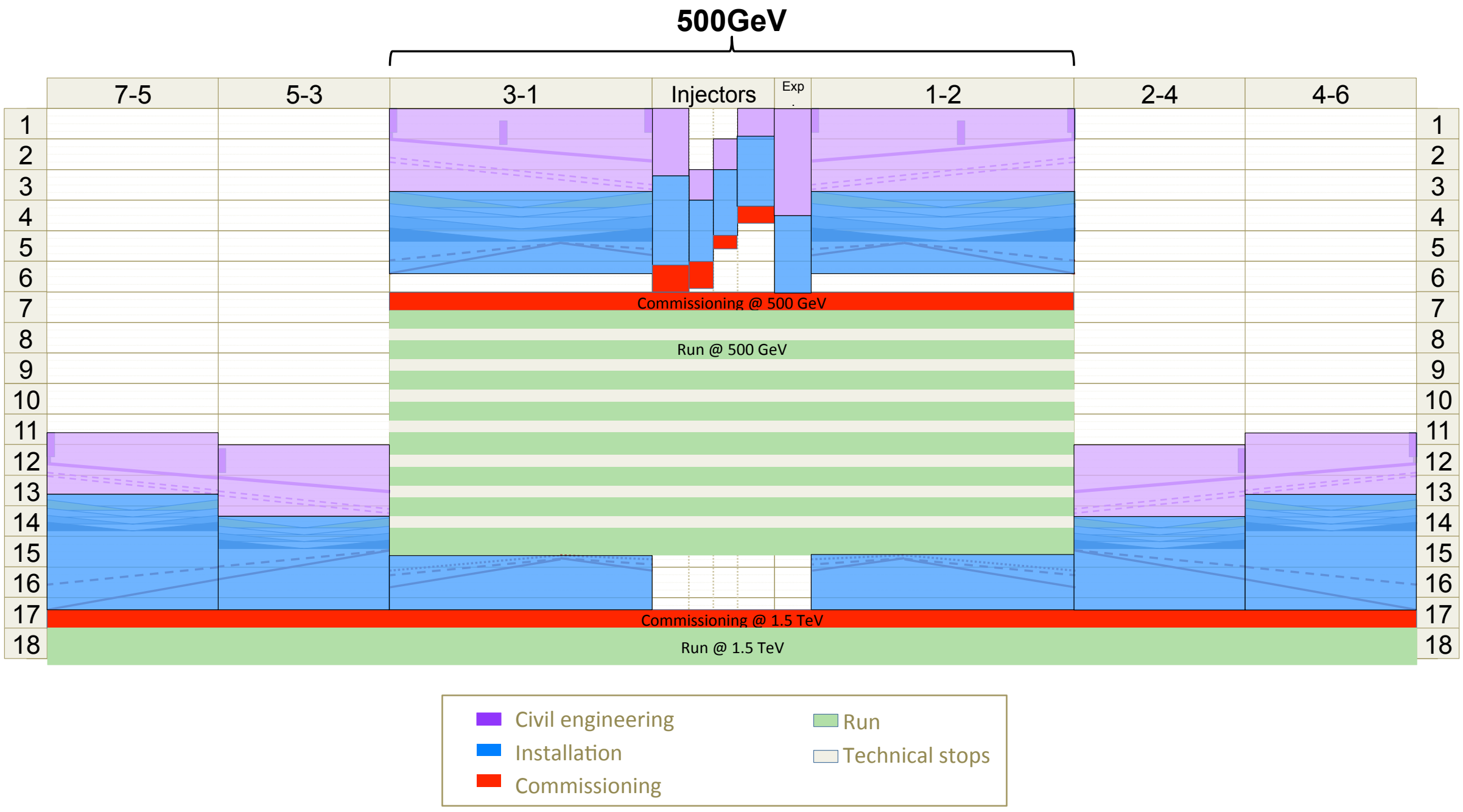

Fig. 5.6: Overall "railway" schedule for the first two stages of scenario B. The same conventions as in Figure 5.3 are used. Construction schedule for main-beam and drive-beam injectors, and for experimental area are shown in the centre.

Table 5.1: Nominal power and efficiency for staging scenarios A and B, where $W_{main\ beam}$ is for the two main beams.

| Staging scenario | $\sqrt{s}$ (TeV) | $\mathcal{L}_{1\%}$ ($cm^{-2}s^{-1}$) | $W_{main\ beam}$ (MW) | $P_{electric}$ (MW) | Efficiency (%) |
|---|---|---|---|---|---|
| A | 0.5 | $1.4 \cdot 10^{34}$ | 9.6 | 272 | 3.6 |
| | 1.4 | $1.3 \cdot 10^{34}$ | 12.9 | 364 | 3.6 |
| | 3.0 | $2.0 \cdot 10^{34}$ | 27.7 | 589 | 4.7 |
| B | 0.5 | $7.0 \cdot 10^{33}$ | 4.6 | 235 | 2.0 |
| | 1.5 | $1.4 \cdot 10^{34}$ | 13.9 | 364 | 3.8 |
| | 3.0 | $2.0 \cdot 10^{34}$ | 27.7 | 589 | 4.7 |

Table 5.2: Residual power without beams for staging scenarios A and B.

| Staging scenario | $\sqrt{s}$ (TeV) | $P_{waiting\ for\ beam}$ (MW) | $P_{shutdown}$ (MW) |
|---|---|---|---|
| A | 0.5 | 168 | 37 |
| | 1.4 | 190 | 42 |
| | 3.0 | 268 | 58 |
| B | 0.5 | 167 | 35 |
| | 1.5 | 190 | 42 |
| | 3.0 | 268 | 58 |

### 5.3.1 Power

The nominal luminosity, total electrical power consumption and overall efficiency are given for staging scenarios A and B in Table 5.1. The electrical power covers all accelerator systems and services, including the experimental area and the detectors. It takes into account the electrical network losses for transformation and distribution on site.

At 500 GeV collision energy, scenario B which has half the nominal luminosity of scenario A, requires 16% less power in the drive beam and about half in the main beam production complex: as a result, the electrical power drawn from the network is lower by 37 MW. The residual power consumption of the accelerator complex without beams is given in Table 5.2 for two modes of operation corresponding to short ("waiting for beams") and long ("shutdown") beam interruptions.

### 5.3.2 Energy

The yearly energy consumption can then be estimated from the values of power consumption in the different operation modes and the assumptions on running periods. Considering 150 days per year of normal operation at nominal power and assuming reduced power for commissioning with beam in the early years at each stage of collision energy, the development of yearly energy consumption can be sketched in Figure 5.7 for the two staging scenarios. Although the yearly consumption at 500 GeV collision energy is smaller in scenario B, the lower luminosity makes it necessary to run two more years for the same physics reach, thus yielding a cumulative energy consumption of 6 TWh, against 5 TWh in scenario A. This difference is more than regained in the second stage, thanks to the higher instantaneous luminosity at 1.5 TeV (scenario B) than at 1.4 TeV (scenario A). It is to be noted that cumulative energy consumption is about equal for reaching and completing the second and third stages of the CLIC programme.

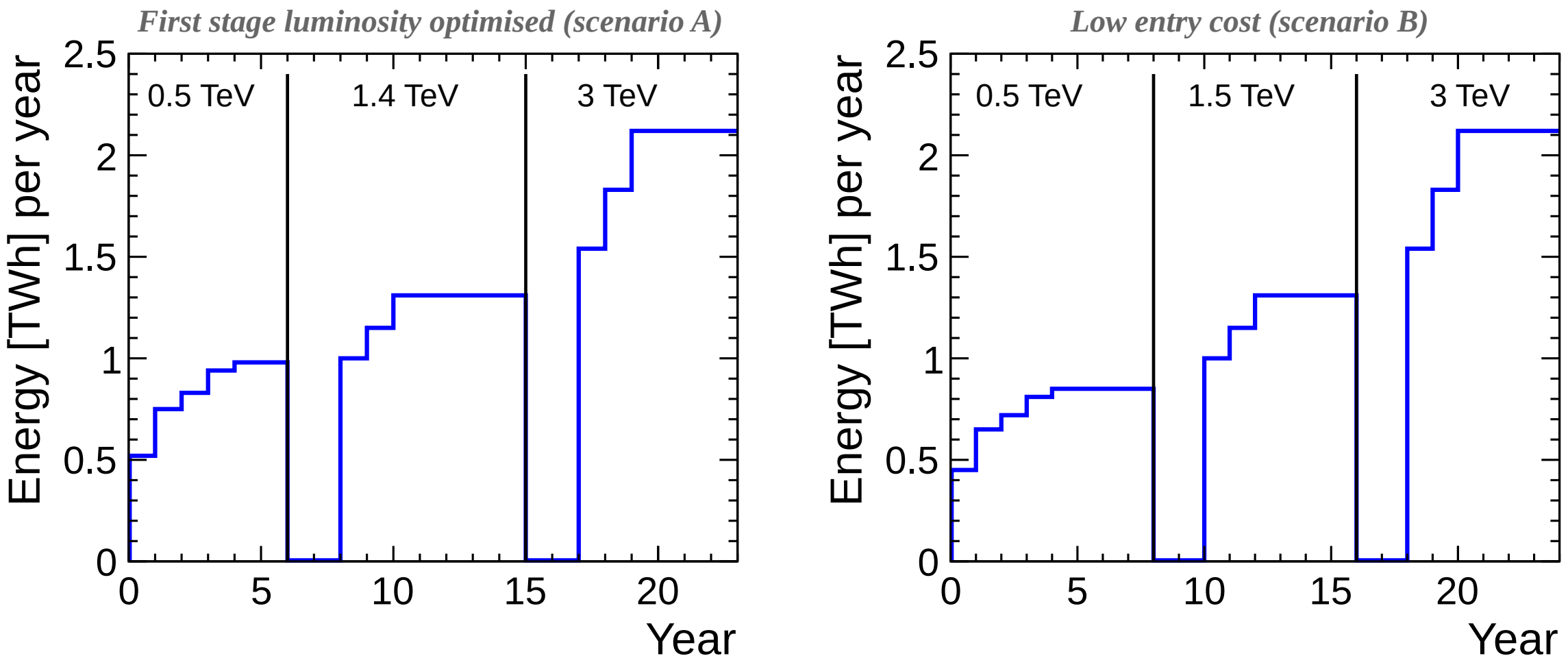

Fig. 5.7: Development of yearly energy consumption for staging scenarios A (left) and B (right).

### 5.3.3 Potential for Power and Energy Savings

The power and energy estimates quoted above are large numbers, particularly for the two later stages. Several paths aiming at saving power and/or energy have been identified and are under investigation.

A first category of actions aims at achieving power sobriety by re-design, trading operation against investment costs. Examples of such actions are reduced current density in normal-conducting magnets and in cables, lower heat loads to the heating, ventilation and air conditioning (HVAC) system by improved water-cooling, or re-optimisation of the accelerating gradient in the main linacs with a different objective function putting stronger emphasis on energy consumption.

Improving efficiency in the use of electrical energy constitutes a second path, e.g. by replacing normal-conducting by "super-ferric" or permanent magnets, or by better network-to-RF power conversion. The ongoing development of efficient high-power klystron modulators directly fed by the high-voltage grid is an example of such actions.

Noting that electrical power from the network is more costly when in high demand and that the power consumption of the CLIC complex strongly varies with the beam operating conditions, adequate energy management appears an interesting path for economical operation, e.g. by optimising low-power configurations in case of beam interruption and by modulating scheduled operation to match daily and seasonal fluctuations in power demand. By avoiding running in the few peak hours per day, CLIC could be operated as a peak-shaving facility, thus contributing to the stability of the electrical network.

Finally, waste-heat recovery appears an interesting option in view of the high power rejected in water. As in other projects, the main issue here remains the low heat rejection temperature. Options for cooling water exiting at higher temperature in specific technical systems should be investigated, as well as possible valorisation of waste heat for concomitant needs, e.g. by heat pumps for residential heating or absorption cooling systems.

## 5.4 Cost

### 5.4.1 Value, Explicit Labour and Cost

The Linear Collider will be a global project constructed and operated in a collaborative manner by its many Parties representing the international community of users. Contributions from the Parties to project construction are likely to take different forms, whether in cash, in kind or in personnel, coming from different countries or regions of the world with different currencies and accounting systems. It is therefore important to discuss project costs independently of a particular accounting system, compatible with this diversity. This is the purpose of the "value and explicit labour" methodology used in the following, in line with the approach of the International Linear Collider (ILC) Reference Design Report [4] and the LHC experiments.

The value of a component or system is defined as the lowest reasonable estimate of the price of goods and services procured from industry on the world market in adequate quality, quantity and delay to meet its specifications. Value therefore reflects both technical content of a supply, and commercial aspects linked to the market structure – in particular monopolistic/oligopolistic cases – as well as to the marketing strategy of suppliers. It however does not consider differences in market openness and protectionism, as the supplies to the global project are expected to be free from local taxes and custom duties. Other obstacles to free trade, e.g. non-tariff barriers and prescription *ab initio* of national/regional returns, are not considered either. Value is expressed in a given currency at a given time (see Section 5.4.4 below).

Explicit labour is defined as the amount and type of personnel provided for project construction by the central laboratory and the collaborating institutes. It does not include personnel in the industrial manufacturing premises, nor industrial support work performed on the basis of task- or result-oriented contracts; the former is included in the value of the supplied components, and the latter counted in the value of industrial services. Explicit labour is expressed in Full Time Equivalent (FTE)·years for different categories of personnel, e.g. scientific and engineering, technical, administrative.

### 5.4.2 Scope of the Study

The study primarily aims at producing estimates of the value for construction of the 500 GeV stage of the CLIC accelerator complex on a site close to CERN, according to staging scenarios A and B as defined in Chapter 3, as well as of the value of the CLIC_ILD and CLIC_SiD particle detectors [5]. A first estimate of explicit labour is also given for the accelerator complex, but not for the detectors. In view of the construction and operation schedules discussed in Sections 5.2 and 5.3, the 500 GeV stage of CLIC will constitute a project *per se*, and further energy stages at 1.5 TeV and 3 TeV centre-of-mass will be the object of separate upgrade projects, requiring additional contracts for civil engineering and procurement of series components. As a consequence, the large-series effects expected on unit costs – "learning curves" and quantity rebates – will be moderate as the production horizon remains limited to the quantities required for the completion of the 500 GeV stage. The study also aims at producing an estimate of the incremental value of CLIC construction per unit of collision energy beyond 500 GeV.

The value estimates cover the "project construction" phase, from approval to start of commissioning with beam. They therefore include the different domains of the CLIC complex – drive-beam and main-beam injectors, main linacs, machine-detector interface, detectors and their infrastructure, and beam disposal systems. Also included are specific tooling required for the production of the components, reception tests and pre-conditioning of the components, and commissioning (without beam) of the technical systems.

Excluded from the value estimates presented are R&D, prototyping and pre-industrialisation costs, acquisition of land and underground rights-of-way, computing, general laboratory infrastructure, e.g. offices and library, general laboratory services, e.g. administration, purchasing, human resource management. Spare parts, even though the specific ones have to be procured in the continuation of series-production contracts, are charged to the operations budget. All procurement is considered to be free from taxes and custom duties, by virtue of the global nature of the project.

### 5.4.3 Methodology, Organisation and Tools

There are basically two types of methods for costing technical projects. The analytical approach resting on the work breakdown structure of the project down to the component level is based on unit costs and quantities combined and aggregated to produce higher-level estimates up to the complete project. It requires a full description of the material and work contents of the different components, as well as the knowledge of their production techniques so as to include the cost of specific tooling (fixed costs) and the yield of the production chain (rejection and reprocessing rates). In the case of large series, learning curves can be applied with adequate learning coefficients, preferably based on previous experience with the same type of product and production techniques [6]. Main risks of the analytical method lie with the cost of forgotten items (incomplete work breakdown structure) and errors in the application of the learning model for series industrial production.

The other type of method is based on cost scaling from similar components or systems, based on suitable estimators and scaling laws. The scaling laws may either be empirical, or reflect "first principles" of applicable physics and engineering. The advantage of the method is that it does not require a detailed knowledge of the work breakdown, but only of relevant estimators characterising the component or system under analysis. The continuity of the cost function also permits variational calculations leading to possible cost optimisation. The main difficulty is to find the correct set of reference projects to scale from, and to establish the domain of validity of the scaling.

Unless the complete work breakdown is established and fully settled down to the component level, cost estimates are based on a hybrid of these two approaches; this is the case of the CLIC value estimate presented here.

For each item of the two upper levels ("domains" and "sub-domains") of the work breakdown structure of the CLIC accelerator complex, coordinators were appointed with the mandate to collect elementary costs from the technical groups performing design work, and to exchange information in an *ad hoc* working group, meeting on a regular basis and gradually addressing all domains and sub-domains as design work progressed. Within the "sub-domains", the work breakdown structure refines into "component", "technical system" and "sub-component" levels: depending upon the type of domain and estimation method, elementary value estimates were entered at either of these levels.

A centralised repository for the value estimates is the CLIC Study Costing Tool [7], developed and maintained by the CERN Advanced Information Systems group. It presents an on-line, updated view of the value estimates, based on the corresponding work breakdown structures which can be entered at any level. The costing tool includes features for fixed and variable costs, currency conversion, escalation and uncertainty, as well as full traceability of input data and production of tabulated reports which can be exported to Microsoft Excel for further processing.

### 5.4.4 Uncertainty, Escalation and Currency Fluctuations

The uncertainty target set for the Conceptual Design Report value estimates of both accelerator complex and detectors is $\pm 30\%$. In order to meet this goal, an analysis of the variance factors for the value of such technical projects was performed, covering all phases from technical design to procurement and contract execution in industry. As a result, the value variance factors could be grouped in three classes, in decreasing order of leverage by the project management.

The first class pertains to the technological maturity and the risk of evolution of configuration, estimated for each component or system by the expert in charge, according to three levels of relative standard deviation $\sigma_{technical}$: 0.1 for equipment of known technology, 0.2 for equipment requiring extrapolation from known technology and 0.3 for equipment requiring specific R&D.

The second class stems from the uncertainty in commercial procurement. A statistical analysis of the offers received for the procurement of accelerator components and technical systems for the LHC, in response to invitations to tender based on precise technical specifications, yielded a probability distribution function which could be modelled as exponential above the threshold of the lowest bid, with a relative standard deviation of about 0.5 [8]. Sampling randomly from such a distribution to obtain a sample of $n$ bids, and buying from the lowest bidder among the sample, yields a distribution of prices for the procured components showing a relative standard deviation of $\sigma_{procurement} = 0.5/n$. This formula, requiring assessment of the number of valid offers expected, was used to characterise the uncertainty in commercial procurement for CLIC components.

The third class of value variance factors is that resulting from the economical and financial context: price escalation and currency exchange rate fluctuations. We consider this to be outside the control of the project management, and assume it will be compensated by the funding agencies. It is therefore not part of the value uncertainty, but must nevertheless be tracked according to the procedure described below.

From the above analysis, we estimate $\sigma_{total}$ by root-mean-square summation of $\sigma_{technical}$ and $\sigma_{procurement}$, thus assumed statistically independent. We then estimate the low and high boundaries of the value estimates by respectively subtracting $\sigma_{technical}$ from and adding $\sigma_{total}$ to the estimate.

The basis for tracking price escalation is the application of economic indices such as published on a periodical basis by national or international economic agencies. These indices, each pertaining to a type or class of products or services, are however expressed in a given currency, and applicable to prices formulated in that particular currency. The first decision is therefore to choose a reference currency for the value estimates: in line with CERN practice, the CLIC value estimates are expressed in Swiss francs (CHF). Consequently, the value estimates must be escalated according to indices published by the Swiss Office Fédéral de la Statistique [9], among which we have chosen two compound indices: the global "Construction" index, updated every semester, is applied to the value estimates of civil engineering, while the global "Arts & Métiers et Industrie" index, updated monthly, is applied to all other technical systems.

Finally, in view of the period in which most of work was performed, and of the turbulent exchange rate fluctuations over the year 2011, the value estimates presented in the following are expressed in Swiss francs of December 2010, with the following average exchange rates:

1 EUR = 1.28 CHF     1 USD = 0.96 CHF     1 JPY = 0.0116 CHF

### 5.4.5 Value Estimates and Cost Drivers of Accelerator Complex

The value estimates were presented to a review panel composed of international experts in February 2012, reporting to the CLIC Steering Committee. The charge to the reviewers was to review the methodology and assumptions of the value estimates, identify incorrect or missing value information, check the consistency of the value estimates with respect to applicable reference work, review the uncertainty estimates, identify main areas of savings and cost mitigation for future work and advise the CLIC study team on matters of value estimate. In its conclusions, the review panel found no major omission and endorsed the methodology used for the value estimates. It also recommended specific points to be corrected, improved or refined. The numbers presented here were updated following the recommendations of the review panel.

The breakdown of value estimate down to the sub-domain level of the work breakdown structure is presented in Table 5.3 for the first stage of the accelerator complex according to scenarios A and B. This is illustrated in graphical form in Figure 5.8: the complexity and extent of the main-beam and drive-beam injector complexes are reflected in the breakdown. Combining the uncertainties estimated with the method described above yields the following values (1 $\sigma$):

- For CLIC 500 GeV in scenario A, 8300 +1900 -1400 MCHF
- For CLIC 500 GeV in scenario B, 7400 +1700 -1300 MCHF

The difference in value between the two scenarios is the price to pay for the higher luminosity of 500 GeV in scenario A: the higher beam current requires additional RF power to be installed in the injectors and larger-aperture accelerating structures operating at lower gradient, thus calling for an additional fifth sector in each of the main linacs.

Table 5.3: Value estimates of CLIC 500 GeV according to staging scenarios A and B.

| | | Value A [MCHF] | Value B [MCHF] |
|---|---|---|---|
| Main beam production | Injectors | 449 | 339 |
| | Damping rings | 383 | 408 |
| | Beam transport | 612 | 456 |
| | **Total** | 1443 | 1203 |
| Drive beam production | Injectors | 1384 | 1248 |
| | Frequency multiplication | 135 | 135 |
| | Beam transport | 260 | 217 |
| | **Total** | 1779 | 1599 |
| Two-beam accelerators | Two-beam modules | 2215 | 2002 |
| | Post-decelerators | 46 | 37 |
| | **Total** | 2260 | 2038 |
| Interaction region | Beam delivery systems | 62 | 62 |
| | Experimental area | 23 | 23 |
| | Post-collision line | 47 | 47 |
| | **Total** | 132 | 132 |
| Civil engineering and services | Civil engineering | 1432 | 1382 |
| | Electricity | 326 | 282 |
| | Survey and alignment | 31 | 31 |
| | Fluids | 494 | 445 |
| | Transport/installation | 100 | 90 |
| | Safety | 20 | 20 |
| | **Total** | 2403 | 2250 |
| Machine control and operational infrastructure | Machine control infrastructure | 226 | 183 |
| | Machine protection | 3 | 3 |
| | Access safety & control system | 20 | 18 |
| | Technical alarm system | 13 | 12 |
| | **Total** | 262 | 216 |
| | **Grand total (rounded)** | 8300 | 7400 |

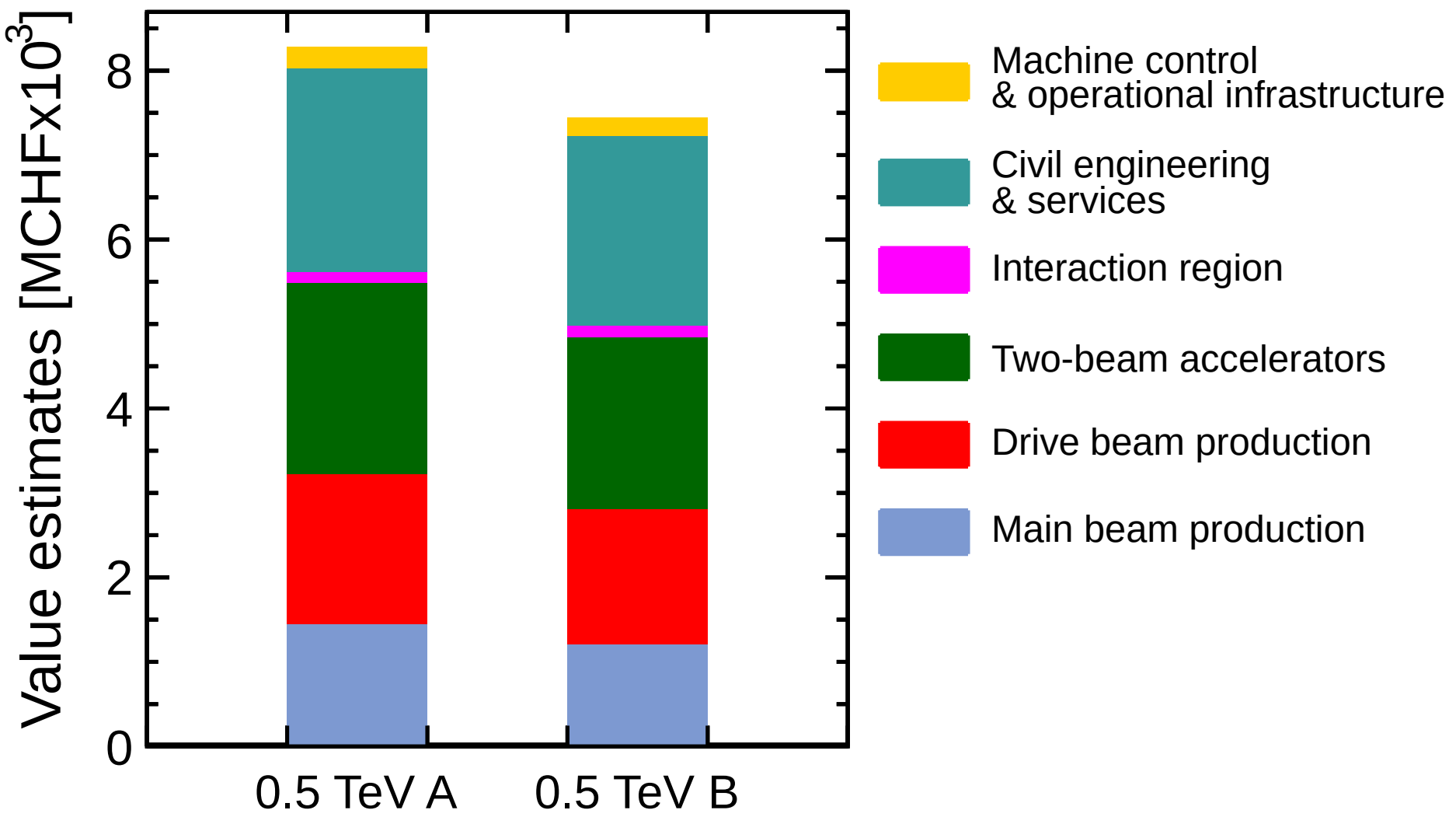

Fig. 5.8: Cost structure of the CLIC accelerator complex at 500 GeV for scenarios A and B.

A first estimate of the cost structure and value of the second stage enables to calculate the incremental value per unit of collision energy. It is about 4 MCHF/GeV for scenario B.

Technical cost drivers have been identified in the CDR study phase, together with, for a number of them, potential cost mitigation alternatives which will need to be addressed in the subsequent phase of the study. Examples of such alternatives are the replacement of the hexapods for the stabilisation of the main beam quadrupoles with beam steering, the doubling in length and thus the halving in number of the support girders for the two-beam acceleration modules, or an alternative technology for the construction of the accelerating structures involving assembled quadrants instead of stacked discs. The overall savings potential through this process is estimated of the order of 10% of the total value, i.e. within the uncertainty presented above. An important structural cost driver however stems from the wide energy-staging range of the CLIC programme, thus imposing over-investments in the first stages, e.g. in infrastructure and services as well as in the injector complex. Revising the collision energy for which the technical design is optimised, while preserving the potential to ultimately reach 3 TeV, is expected to provide the main lever for further cost reduction.

### 5.4.6 Labour Estimates for Construction of the Accelerator Complex

A first estimate of the explicit labour needed for construction of the CLIC accelerator complex was obtained by assuming a fixed ratio between personnel and material expenditure for projects of similar nature and size, and scaling with respect to the closest such project realised today, namely the LHC accelerator at CERN, which required some 7000 FTE·years for a material cost of 3690 MCHF (December 2010), i.e. a ratio of about 1.9 FTE·year/MCHF. About 40% of this labor was scientific and engineering personnel, and the remaining 60% technical and execution.

From this approach, construction of the first stage of the CLIC accelerator complex would require 15700 FTE·years of explicit labour according to scenario A, and 14100 FTE·years according to scenario B. It is worth noting that in spite of this very crude approach, these numbers are not too far from those taken in the ILC Reference Design Report [4], yielding a ratio of explicit labour to material of about 1.7 FTE·year/MCHF.

Table 5.4: Value estimate of the CLIC detectors.

| | CLIC_ILD (MCHF) | CLIC_SiD (MCHF) |
|---|---|---|
| Vertex | 13 | 15 |
| Tracker | 51 | 17 |
| Electromagnetic calorimeter | 197 | 89 |
| Hadronic calorimeter | 144 | 86 |
| Muon system | 28 | 22 |
| Coil and yoke | 117 | 123 |
| Other | 11 | 12 |
| **Total (rounded)** | 560 | 360 |

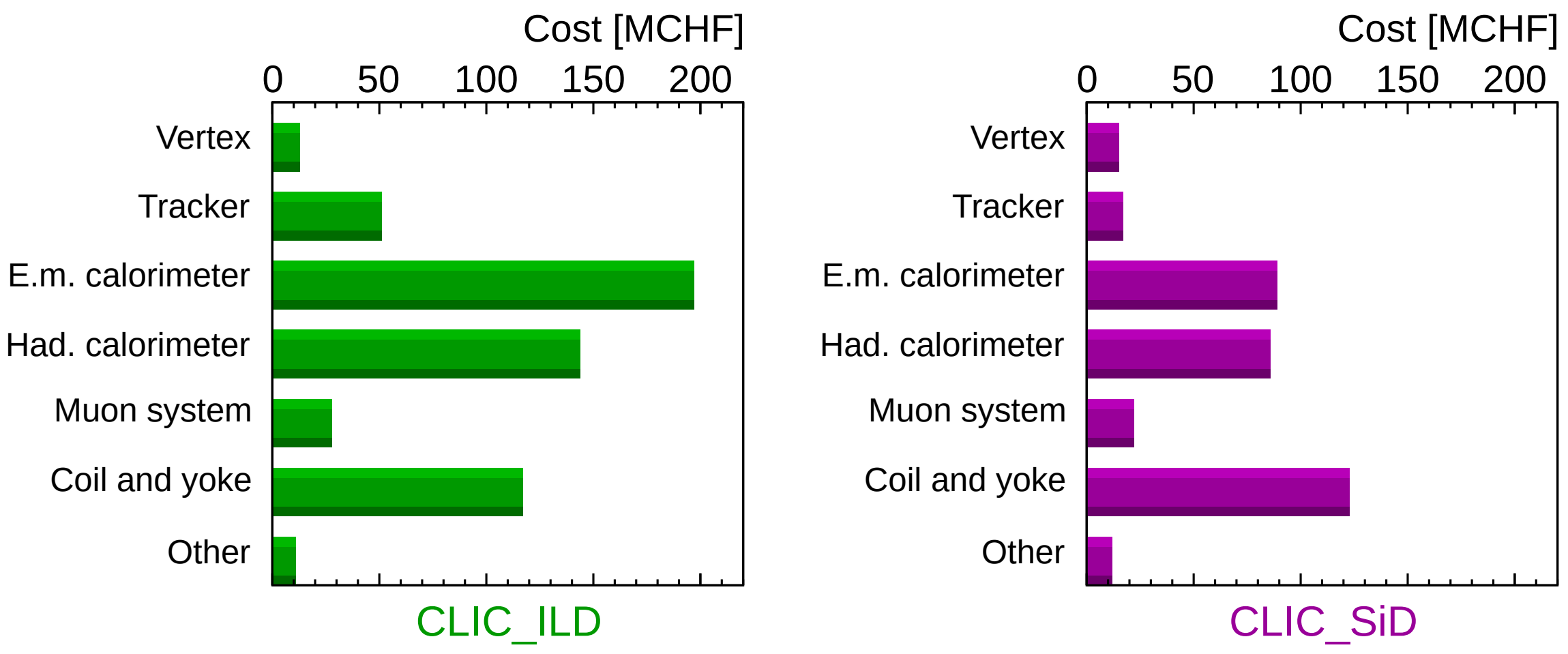

Fig. 5.9: Cost structure of the CLIC detectors.

### 5.4.7 Value Estimates and Cost Drivers of Detectors

The methodology for estimating the value of the CLIC detectors [5] is similar to that used for the accelerator complex, based on work breakdown structures with more or less granularity[1]. The target for uncertainty is also $\pm 30\%$. There are however a few differences in the approaches, stemming from the specificities of the detectors and their construction. The value of a detector depends critically on the unit costs of a limited number of specific materials and commodities, for which fixed values have been agreed for the estimate (see Annex C of [10]). The use of general industrial indices for escalation may then imperfectly reflect the real situation, particularly in case of large price variations in a particular commodity. In addition, explicit labour has not been estimated for the detectors. However, based on the experience with the ATLAS and CMS projects, the construction and testing efforts may be assumed at 500 FTE for each year of construction of a CLIC detector. Finally, at the interface between detector and experimental area, the mobile platforms carrying the detectors, the "anti-solenoids" for compensation of stray magnetic field and the proximity equipment associated with the final-focus quadrupoles are included in the

[1] The preliminary CLIC detector value estimated were extrapolated, for their major part, from the ILC Letter of Intent (LoI) cost estimates, taking the significant changes (technology, dimensions) for CLIC into account and using modified unit costs. Therefore they cannot be directly compared with the ILC estimates.

"interaction region" part of the value estimate for the accelerator complex, and thus do not appear here.

The high-level breakdown of the value estimates of the CLIC_ILD and CLIC_SiD detectors appears in Table 5.4 and is illustrated in Figure 5.9, showing some differences in the absolute value and cost structure of the two detectors: the larger dimensions in the tracker volume of CLIC_ILD lead to large electromagnetic and hadronic calorimeters, thus to a significantly higher value. For both detectors the main cost drivers are the cost of silicon sensors for the Electromagnetic Calorimeter (ECAL), and of tungsten for the Hadronic Calorimeter (HCAL). Studies are continuing to investigate alternative designs of the calorimeters, including a possible change of technology for the active planes, as options towards overall cost reduction.

# Chapter 6

# CLIC Physics Potential

To demonstrate that experiments at CLIC are capable of delivering the required performance for detailed Standard Model (SM) and Beyond the Standard Model (BSM) studies at the energy frontier, as outlined in Chapter 2, several benchmark physics studies were performed in the scenario of a staged construction of the accelerator. These complement the studies performed in the context of the CLIC Physics and Detectors volume of the CDR [1], which focuses mainly on the experimentally most challenging case of 3 TeV collisions. Rather than providing a complete overview of the physics capabilities of CLIC at various energies, the additional processes studied for the staged energy scenario are selected to illustrate the performance for precision SM measurements as well as the capabilities for the exploration of New Physics in one specific model. The processes studied in this context are Higgs physics at all energy stages, top physics at the lowest energy stage and various Supersymmetric (SUSY) processes in a specific SUSY model (*model III*, see Chapter 2) as an example for BSM physics at higher CLIC energies. In addition, the sensitivity to a high-mass $Z'$ as an alternative BSM model is investigated. A description of each of the benchmark processes is given in [2]. The results of the benchmark studies presented here as well as of the studies included in [1] are summarised in tables at the end of this chapter.

## 6.1 Benchmark Studies

The benchmark studies are performed using detailed GEANT4 [3, 4] simulations of the CLIC detector concepts introduced in Chapter 4. The presented studies for a given channel are each performed only in one of the two concepts. They use realistic experimental conditions including the luminosity spectrum at the different collision energies and the overlay of pile-up from $\gamma\gamma \rightarrow$ hadrons background events taking into account the time structure of the CLIC beams. The results are based on full event reconstruction including tracking, the application of particle flow algorithms with timing cuts and flavour tagging.

Events are generated with the WHIZARD [5, 6] or the PYTHIA [7] generator. Unpolarised beams are assumed for the SM and SUSY benchmarks, while the impact of electron and positron polarisation is assessed within the $Z'$ study. Parton showering and hadronisation is performed with PYTHIA. In the event generation, the CLIC luminosity spectrum for the appropriate energy, generated with GUINEAPIG [8], as well as initial state radiation are taken into account. The $\gamma\gamma \rightarrow$ hadrons background events are generated with GUINEAPIG and are hadronised using PYTHIA. These events are overlaid on the physics events prior to event reconstruction. Further details on the methodology of the benchmark studies are given in [1].

For these studies, realistic scenarios for the integrated luminosity are considered. At 350 GeV and at 500 GeV, 500 fb$^{-1}$ are assumed, with up to 100 fb$^{-1}$ dedicated to a scan of the top pair production threshold. At higher energies with correspondingly higher instantaneous luminosities, integrated luminosities of 1.5 ab$^{-1}$ at 1.4 TeV and 2 ab$^{-1}$ at 3 TeV are assumed. As a general rule, charge conjugation is always implied.

## 6.2 Higgs Physics

A linear $e^+e^-$ collider offers excellent conditions for precision studies of a low-mass Standard Model Higgs, at various energy stages. The 125 GeV mass of the Higgs-like boson recently observed at the LHC [9, 10] provides a rich spectrum of production and decay modes that can be measured with high precision at CLIC, as discussed in Section 2.2.2 and as shown in Figure 2.1. At energies above the production threshold and up to 500 GeV, the measurement of the $Z$ recoil in $HZ$ production offers a model-independent coupling and mass determination, a unique feature of lepton colliders. These measurements, performed without consideration of the Higgs decay itself, complement direct branching fraction measurements at these energies. At higher energies, the top Yukawa coupling and the tri-linear Higgs self-coupling become accessible using the $ttH$ final state and double Higgs production, respectively. The high luminosity at the highest energies, combined with the rising $WW$ fusion cross-section, enables the determination of the branching fraction of rare processes such as the decay into muons at 3 TeV.

In this report, the branching ratio studies at 3 TeV reported in [1] are complemented with the investigation of the potential for model-independent cross-section and mass measurements in the $ZH$ channel at 350 GeV, mass and cross-section measurements at 500 GeV, the measurement of the branching fraction of $H \to \tau^+\tau^-$ at 1.4 TeV, and by a study of the achievable precision for a measurement of the triple Higgs coupling in $WW$ fusion at 1.4 TeV and at 3 TeV. All studies are performed assuming a Standard Model Higgs of mass 120 GeV. Since cross-sections and decay branching ratios are not dramatically different for a mass of 125 GeV, the conclusions drawn from the present studies also apply to the boson observed at the LHC.

### 6.2.1 Model-independent Cross-Section and Mass Measurement

To perform model-independent measurements in the reaction $e^+e^- \to ZH$, the $Z$ boson is reconstructed in its decay to muon and electron pairs. The Higgs boson is identified through the measurement of the mass of the system recoiling against the $Z$ boson, defined by $M^2_{recoil} = s + M^2_Z - 2E_Z\sqrt{s}$. From the distribution of the recoil mass, the $HZ$ production cross-section and the Higgs mass are determined with a combined signal and background fit. The width of the distribution, and therefore the achievable precision, is strongly influenced by the momentum resolution for leptons.

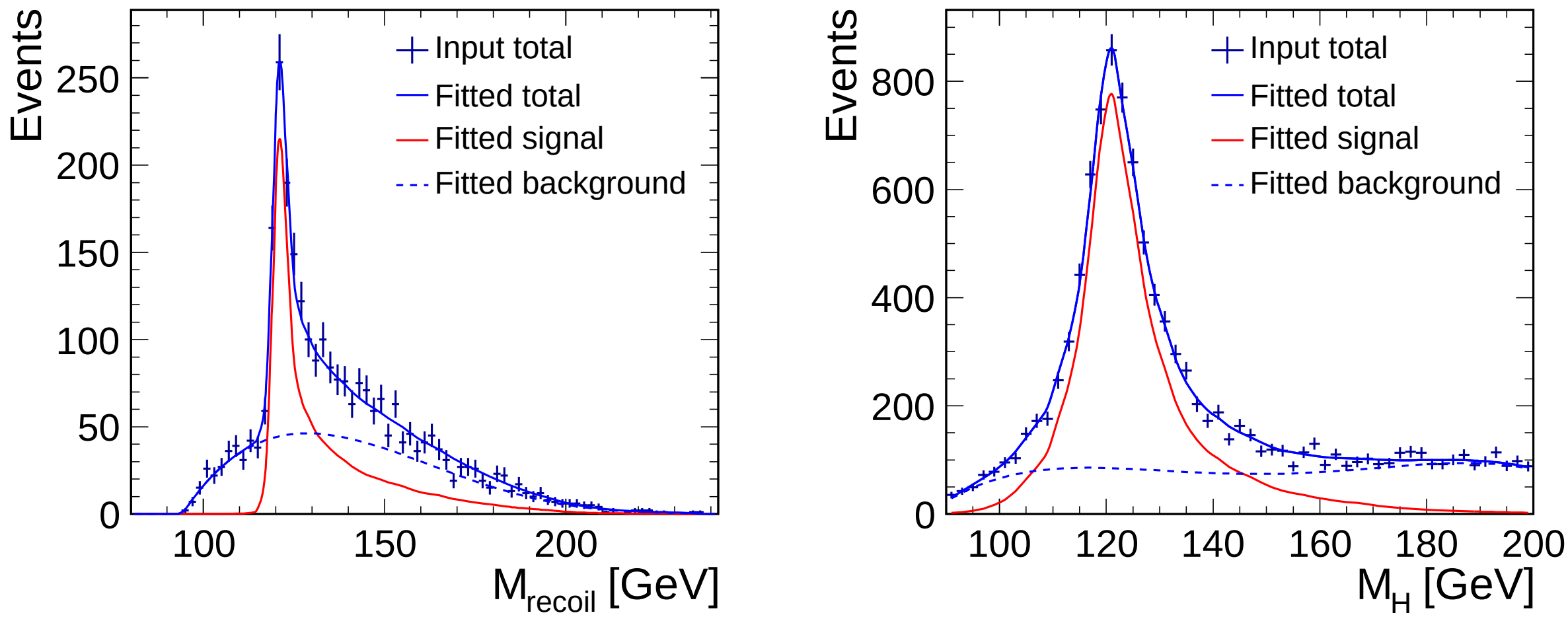

Fig. 6.1: Recoil mass distribution in $e^+e^- \to ZH \to \mu^+\mu^-X$ at 350 GeV for an integrated luminosity of 500 fb$^{-1}$. A fit to signal and background is also shown (left). Distribution of the reconstructed Higgs mass in $e^+e^- \to ZH \to q\bar{q}q\bar{q}$ at 500 GeV with explicit reconstruction of the Higgs boson (right).

Figure 6.1 left shows the recoil mass distribution in the process $e^+e^- \to ZH \to \mu^+\mu^-X$ at 350 GeV for an integrated luminosity of 500 fb$^{-1}$. From this distribution, the mass and the cross-section are extracted with a fit based on a simplified kernel estimation. The background is modelled by fitting a background function with a predetermined shape given by a $4^{th}$ order polynomial to a background-only function to determine the shape of the distribution. The normalisation of the background is a free parameter in the fit. The cross-section can be determined with a precision of 4.9%, and the Higgs mass with a statistical uncertainty of 131 MeV. In the $Z$ decay to $e^+e^-$, the uncertainties are larger, with 7.9% on the cross-section after the application of a bremsstrahlung correction for the final state electrons. More details on this analysis, as well as on the one at 500 GeV described in the following, can be found in [11].

### 6.2.2 Mass and Cross-Section Measurements at 500 GeV

At an energy of 500 GeV, the uncertainties of the recoil mass measurement are considerably larger. This is, on the one hand, due to a reduced momentum resolution at the higher lepton energies and, on the other hand, due to a reduction of the $ZH$ cross-section and overall less favourable background conditions. At this energy, the mass and cross-section measurement is studied by explicitly reconstructing the final-state Higgs boson in its two-quark decay, predominantly into $b\bar{b}$. Two final states have been investigated: the four-jet final state, originating from the process $e^+e^- \to ZH \to q\bar{q}q\bar{q}$, shown in Figure 6.1 right, and the two jet two neutrino final state, which receives contributions both from Higgsstrahlung, $e^+e^- \to ZH \to \nu\bar{\nu}q\bar{q}$, and $WW$ fusion, $e^+e^- \to H\nu\bar{\nu} \to \nu\bar{\nu}q\bar{q}$. With an integrated luminosity of 500 fb$^{-1}$, a statistical precision of 100 MeV on the mass is reached for both final states, with a precision of 1.6% and 1.0% on $\sigma \times$ BR for the $q\bar{q}q\bar{q}$ and the $\nu\bar{\nu}q\bar{q}$ final state, respectively, considering final-state dependent selection efficiencies. From a three component fit of the $p_{\mathrm{T}}$ distribution of the di-jet system in the $\nu\bar{\nu}q\bar{q}$ final state the cross-section ratio of $WW$ fusion and Higgsstrahlung is determined with a statistical precision of 5.1%, thus determining the ratio of the coupling $g_{HZZ}/g_{HWW}$ to 2.5%.

### 6.2.3 Higgs Self-Coupling

The measurement of the Higgs tri-linear self-coupling is a key component of a complete study of the Higgs mechanism, since it provides the possibility for a direct exploration of the Higgs potential. Above an energy of 1 TeV, the dominating process for double Higgs production is the $WW$ fusion process $e^+e^- \to HH\nu_e\bar{\nu}_e$, shown in Figure 6.2. This process is currently being studied both at 1.4 TeV and at 3 TeV. The low cross-sections of 0.15 fb at 1.4 TeV and 0.6 fb at 3 TeV make this a very challenging analysis, where background rejection is of key importance.

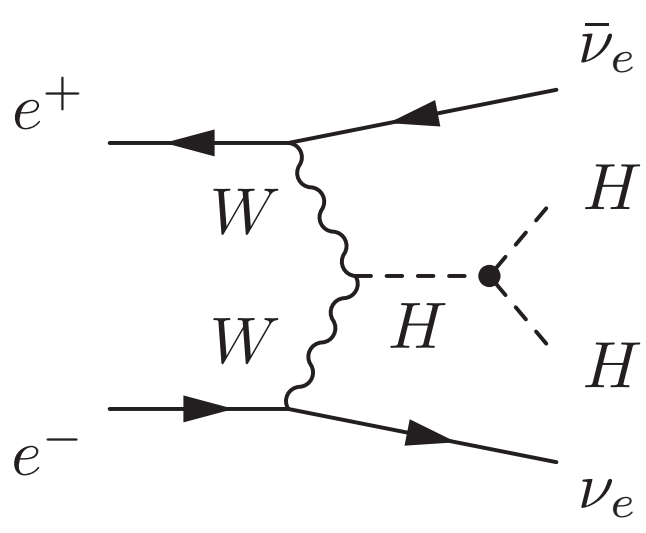

Fig. 6.2: Double Higgs production in $WW$ fusion, providing sensitivity to the Higgs tri-linear self-coupling.

The tri-linear Higgs coupling is determined via a measurement of the cross section of double Higgs production, using WHIZARD to derive the factor relating cross-section and coupling, correcting for other diagrams contributing to the same final state without a triple Higgs vertex. At present, the studies are performed assuming unpolarised beams. With 80% left-handedly polarised electrons and unpolarised positrons the signal cross-section increases by approximately 70% to 80%. Preliminary results indicate that a measurement of the coupling with a statistical precision of 20% or better will be possible at both centre-of-mass energies, with a potential for higher precision at 3 TeV. The full results will be documented in [12].

### 6.2.4 Higgs Branching Fractions and SUSY Higgs Sectors at Higher Energy

The increasing cross-section of Higgs production in $WW$ fusion, coupled with the possibility to acquire large integrated luminosities, provides large samples of Higgs bosons at 1.4 TeV and 3 TeV, with approximately 1 million produced bosons at 3 TeV. This enables the precise measurement of branching fractions extending to rare decays. In [1], the study of a measurement of $\sigma \times$ BR of $H \to b\bar{b}$, $H \to c\bar{c}$ and $H \to \mu^+\mu^-$ was performed at 3 TeV, giving statistical precisions of 0.22%, 3.2% and 15.7%, respectively. In addition, a preliminary study of $H \to \tau^+\tau^-$ [13] was performed at 1.4 TeV considering only hadronic $\tau$ decays, achieving a statistical precision of 3.7% on $\sigma \times$ BR. This analysis re-uses all background samples generated for the study of staus discussed in Section 6.4.1.2, which include the signal channel $H \to \tau^+\tau^-$. These samples were produced with cuts already imposed during the production phase. These cuts are too strict for the present analysis. It is thus expected that better results can be achieved with more appropriate cuts at production level.

The production of heavy Higgs bosons $H^0$, $A^0$ and $H^\pm$ has been studied in SUSY *models I* and *II* at 3 TeV in [1]. Despite challenging final states with up to 8 jets in the decay of the $H^\pm$ to top and bottom quarks high signal purities and precise mass and width measurements at the level of 0.3% and 20–30% are achieved, respectively, for heavy Higgs masses in the 740–900 GeV range.

## 6.3 Top Quark Physics

As the heaviest SM particle, the top quark is of particular interest since it most strongly couples to the Higgs field and may provide sensitivity to BSM physics. Experiments at $e^+e^-$ colliders offer the possibility for a wide variety of studies involving top quarks, as outlined in Chapter 2. Among those is the precise determination of the top quark mass, which is possible with two different techniques: through the direct reconstruction of top quarks from their decay products at energies above the production threshold, and through a scan of the top-pair production threshold. The latter technique has the advantage of providing the mass measurement in a theoretically well-defined scheme, while the former measurement can be performed essentially at arbitrary energies above threshold, however with potentially significant uncertainties when transferring the measured invariant mass to a theoretically meaningful value.

Here, we investigate the potential for the determination of the top quark mass from a measurement of the top-pair production cross-section at several energies around the threshold, with a total integrated luminosity of up to 100 fb$^{-1}$. For the correct description of the cross-section near threshold, the inclusion of high-order QCD contributions is necessary. Since no appropriate event generator is available at present, the study follows the strategy of earlier studies performed for the TESLA collider [14] by factorising the simulation study into the determination of event selection efficiency and background contamination and the calculation of the top-pair production threshold. The study is documented in more detail in [15].

The event selection efficiency and the background contributions, mainly from di- and tri-boson production, are determined using events generated with PYTHIA at a collision energy of 352 GeV with a top mass of 174 GeV. These events are fully simulated including pile-up from $\gamma\gamma \to$ hadrons background. The top-pair signal cross-section is determined using full NNLO calculations provided by TOPPIK [16, 17], with 174 GeV in the 1S mass scheme used as input parameter, corrected for initial state radiation and the CLIC beam energy spectrum. In the analysis, the signal yield as a function of collision energy is determined from this calculated cross-section, using the event selection efficiency obtained from the full simulations. Top pair events are identified in the fully hadronic decay mode $t\bar{t} \to W^+bW^-\bar{b} \to q\bar{q}q\bar{q}b\bar{b}$ and in the semi-leptonic mode $t\bar{t} \to W^+bW^-\bar{b} \to q\bar{q}\ell^\pm\nu_\ell b\bar{b}$, $(l = e, \mu)$, with background rejection provided by a kinematic fit imposing constraints based on top-pair production and by using a binned likelihood method.

The top mass is determined with a template fit to the background-subtracted cross-section measurements, with templates for the cross-section for different top masses calculated using TOPPIK. Figure 6.3 left shows the background-subtracted cross-section obtained with a threshold scan in 1 GeV steps with an integrated luminosity of 10 fb$^{-1}$ per data point, together with the input cross-section provided by TOPPIK, as well as the cross-sections for top masses shifted by ±200 MeV. The top-pair production cross-section is also sensitive to the strong coupling constant, in particular in the region above threshold. Using the first six data points, the 1S mass is determined with a statistical precision of 21 MeV, with a systematic uncertainty of 20 MeV from the current world-average uncertainty of $\alpha_s$. Additional systematics from theory uncertainties are 15 MeV (45 MeV), assuming an overall normalisation uncertainty of 1% (3%). The top mass and the strong coupling constant are also determined simultaneously using the full ten points of the scan. This results in a statistical uncertainty of 33 MeV on the mass and 0.0009 on the strong coupling constant, with the correlation of the two variables as well as the uncertainties illustrated in Figure 6.3 right. The systematic uncertainties from theory are 6 MeV (13 MeV) on the mass and 0.0009 (0.0023) on $\alpha_s$ for a normalisation uncertainty of 1% (3%). Further improvements of the precision of these measurements can be achieved by considering additional observables, which are not included in this analysis.

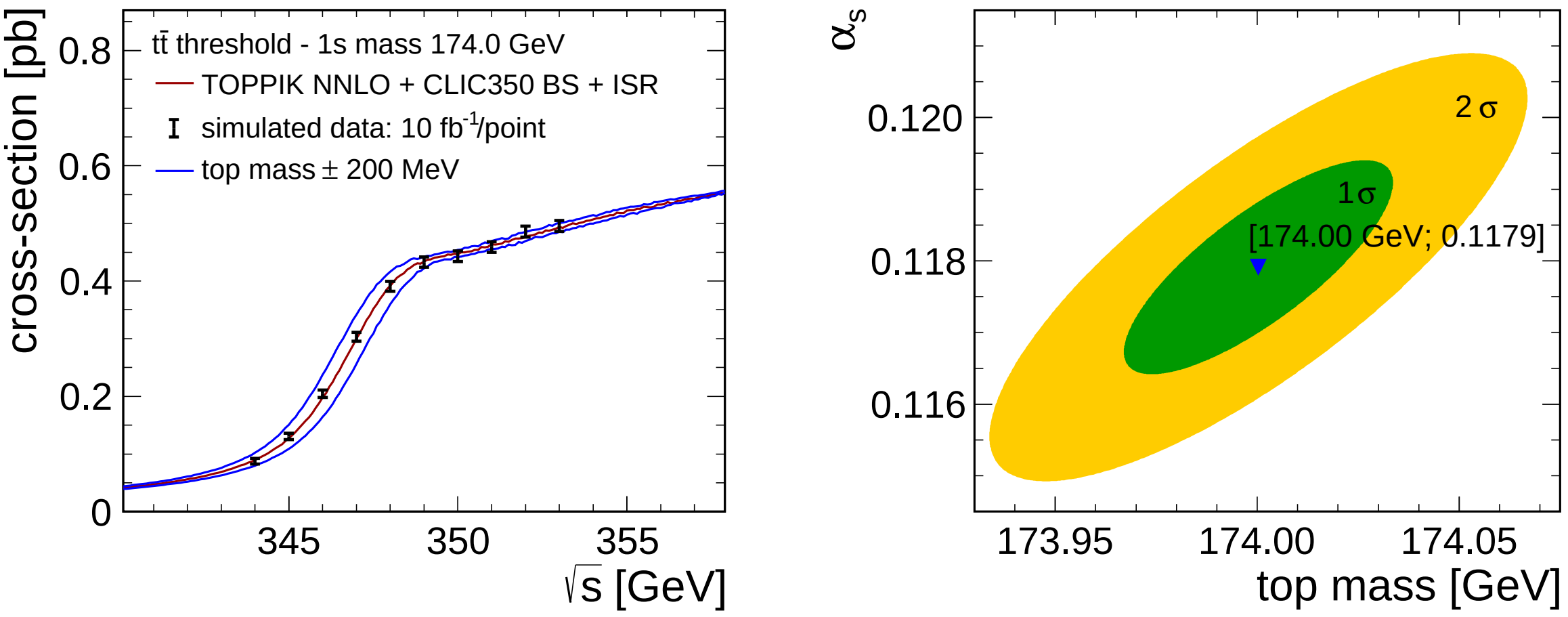

Fig. 6.3: Background-subtracted cross-section for 10 fb$^{-1}$ per data point, together with the cross-section for the generator mass of 174 GeV as well as for a shift in mass of ±200 MeV (left). Simultaneous fit of the top mass and the strong coupling constant, showing the correlation of the two variables and the achieved precision (right).

This study complements a previous CLIC study of top mass measurements at 500 GeV, showing that the invariant mass of the top quark can be determined with a precision of better than 100 MeV with 100 fb$^{-1}$ in fully hadronic and semi-leptonic decays of the top pairs [1], and demonstrates that precision top measurements are possible at CLIC both at and above threshold.

## 6.4 Supersymmetry

For the study of the physics performance for SUSY, a specific mSUGRA model (*model III*) with non-universal squark masses was selected, as presented in Chapter 2. This model, which is compatible with current LHC data, has heavy first and second generation squarks with masses in the 2 TeV range, third generation squarks with masses around 1 TeV, and sleptons as well as the lightest two neutralinos and the lightest chargino within the reach of CLIC at 1.4 TeV. The studies presented here cover sleptons from all three generations accessible via pair production at 1.4 TeV. These studies complement the SUSY studies at 3 TeV in *model I* and *model II* reported in [1]. In all cases, the goal of the analysis is to determine the mass and the production cross-section of the respective sparticles.

### 6.4.1 Sleptons

The reconstruction of sleptons primarily requires highly efficient lepton identification and precise energy and momentum measurements in an experimental environment characterised by a high level of hadronic background. While this background has no effect on final states with muons and relatively small effects on the reconstruction efficiency for final states with electrons, it is a key factor in the reconstruction of $\tau$ leptons. In the following, we study the precision achievable for mass and cross-section measurements for first, second and third generation sleptons. The former two also provide a precise measurement of the $\tilde{\chi}_1^0$ and $\tilde{\chi}_1^\pm$ masses, which can serve as input for other SUSY studies at a linear collider.

#### 6.4.1.1 First and Second Generation Sleptons

In SUSY *model III* the following processes for the production of first and second generation sleptons are investigated at 1.4 TeV:

- $e^+e^- \to \tilde{\mu}_R^+\tilde{\mu}_R^- \to \mu^+\mu^-\tilde{\chi}_1^0\tilde{\chi}_1^0$
- $e^+e^- \to \tilde{e}_R^+\tilde{e}_R^- \to e^+e^-\tilde{\chi}_1^0\tilde{\chi}_1^0$
- $e^+e^- \to \tilde{\nu}_e\tilde{\nu}_e \to e^+e^-\tilde{\chi}_1^\pm\tilde{\chi}_1^\pm \to e^+e^-W^+W^-\tilde{\chi}_1^0\tilde{\chi}_1^0$

For the first two processes, the branching fraction is 100%, while it is 53% for the decay of the $\tilde{\nu}_e$. The particle masses are around 559 GeV for the charged sleptons, and 644 GeV for the sneutrino. For all channels, the masses of the slepton and the neutralino or chargino are determined from the upper and lower edge of the energy distribution of the reconstructed final-state leptons.

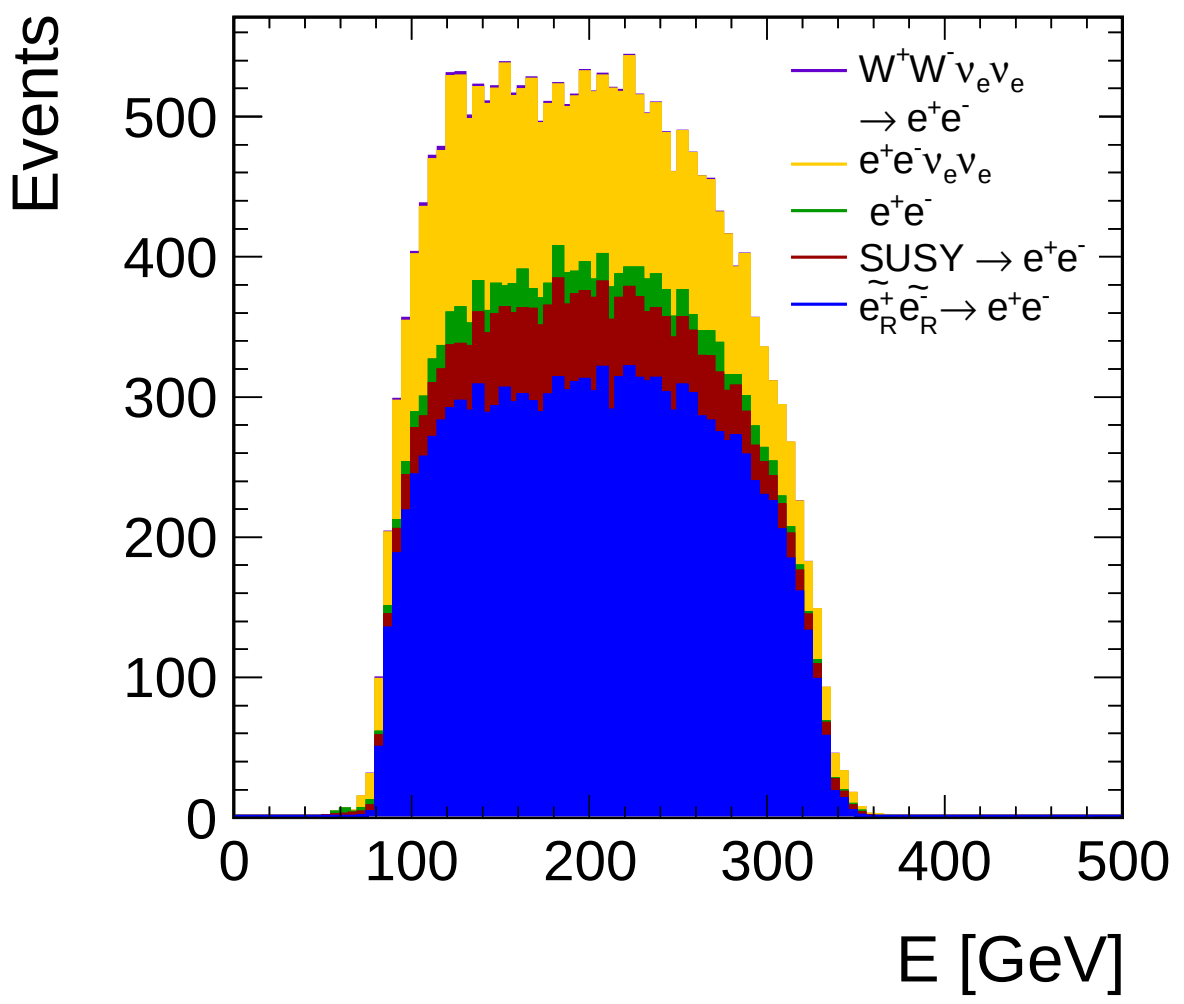

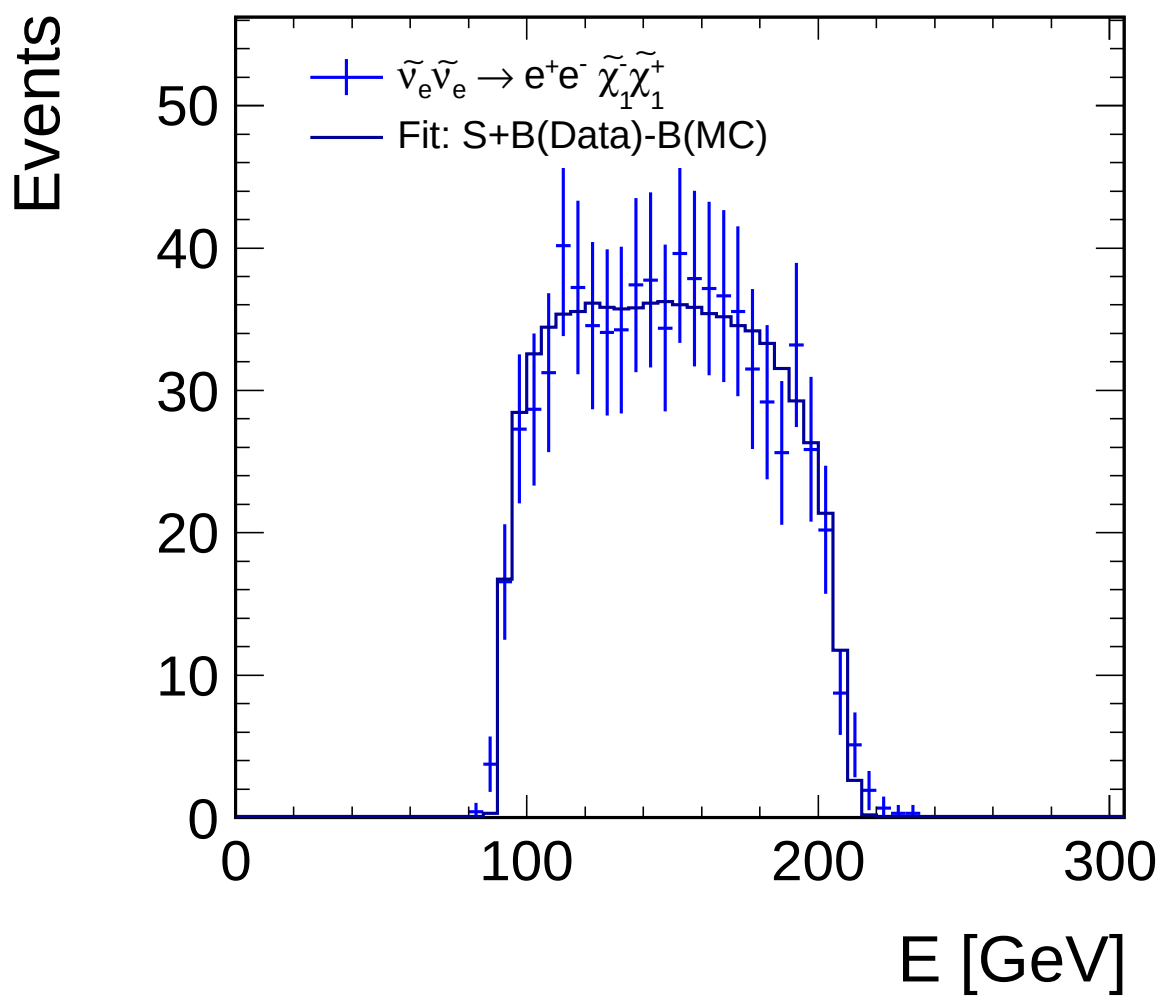

Fig. 6.4: Reconstructed charged lepton energy for the process $e^+e^- \to \tilde{e}_R^+\tilde{e}_R^- \to e^+e^-\tilde{\chi}_1^0\tilde{\chi}_1^0$ together with relevant background processes after event selection with a BDT. Signal and background histograms are stacked (left). Background-subtracted distribution of the reconstructed electron energy for the process $e^+e^- \to \tilde{\nu}_e\tilde{\nu}_e \to e^+e^-\tilde{\chi}_1^\pm\tilde{\chi}_1^\pm \to e^+e^-W^+W^-\tilde{\chi}_1^0\tilde{\chi}_1^0$ (right).

Signal events are identified by high-energy leptons, and discriminated from SM and SUSY background using a boosted decision tree (BDT) based on variables of the di-lepton system. The signal selection efficiency is 90% for the di-muon and the di-electron plus four jets final states, and 80% for the di-electron final state. Figure 6.4 left shows the reconstructed lepton energy for $\tilde{e}_R$ pair production together with the relevant background processes after the application of the BDT-based selection cuts. The $\tilde{e}_R$ and the $\tilde{\chi}_1^0$ masses as well as the cross-section are determined from a $\chi^2$ fit of the background-subtracted cross-section, taking the influence of Initial State Radiation (ISR) and the CLIC beam energy spectrum into account. Figure 6.4 right shows the background-subtracted energy distribution of electrons and positrons for the case of $\tilde{\nu}_e$ production, which is used to determine the $\tilde{\nu}_e$ and the $\tilde{\chi}_1^\pm$ masses. The masses are determined with a statistical accuracy of 0.1% for the $\tilde{\mu}_R$, the $\tilde{e}_R$, and the $\tilde{\chi}_1^0$. For the $\tilde{\nu}_e$ and the $\tilde{\chi}_1^\pm$, the accuracy of the mass determination is 2.5% and 2.7%, respectively. Systematic errors from the event selection and background subtraction are at the level of 0.1% to 0.2% for the slepton masses, and 0.2% to 0.4% for the neutralino and chargino mass. The cross-sections for $e^+e^- \to \tilde{\mu}_R^+\tilde{\mu}_R^-$, $e^+e^- \rightarrow \tilde{e}_R^+\tilde{e}_R^-$ and $e^+e^- \rightarrow \tilde{\nu}_e\tilde{\nu}_e$ are measured with a statistical uncertainty of 2.7%, 1.1% and 3.6%, respectively. More details on the study are given in [18].

For higher-mass sleptons in the TeV range, an identical study was performed for a 3 TeV CLIC collider, achieving comparable precision [1]. This demonstrates that such measurements are very robust at CLIC, provided the particles are produced with sufficient cross-sections at the available centre-of-mass energy.

#### *6.4.1.2 Staus*

The potential for $\tilde{\tau}$ measurements is studied with the process $e^+e^- \rightarrow \tilde{\tau}_1\tilde{\tau}_1$, as presented in detail in [13]. In *model III*, the $\tilde{\tau}_1$ with a mass of 517 GeV decays with a branching ratio of 99% to $\tau$ and $\tilde{\chi}_1^0$, making $\tau$ identification the key performance criterion for this analysis. The $\tau$ particles are reconstructed using a seeded cone-based jet clustering algorithm which places additional constraints on the seed's $p_T$, the invariant mass, and the isolation of the $\tau$ candidate [19]. Only $\tau$-leptons decaying to hadrons with a branching fraction of 65% are considered in the analysis. For this, $\tau$ jets with either one or three charged tracks and no leptons are selected. Signal events are discriminated from SM and SUSY background by means of a boosted decision tree using event shape variables as well as kinematic information from the $\tau$ candidates and overall event energy information.

The low multiplicity of charged tracks in the $\tau$ decay makes this analysis particularly sensitive to additional hadrons originating from $\gamma\gamma \rightarrow$ hadrons background. While the background does not affect the energy reconstruction of the $\tau$ candidates, it strongly affects the identification efficiency. With tight timing cuts following the particle flow reconstruction, the impact of the background can be largely eliminated. Figure 6.5 left shows the distribution of the reconstructed $\tau$ energy after all cuts for signal and background.

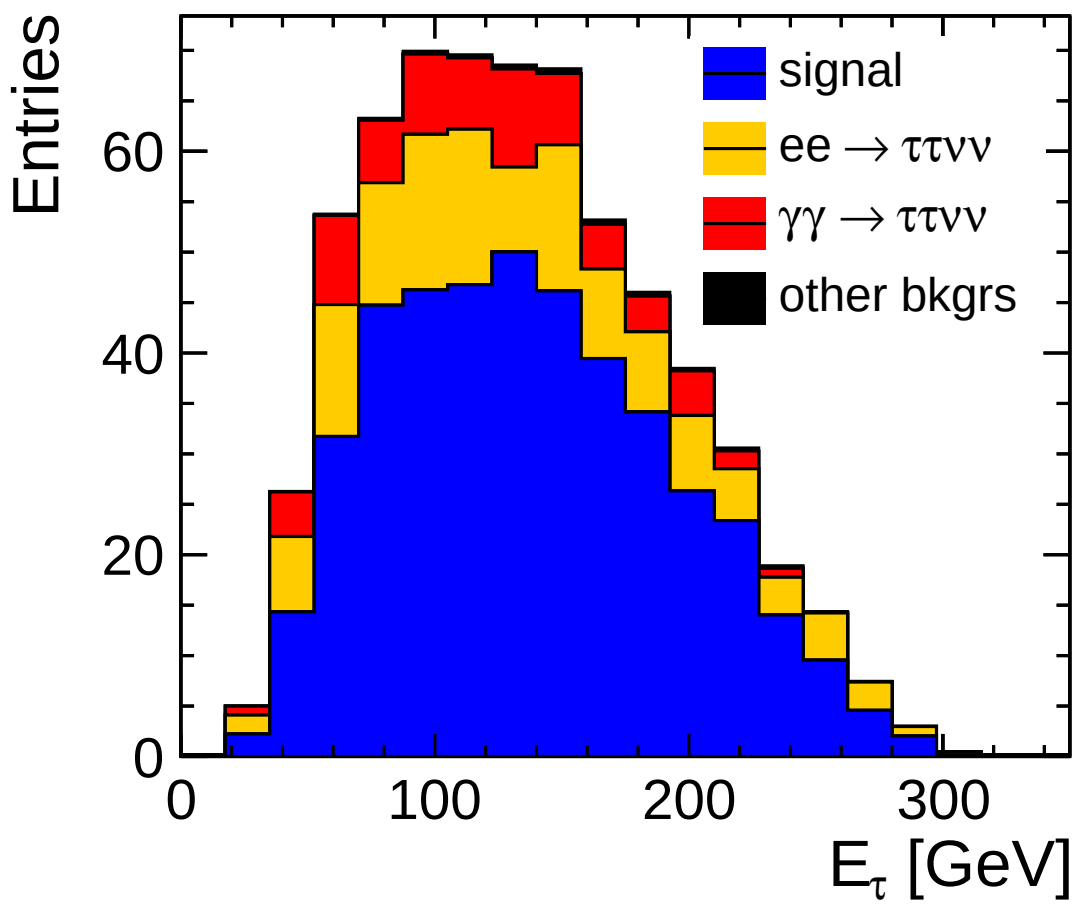

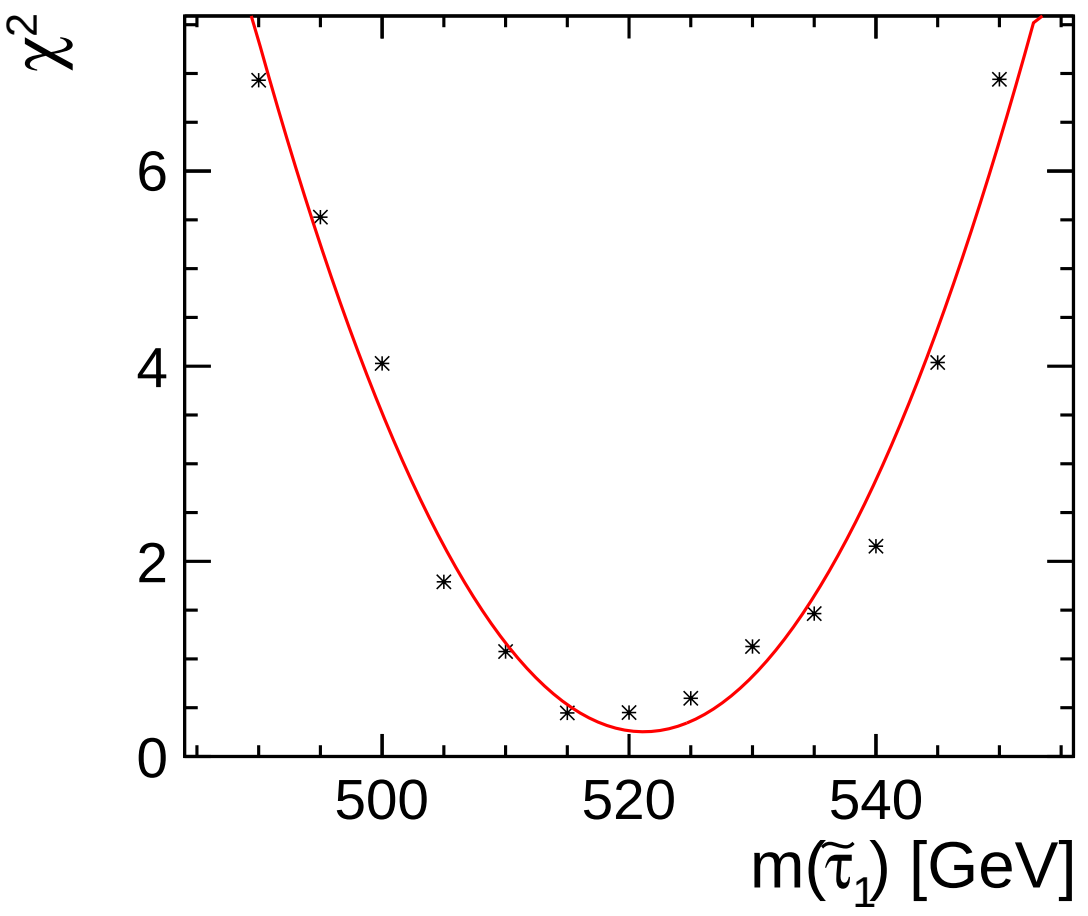

Fig. 6.5: Reconstructed $\tau$ energy after event selection with a BDT. Signal and background histograms are stacked (left). $\chi^2$ values for templates with different $\tilde{\tau}$ mass assumptions compared to the reconstructed energy distribution. The measured $\tilde{\tau}$ mass is given by the minimum of the distribution. The generated $\tilde{\tau}$ mass is 517 GeV (right).

The mass and production cross-section of the $\tilde{\tau}_1$ are determined from a two-dimensional template fit to the energy distribution of the reconstructed $\tau$-leptons using templates with varying $\tilde{\tau}_1$ masses. With this fit, the $\tilde{\tau}_1$ mass is determined with a statistical precision of 2%, and the cross-section is determined with a statistical precision of 7.5%. For the fit, the $\tilde{\chi}_1^0$ mass is assumed to be known from the measurement of first and second generation sleptons discussed above. The uncertainty on the $\tilde{\chi}_1^0$ mass has a negligible effect compared to the statistical uncertainties of the $\tilde{\tau}_1$ mass determination. The mass assumption used in the training of the BDT used for signal selection results in a systematic uncertainty of 2.6 GeV (0.5%) on the mass, which is small compared to the statistical uncertainty.

### 6.4.2 Gauginos

Due to their decay into gauge bosons and Higgs bosons and the lightest neutralino, the measurement of heavier charginos and neutralinos places particular emphasis on the jet energy resolution and the di-jet mass reconstruction of the experiment, needed to separate $W$, $Z$ and Higgs bosons. Here, the processes

- $e^+e^- \rightarrow \tilde{\chi}_1^+\tilde{\chi}_1^- \rightarrow W^+W^-\tilde{\chi}_1^0\tilde{\chi}_1^0$ and
- $e^+e^- \rightarrow \tilde{\chi}_2^0\tilde{\chi}_2^0 \rightarrow h(Z)h(Z)\tilde{\chi}_1^0\tilde{\chi}_1^0$

are studied in hadronic final states. $\tilde{\chi}_2^0$ and $\tilde{\chi}_1^\pm$ both have a mass of 487 GeV in SUSY *model III*, and in the case of the $\tilde{\chi}_2^0$, the decay to Higgs bosons is dominating with a branching ratio of 94.6%. A particular challenge of this latter channel is that, due to the mass of 357 GeV of the $\tilde{\chi}_1^0$, the Higgs bosons are produced almost at rest.

Events are reconstructed by clustering into four jets using tight timing and momentum cuts on the particle flow object (PFO) to minimise the influence of $\gamma\gamma \rightarrow$ hadrons background. The pairing of the jets into boson candidates is performed by minimising the sum of the squared differences between the measured di-jet invariant masses and the nominal $W$ or Higgs mass. The rejection of SM and SUSY background is provided by a boosted decision tree using a variety of event shape and kinematic variables, none of which are strongly correlated with the energy of the $W$ and Higgs bosons. For the $\tilde{\chi}_1^\pm$ and the $\tilde{\chi}_2^0$ study, a selection efficiency of 68% and 72%, respectively, is achieved, with the cuts on the BDT classifier optimised for maximum significance. Figure 6.6 left shows the di-jet invariant mass distribution for signal and background for the $\tilde{\chi}_1^\pm$ channel, demonstrating the clean separation from the $\tilde{\chi}_2^0$ through mass resolution.

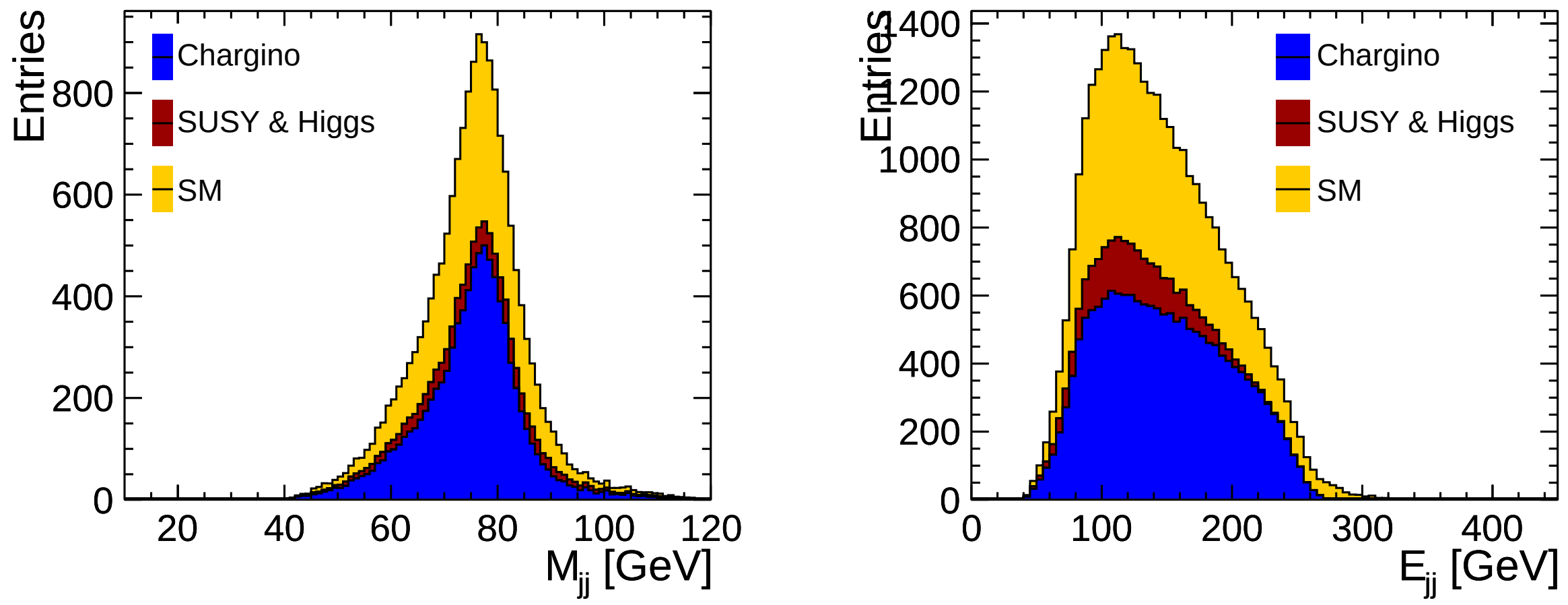

Fig. 6.6: Di-jet invariant mass distribution for the selected signal and background events in the $e^+e^- \rightarrow \tilde{\chi}_1^+\tilde{\chi}_1^- \rightarrow W^+W^-\tilde{\chi}_1^0\tilde{\chi}_1^0 \rightarrow q\bar{q}q\bar{q}\tilde{\chi}_1^0\tilde{\chi}_1^0$ channel, showing good identification of the $W$ bosons in their hadronic decay (left). Energy of di-jet systems (right).

The $\tilde{\chi}_1^\pm$ and $\tilde{\chi}_2^0$ masses and the corresponding production cross-sections are simultaneously determined from template fits of the di-jet energy distribution of the $W$ and Higgs candidates, as shown for the $\tilde{\chi}_1^\pm$ in Figure 6.6 right, using fully simulated templates including background. The mass of the $\tilde{\chi}_1^\pm$ is determined with a statistical precision of 0.2%, and the mass of the $\tilde{\chi}_2^0$ is determined with a statistical precision of 0.1%. The corresponding cross-sections are measured with an accuracy of 1.3% and 1.2%, respectively. The $\tilde{\chi}_1^\pm$ and $\tilde{\chi}_2^0$ masses and pair production cross-sections can not be extracted simultane-ously with the mass of the lightest neutralino due to the large correlations between the three observables. The $\tilde{\chi}_1^0$ mass is taken from the first and second generation slepton measurements. An assumed uncertainty of 1 GeV on the $\tilde{\chi}_1^0$ mass leads to a 0.3% error on the $\tilde{\chi}_1^\pm$ and $\tilde{\chi}_2^0$ mass. Further details on the analysis are given in [20].

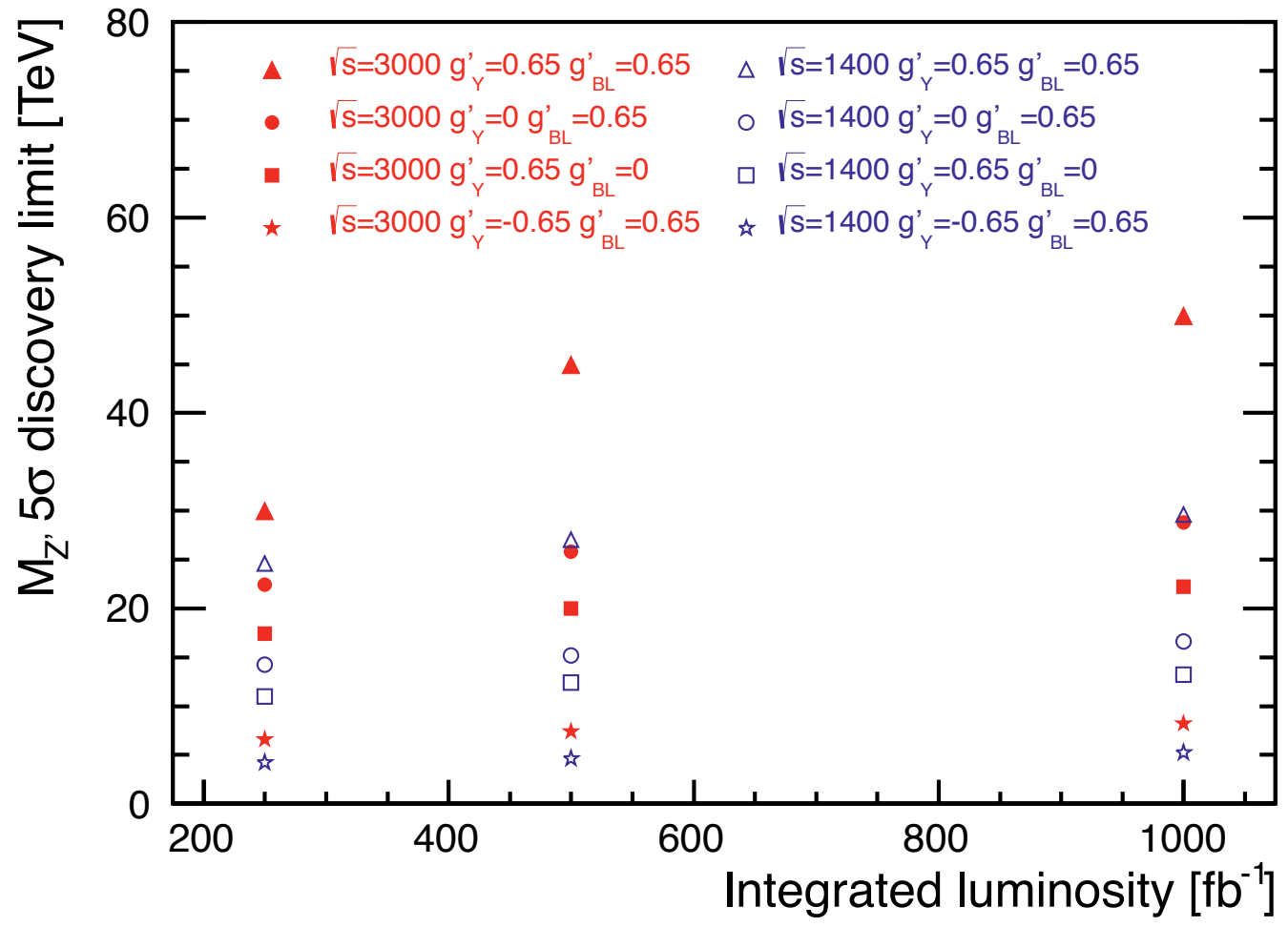

Fig. 6.7: $M_{Z'}$ $5\sigma$ discovery limit as function of the integrated luminosity for different values of the couplings $g'_Y$ and $g'_{BL}$. The limits shown are determined from the combined observables $\sigma$ and $A_{FB}$ at 3 TeV and 1.4 TeV [21].

This study complements a study of chargino and neutralino poduction at 3 TeV in SUSY *model I* and *model II* [1], where the masses of the particles are approximately 200 GeV heavier than in the present case. In that case, per cent-level accuracies on the masses are achieved, showing that also the gaugino sector is accessible over a wide parameter range. The achievable accuracy depends on the detailed kinematics, and with that on the mass splittings between particles.

#### 6.4.2.1 Coloured States at 3 TeV

In addition to the extensive studies of the slepton and gaugino sector, also the potential to study TeV scale right-squarks, which have a very generic New Physics signature with two energetic jets and missing energy, was studied [1]. For these particles, sub-percent accuracy was obtained for the mass when assuming knowledge of the mass of the lightest neutralino, showing that CLIC can also explore coloured states up to the production limit.

## 6.5 Heavy Gauge Bosons

Many theories of physics beyond the SM call for an additional Abelian gauge group spontaneously broken near the TeV scale. The most minimal such model that is anomaly free with respect to the SM particle content has a neutral $Z'$ gauge boson whose couplings to the SM fermions are proportional to a linear combination of the fermion's hypercharge and $B-L$ charge (baryon number minus lepton number): $Q'_f = g'_Y Y_f + g'_{BL}(B-L)_f$. Various choices of the linear combination coefficients $g'_Y$ and $g'_{BL}$ have been selected, and the resulting $Z'$ phenomenology has been studied with full detector simulation and realistic beam conditions for $\sqrt{s} = 1.4$ TeV and $\sqrt{s} = 3$ TeV [21]. A combination of the total cross-section and the forward-backward asymmetry in $e^+e^- \to \mu^+\mu^-$ is analysed to search for deviations caused by physics beyond the SM. Figure 6.7 shows the $5\sigma$ discover reach for $M_{Z'}$ as a function of the integrated luminosity for different values of the couplings $g'_Y$ and $g'_{BL}$. One sees that the reach is variable depending on the couplings, but there are opportunities to detect the effects of a weakly interacting $Z'$ boson with over 50 TeV mass, well beyond the search capacity of LHC.

## 6.6 Summary

Several physics benchmark studies have been performed to demonstrate the performance of experiments at CLIC at different energy stages. These complement the wide range of studies performed at 3 TeV in the framework of the Physics and Detector volume of the CLIC conceptual design report [1]. All studies were carried out using full detector simulation and reconstruction, including the pile-up from $\gamma\gamma \rightarrow$ hadrons interactions.

The results of these studies demonstrate that a linear collider based in CLIC technology is capable of a detailed exploration of the Higgs sector in various processes over the full energy range of the CLIC programme, of a precise measurement of the top quark mass in a threshold scan as well as above threshold, and of direct measurements of the properties of BSM particles, exemplified here by the study of several SUSY models. In addition, precision measurements at CLIC also provide sensitivity to detect the presence of new heavy gauge bosons far beyond the direct discovery reach of present-day colliders. The results are summarised in Table 6.1 for the SM Higgs studies, in Table 6.2 for the top quark studies, and in Tables 6.3 and 6.4 for the SUSY studies at 1.4 TeV and 3 TeV, respectively.

Table 6.1: Summary of results obtained in the Higgs studies for $m_H$ =120 GeV. All analyses at centre-of-mass energies of 350 GeV and 500 GeV assume an integrated luminosity of 500 fb$^{-1}$, while the analyses at 1.4 TeV (3 TeV) assume 1.5 ab$^{-1}$(2 ab$^{-1}$).

| **Higgs studies for $m_H$ =120 GeV** | | | | | | | |
|---|---|---|---|---|---|---|---|
| $\sqrt{s}$ (GeV) | Process | Decay mode | Measured quantity | Unit | Generator value | Stat. error | Comment |
| 350 | SM Higgs production | $ZH \rightarrow \mu^+\mu^- X$ | $\sigma$ | fb | 4.9 | 4.9% | Model independent, using $Z$-recoil |
| | | | Mass | GeV | 120 | 0.131 | |
| 500 | | $ZH \rightarrow q\bar{q}q\bar{q}$ | $\sigma\times$ BR | fb | 34.4 | 1.6% | $ZH \rightarrow q\bar{q}q\bar{q}$ mass reconstruction |
| | | | Mass | GeV | 120 | 0.100 | |
| 500 | | $ZH, H\nu\bar{\nu} \rightarrow \nu\bar{\nu}q\bar{q}$ | $\sigma\times$ BR | fb | 80.7 | 1.0% | Inclusive sample |
| | | | Mass | GeV | 120 | 0.100 | |
| 1400 | $WW$ fusion | $H \rightarrow \tau^+\tau^-$ | $\sigma\times$ BR | fb | 19.8 | <3.7% | |
| 3000 | | $H \rightarrow b\bar{b}$ | | | 285 | 0.22% | |
| | | $H \rightarrow c\bar{c}$ | | | 13 | 3.2% | |
| | | $H \rightarrow \mu^+\mu^-$ | | | 0.12 | 15.7% | |
| 1400 | $WW$ fusion | | Higgs tri-linear coupling $g_{HHH}$ | | | ∼20% | |
| 3000 | | | | | | ∼20% | |

Table 6.2: Summary of full detector-simulation results obtained under realistic CLIC beam conditions in the top quark studies. The first (second) threshold scan contains 6 points (10 points) separated by 1 GeV and with 10 $fb^{-1}$ of luminosity at each point.

| Top studies | | | | | | |
|---|---|---|---|---|---|---|
| $\sqrt{s}$ (GeV) | Technique | Measured quantity | Integrated luminosity ($fb^{-1}$) | Unit | Generator value | Stat. error |
| 350 | Threshold scan | Mass | $6 \times 10$ | GeV | 174 | 0.021 |
| | | Mass<br>$\alpha_S$ | $10 \times 10$ | GeV | 174<br>0.118 | 0.033<br>0.0009 |
| 500 | Invariant mass | Mass | 100 | GeV | 174 | 0.060 |

Table 6.3: Summary table of the CLIC SUSY benchmark analyses results obtained with full detector simulations with background overlaid. All studies are performed at a centre-of-mass energy of 1.4 TeV and for an integrated luminosity of 1.5 $ab^{-1}$.

| $\sqrt{s}$ (TeV) | Process | Decay mode | SUSY model | Measured quantity | Unit | Generator value | Stat. error |
|---|---|---|---|---|---|---|---|
| 1.4 | Sleptons production | $\tilde{\mu}_R^+\tilde{\mu}_R^- \rightarrow \mu^+\mu^-\tilde{\chi}_1^0\tilde{\chi}_1^0$ | III | $\sigma$<br>$\tilde{\ell}$ mass<br>$\tilde{\chi}_1^0$ mass | fb<br>GeV<br>GeV | 1.11<br>560.8<br>357.8 | 2.7%<br>0.1%<br>0.1% |
| | | $\tilde{e}_R^+\tilde{e}_R^- \rightarrow e^+e^-\tilde{\chi}_1^0\tilde{\chi}_1^0$ | | $\sigma$<br>$\tilde{\ell}$ mass<br>$\tilde{\chi}_1^0$ mass | fb<br>GeV<br>GeV | 5.7<br>558.1<br>357.1 | 1.1%<br>0.1%<br>0.1% |
| | | $\tilde{\nu}_e\tilde{\nu}_e \rightarrow \tilde{\chi}_1^0\tilde{\chi}_1^0 e^+e^-W^+W^-$ | | $\sigma$<br>$\tilde{\ell}$ mass<br>$\tilde{\chi}_1^\pm$ mass | fb<br>GeV<br>GeV | 5.6<br>644.3<br>487.6 | 3.6%<br>2.5%<br>2.7% |
| 1.4 | Stau production | $\tilde{\tau}_1^+\tilde{\tau}_1^- \rightarrow \tau^+\tau^-\tilde{\chi}_1^0\tilde{\chi}_1^0$ | III | $\tilde{\tau}_1$ mass<br>$\sigma$ | GeV<br>fb | 517<br>2.4 | 2.0%<br>7.5% |
| 1.4 | Chargino production | $\tilde{\chi}_1^+\tilde{\chi}_1^- \rightarrow \tilde{\chi}_1^0\tilde{\chi}_1^0 W^+W^-$ | III | $\tilde{\chi}_1^\pm$ mass<br>$\sigma$ | GeV<br>fb | 487<br>15.3 | 0.2%<br>1.3% |
| | Neutralino production | $\tilde{\chi}_2^0\tilde{\chi}_2^0 \rightarrow h/Z^0 h/Z^0\tilde{\chi}_1^0\tilde{\chi}_1^0$ | | $\tilde{\chi}_2^0$ mass<br>$\sigma$ | GeV<br>fb | 487<br>5.4 | 0.1%<br>1.2% |

Table 6.4: Summary table of the CLIC SUSY benchmark analyses results obtained with full detector simulations with background overlaid. All studies are performed at a centre-of-mass energy of 3 TeV and for an integrated luminosity of 2 ab$^{-1}$.

| $\sqrt{s}$ (TeV) | Process | Decay mode | SUSY model | Measured quantity | Unit | Generator value | Stat. error |
|---|---|---|---|---|---|---|---|
| 3.0 | Sleptons production | $\tilde{\mu}_R^+\tilde{\mu}_R^- \to \mu^+\mu^-\tilde{\chi}_1^0\tilde{\chi}_1^0$ | II | $\sigma$ | fb | 0.72 | 2.8% |
| | | | | $\tilde{\ell}$ mass | GeV | 1010.8 | 0.6% |
| | | | | $\tilde{\chi}_1^0$ mass | GeV | 340.3 | 1.9% |
| | | $\tilde{e}_R^+\tilde{e}_R^- \to e^+e^-\tilde{\chi}_1^0\tilde{\chi}_1^0$ | | $\sigma$ | fb | 6.05 | 0.8% |
| | | | | $\tilde{\ell}$ mass | GeV | 1010.8 | 0.3% |
| | | | | $\tilde{\chi}_1^0$ mass | GeV | 340.3 | 1.0% |
| | | $\tilde{e}_L^+\tilde{e}_L^- \to \tilde{\chi}_1^0\tilde{\chi}_1^0 e^+e^-hh$<br>$\tilde{e}_L^+\tilde{e}_L^- \to \tilde{\chi}_1^0\tilde{\chi}_1^0 e^+e^-Z^0Z^0$ | | $\sigma$ | fb | 3.07 | 7.2% |
| | | $\tilde{\nu}_e\tilde{\nu}_e \to \tilde{\chi}_1^0\tilde{\chi}_1^0 e^+e^-W^+W^-$ | | $\sigma$ | fb | 13.74 | 2.4% |
| | | | | $\tilde{\ell}$ mass | GeV | 1097.2 | 0.4% |
| | | | | $\tilde{\chi}_1^\pm$ mass | GeV | 643.2 | 0.6% |
| 3.0 | Chargino production | $\tilde{\chi}_1^+\tilde{\chi}_1^- \to \tilde{\chi}_1^0\tilde{\chi}_1^0 W^+W^-$ | II | $\tilde{\chi}_1^\pm$ mass | GeV | 643.2 | 1.1% |
| | | | | $\sigma$ | fb | 10.6 | 2.4% |
| | Neutralino production | $\tilde{\chi}_2^0\tilde{\chi}_2^0 \to h/Z^0 h/Z^0 \tilde{\chi}_1^0\tilde{\chi}_1^0$ | | $\tilde{\chi}_2^0$ mass | GeV | 643.1 | 1.5% |
| | | | | $\sigma$ | fb | 3.3 | 3.2% |
| 3.0 | Production of right-handed squarks | $\tilde{q}_R\tilde{q}_R \to q\bar{q}\tilde{\chi}_1^0\tilde{\chi}_1^0$ | I | Mass | GeV | 1123.7 | 0.52% |
| | | | | $\sigma$ | fb | 1.47 | 4.6% |
| 3.0 | Heavy Higgs production | $H^0A^0 \to b\bar{b}b\bar{b}$ | I | Mass | GeV | 902.4 | 0.3% |
| | | | | Width | GeV | | 31% |
| | | $H^+H^- \to t\bar{b}b\bar{t}$ | | Mass | GeV | 906.3 | 0.3% |
| | | | | Width | GeV | | 27% |

# Chapter 7

# Strategy and Objectives for the CLIC Programme

## 7.1 Overall Time Line Towards Construction

The recent observation at LHC of a Higgs-like particle with mass around 125 GeV has established a strong physics case for a future linear collider. In addition there is still a major discovery potential in running the LHC at 14 TeV and with increased luminosity, particularly for BSM physics being assessed through more precise Higgs measurements or searches where the energy reach is paramount. After a few years of running the LHC at full energy (i.e. by 2016–2017) it is expected that one can make an informed decision about a next accelerator at the energy frontier with physics capabilities beyond and complementary to the LHC. In the period until then it is important to pursue technical developments and comparative physics studies for the major candidates for such a future facility, being a linear collider or an energy-upgraded LHC.

After this basic decision, an initial project preparation phase will be needed (2017–2022) before full construction can start for an accelerator like CLIC. During this phase the technical developments will focus on large-scale production and system verifications, leading to a final technical design with reduced risks. As this phase calls for an increased deployment of resources and is likely to coincide with LHC upgrade activities, it will require broadly coordinated resource planning. Initiating the construction of a new accelerator around 2022–2023 is compatible with a start of operation at the time of the estimated end of LHC running, i.e. around 2030.

For these reasons the CLIC project time line comprises two phases, covering the years 2012–2016 and 2017–2022. The main objectives for the CLIC programme can be summarised as follows:

– **2012:** Finalise the CLIC Conceptual Design Reports, establish feasibility, and provide input for the update of the European Strategy for Particle Physics in 2012–2013 [1];
– **2012–2016:** Project Development Phase, producing a Project Implementation Plan for a CLIC construction project by 2016;
– **2016–2017:** Decision about the next project at the energy frontier;
– **2017–2022:** Project Implementation Phase, including an initial Project Preparation Phase to lay the ground work for full construction;
– **2022–2023:** CLIC construction start-up;
– **2023–2030:** Construction of the first CLIC energy stage, making use of a significant fraction of the hardware developed during the Project Implementation Phase;
– **From 2030:** Commissioning of CLIC.

This CLIC project time line is depicted in Figure 7.1. The two upcoming phases are described in more detail in the following two sections.

## 7.2 2012–2016: Towards a Project Implementation Plan

The overall objective for the next phase of the CLIC accelerator project, a period referred to as Project Development Phase, is to develop a Project Implementation Plan by 2016. As part of the development towards this Project Implementation Plan a series of specific studies has been defined for the period 2012–2016, as described below. The plan will be adapted to the physics scenario emerging from LHC data during this time and will describe the implementation steps needed beyond 2016. The current CLIC parameters have been optimised for a 3 TeV implementation. Therefore an important part of the work in this phase will be to optimise the design also for the initial energy stages. In case of a growing interest for an initial lower energy Higgs factory a klystron-based implementation [2] of such an option can also be considered.

**2012-16 Development Phase**
Develop a Project Plan for a staged implementation in agreement with LHC findings; further technical developments with industry, performance studies for accelerator parts and systems, as well as for detectors.

**2017-22 Preparation Phase**
Finalise implementation parameters, Drive Beam Facility and other system verifications, site authorisation and preparation for industrial procurement.
Prepare detailed Technical Proposals for the detector-systems.

**2023-2030 Construction Phase**
Stage 1 construction of a 500 GeV CLIC, in parallel with detector construction.
Preparation for implementation of further stages.

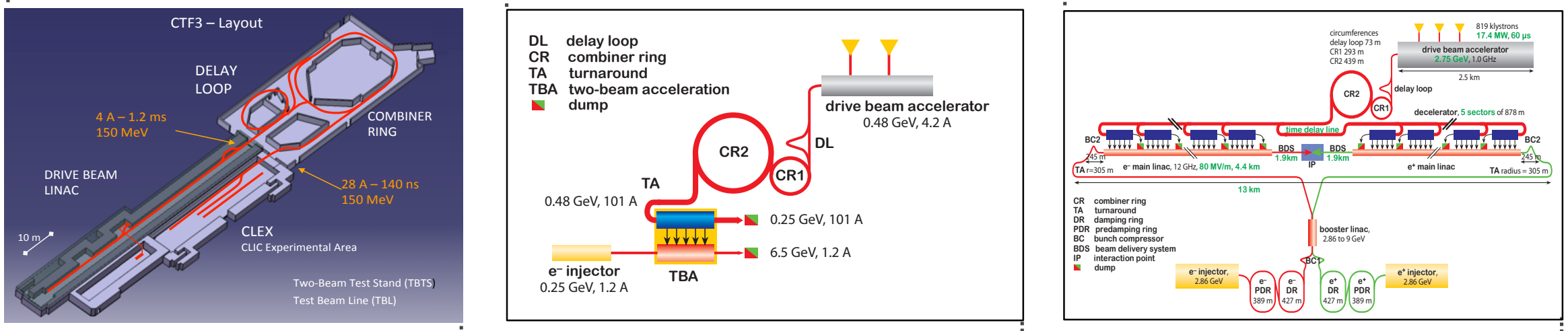

**2016-17 Decisions**
On the basis of LHC data and Project Plans (for CLIC and other potential projects), take decisions about next project(s) at the Energy Frontier.

**2022-23 Construction Start**
Ready for full construction and main tunnel excavation.

**2030 Commissioning**
From 2030, becoming ready for data-taking as the LHC programme reaches completion.

Fig. 7.1: Top row: An outline of the CLIC project time line with main activities leading up to and including the first stage construction. Middle row: illustrations of the CTF3 facility (one of several testing facilities of importance to the project development), a new large drive beam facility with final CLIC elements which is also needed for acceptance tests, and a 500 GeV implementation. Bottom row: Main decision points and activities.

In a similar way specific physics and detector studies are foreseen. They focus on three areas: physics studies, detector optimisation and technology demonstrators.

Both the CLIC accelerator study and the CLIC physics and detector study are organised as Collaborations governed by Collaboration/Institute Boards, and coordinated by Steering Groups and an overall CLIC Steering Committee, and are hosted by CERN.

### 7.2.1 Accelerator Activities

During the past years objectives for the CLIC programme in the post-CDR phase were extensively discussed within the CLIC/CTF3 collaboration. The goals for the CLIC programme for the period 2012–2016 were laid down in the CERN Medium Term Plan (MTP) and approved by the CERN Council in 2010. While representing a scale back on previous plans to produce a TDR already by 2016, the MTP foresees sufficient funding from CERN to cover the development towards a complete Project Implementation Plan by 2016. The CLIC programme objectives are also compatible with the estimated availability of resources among the collaboration partners during this period. This will put the CLIC project in a position to be ready by 2016, i.e. after two years of LHC data-taking at full energy, for a decision on a future facility at the energy frontier. The main input to this plan is:

– The evolution of the physics findings at LHC and other relevant data;
– Findings from the CDR and further technical studies for key elements or during system tests;
– Results of detailed implementation studies for a staged project including costing, power, site-studies and schedules;
– A Governance Model as developed with partners.

The detailed planning of the work for the period 2012–2016 has been set up by the CLIC/CTF3 collaboration, currently comprising the 44 institutes listed in Appendix A. The plan sets detailed technical objectives and describes the corresponding deployment of personnel and material contributions based on 75 expressions of interests from individual groups within the collaboration (see CLIC workshop November 2011 [3]). The plan also encompasses the CLIC specific activities in 20 groups within CERN. The plan was reviewed in March 2012 with particular attention on the availability of funds over the full period, implementing an overall 30% reduction on the material fund deployment. The work-packages are shown in Appendix B.

### 7.2.2 Physics and Detector Studies

In the period 2012–2016 the CLIC physics and detector study will concentrate on three main activities: further exploration of the physics potential, detector optimisation studies and technical demonstrators that meet the required detector performances. An outline of these activities is given below. For a more complete description, see Chapter 13 in [4].

**Exploration of the physics potential**

Following up on 8 TeV and 14 TeV LHC results, the CLIC physics potential will be explored further on the basis of detailed benchmarking studies. The studies will address the following main areas:

- Precision measurements of standard model physics (e.g. Higgs, top);
- Exploration of the discovery reach for New Physics;
- Sensitivity to effects of New Physics via high-precision measurements.

As a result, and in collaboration with the CLIC accelerator study, the scenarios for machine energy staging and operation will optimised.

**Detector optimisation**

The detector concept designs will be further refined and improved. In particular, the CDR studies have revealed some regions with high cell occupancies, which can be mitigated by careful design and adapted technology choices. The two options consisting of placing the final focusing elements either inside or outside of the detector volume will be studied in detail, and their respective benefits and losses on the physics potential will be quantified. The detector simulation models and the software tools for event reconstruction will be optimised further and will include improved knowledge of detector responses following hardware tests.

**Technology demonstrators**

The CDR studies have provided a good understanding of the detector requirements, taking into account the physics goals as well as the experimental conditions (see Chapter 4). The detector requirements for CLIC have much in common with those of the detectors under development for ILC, but they are more demanding in several areas. Hardware demonstrators are therefore required to validate these CLIC-specific aspects. Although challenging, the CLIC detector technologies are considered feasible following an R&D phase of 5 years. The list of hardware demonstrators comprises:

- **Vertex detector** demonstration module that meets the requirements of high precision, 10 ns time-stamping and ultra-low mass;
- **Main tracker** demonstration modules that meet technical requirements, including manageable occupancies in the event reconstruction;
- **Calorimeter** demonstration modules that meet requirements and address control of cost, complemented with technological prototypes;
- **Electronics** demonstrators that meet technical requirements, in particular in view of power pulsing;
- **Magnet system** demonstrator of reinforced conductor, safety systems and movable service lines;
- **Engineering and detector integration** studies that are harmonised with the various hardware demonstrators.

The above studies will be carried out in the framework of a cooperation of institutes and groups, the *CLIC physics and detector study*, governed by an Institute Board. The CLIC physics and detector study will operate in close collaboration and synergy with the ILC physics and detector activities.

## 7.3 2017–2022: Towards CLIC Construction

Following two years of LHC data taking at 14 TeV and further technical work on accelerator and detectors for future machine options, the timescale 2016–2017 stands out as an appropriate moment for deciding on a future accelerator project at the energy frontier. As mentioned above, the principal CLIC objective in the period up to 2016 is to establish a Project Implementation Plan that can be used as a basis for the decision to construct a linear collider based on CLIC technology. Subsequently, an initial Project Preparation Phase is foreseen, covering 2017–2022. It will make use of resources across the collaboration to prepare for the construction of the first CLIC energy stage, which could be initiated around 2022–2023.

### 7.3.1 Accelerator Project Preparation Phase

Following a decision on the scope and technology for the future accelerator, and before launching major construction contracts, it is essential to optimise the component performances and to reduce their cost. In addition, a number of key system performances need to be addressed to minimise the risk of the CLIC project implementation. This will allow to rationalise requirements by understanding the interplay between safety margins and can therefore reduce the overall project cost. The drive beam and luminosity performances, in particular, are best addressed in larger system tests.

In parallel one needs to compose the final construction consortia and collaborations, develop the industrial capabilities, refine the estimates of the required resources, optimise the resource planning, and carry out a detailed environmental impact study in agreement with the foreseen implementation. The corresponding governance structures have to be established and to become operational.

Hence the specific goals of this period are:

- Finalisation of the CLIC technical design taking into account the results of the technical studies completed in the previous phase and the chosen energy staging scenario based on LHC results;
- Specific industrial developments and contract preparations for items that are driving the cost, power consumption and accelerating gradient;
- Constructing a significant part of the drive beam facility using prototypes of the hardware components of the full project. This facility produces the full CLIC drive beam current and is also essential for the reception tests of the main linac modules. In addition it will allow to optimise and validate the drive beam components and related beam performances. The components will eventually become part of the final project. This project is called CLIC0 as it also forms the start of the actual project implementation and of the industrialisation of key parts of the CLIC accelerator;
- Systems tests aimed at benchmarking and validating the CLIC luminosity performance and the validation of the main beam components. Such tests can be carried out at facilities that have small emittance beams, possibly primarily targeting other applications, as for example existing light sources. By installing a number of CLIC accelerator structures at such facilities, beam-transport can be tested over a significant length. Several such options are currently explored and will be developed further in the coming years. It is considered very beneficial to operate such a facility, or a combination of facilities, as early as possible during the Project Preparation Phase. Existing facilities like ATF and FACET will allow some of such tests to take place at an earlier stage.

The Project Preparation Phase will lay the ground work for the full construction start-up. It aims at the finalisation of the detailed technical design, the construction planning, the main industrial contracts and the governance to be ready by 2022–2023. The system test facilities mentioned above will continue operating beyond that time if considered useful for the project, but many of their parts can also be re-used for the final drive beam.

### 7.3.2 Preparation Phase for the Detector Construction

During the preparation phase for the detector construction (2017–2022) large prototype studies, industrialisation studies and detailed integration studies will be carried out to provide solid proof of construction feasibility and cost optimisation prior to the actual detector construction. This phase also needs to address all the detailed technical issues, including the technically less critical items, as they have to become an integral part of a reliable cost book and construction schedule prior to a formal project approval. Following the example of the LHC experiments a detailed Technical Proposal will be prepared. Subsequently detailed Technical Design Reports (TDR) will be composed for the individual subsystems to enable their construction and to engage spending on the final hardware. This process will be completed first for the largest and most time-critical systems for which procurement and construction will start by 2023. The TDR process can still be preceded by several years of R&D and prototyping for subsystems where access to the latest technologies is critical and where assembly and integration schedules are more relaxed. Formal collaboration agreements, including technical and financial commitments for the construction and commissioning of the detectors, are also concluded during the period 2017–2022.

## 7.4 Summary of the Key Project Development Steps

The CLIC accelerator project aims to present a Project Implementation Plan by 2016. The CLIC/CTF3 collaboration has set up a technical planning to reach this goal. The detailed work plan focuses on technical studies, industrial collaboration and system developments, along with implementation studies for construction and operation of CLIC in a few energy stages. It is compatible with the expected resources within the collaboration during the period 2012–2016. In a similar way objectives for the period 2012–2016 have been defined for the CLIC physics and detector study. They focus on physics studies, detector optimisation and the development of technology demonstrators, and are considered realistic within a 5-year period.

By 2016–2017 both the LHC physics results and the technical developments for future accelerators and detectors at the energy frontier, such as linear colliders and an energy-upgraded LHC, are expected to have reached a maturity that will allow for a decision concerning the most appropriate future collider with physics capabilities beyond and complementary to the LHC.

If CLIC is chosen, the project implementation will require an initial Project Preparation Phase of $\sim$6 years (2017–2022) focusing on industrial build-up, larger system verifications, risk and cost reduction, as well as developing Technical Proposals for the detectors. The governance structure and the international collaboration agreements for the construction and operation will be set up during this time.

The final site authorisations will also be established during this period. Preliminary site studies show that CLIC can be implemented underground near CERN, with the central main and drive beam injector complex on the CERN domain, as shown in Figure 7.2. The site specifications do not constrain the implementation to this location.

The construction of the first CLIC energy stage could be launched around 2022–2023. Further stages, in particular the second energy stage, can be launched quickly thereafter with construction starting after 2–3 years of operation at the first stage, drawing on all experiences and lessons acquired at that time. Such a time line would ensure that CLIC would be ready for operation by 2030 when LHC is expected to finish, and a second stage can be ready when the initial CLIC machine reaches 500–700 $fb^{-1}$.

Faster implementations, for example motivated by a growing interest in a lower energy Higgs factory, can be considered and studied in the framework of a klystron-based initial stage [2].

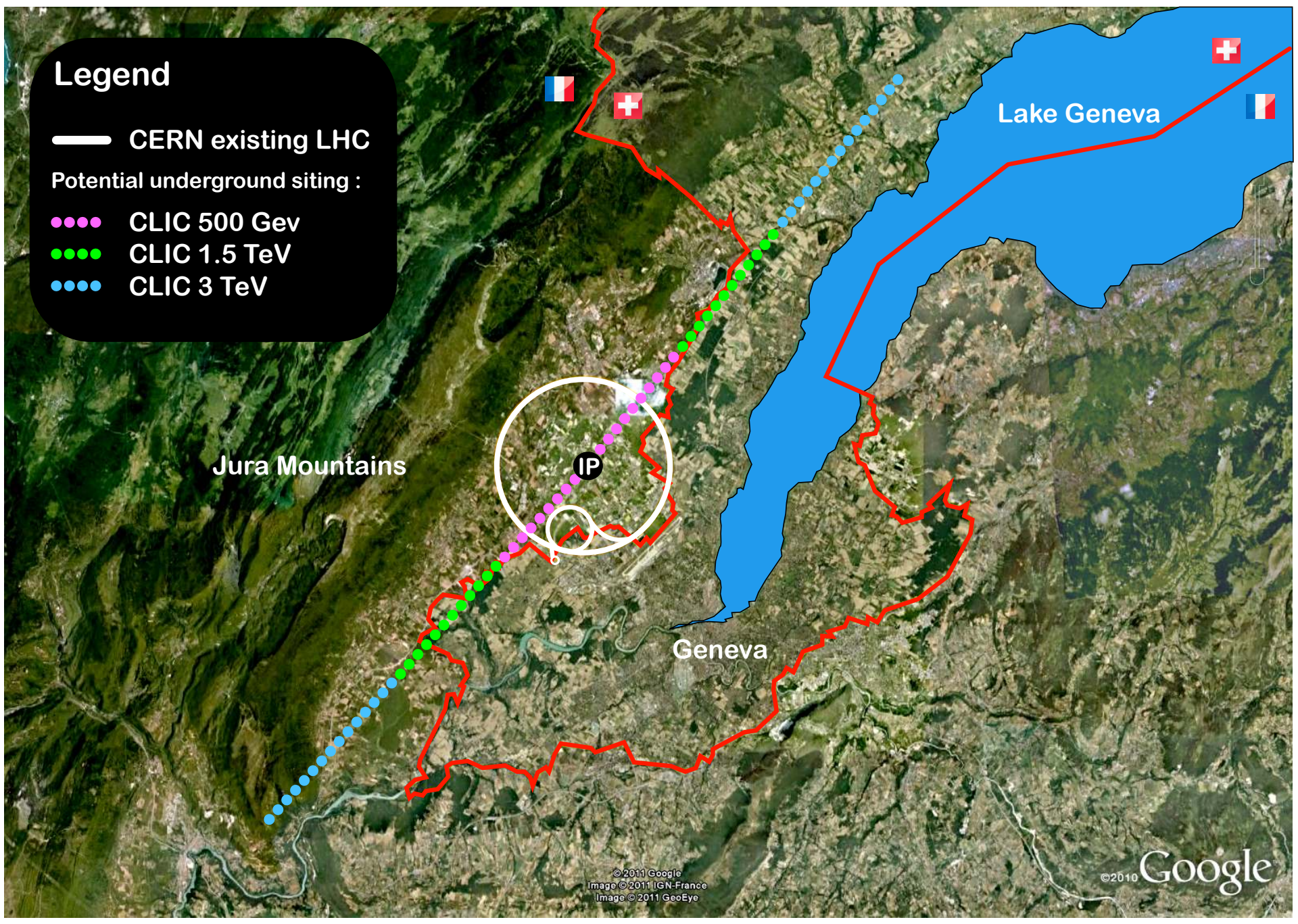

Fig. 7.2: CLIC footprints near CERN, showing various implementation stages [5].

# Summary

The CLIC project has now successfully completed its Conceptual Design Report phase. The feasibility of the accelerator technology has been demonstrated, based on beam simulations, component development and engineering studies, which have been validated through a wealth of hardware tests, ranging from single components to large systems. The tests have culminated in successful two-beam acceleration tests at the CTF3 facility reaching acceleration gradients well above the 100 MV/m design target, using accelerating structures that fulfill the stringent breakdown rate requirements. Numerous developments and studies likewise demonstrate the feasibility of creating, accelerating and colliding the CLIC beams with the parameters needed to deliver the required high luminosities. On the experiment side, the two detector designs CLIC_ILD and CLIC_SiD, have demonstrated to fulfill the requirements for high-precision physics measurements under CLIC experimental conditions. These detector concepts are based on realistic technologies, that have either been demonstrated or are considered achievable following a 5-year R&D programme.

CLIC offers the possibility to study $e^+e^-$ collisions at centre-of-mass energies ranging from a few hundred GeV to 3 TeV, with peak luminosities of a few $10^{34}$ cm$^{-2}$s$^{-1}$. This opens the door to an impressive physics potential including:

– Precision measurements of standard model physics (e.g. Higgs, top);
– Direct searches for pair production of new particles with mass up to 1.5 TeV;
– Sensitivity to effects of New Physics at much higher mass scales via high-precision measurements.

Following the recent LHC observation of a $\sim$125 GeV particle compatible with the Higgs boson, and anticipating further LHC results at 8 TeV and 14 TeV, the CLIC physics potential will be best explored through a staged construction and operation strategy. The choice of the first energy stage will be driven by the physics objectives of Higgs and top studies. The centre-of-mass energies for the higher energy stages would be determined by further precision Higgs measurements, like the top-Yukawa coupling and the Higgs self-coupling, and by the physics scale of possible New Physics based on input from LHC or other sources.

As an example, the implementation of two possible scenarios, each consisting of three energy stages, 500 GeV, 1.4 (1.5) TeV and 3 TeV, has been presented in this report. These scenarios have been developed by taking into consideration the physics goals, including luminosity requirements, as well as technical and financial aspects. At each energy stage the centre-of-mass energy can be tuned to lower values within a range of a factor three with limited loss on luminosity performances.

These staging scenarios serve as examples for studying a number of central implementation issues, including construction models and siting, operation schedules, luminosity development, power and energy consumption and also cost. The physics potential for such scenarios has been illustrated following detailed detector simulations under realistic beam conditions. These simulation studies have focused on precision Higgs and top measurements and have also explored various examples of the CLIC sensitivity for New Physics models involving high-mass states with very small production cross-sections. For a demonstration of the latter, measurements of the production of high-mass sleptons, gauginos, squarks and heavy Higgs superpartners in three example SUSY models have been used as examples. These studies show that very high precisions can be achieved at CLIC for a wealth of physics measurements and over a unique span of centre-of-mass energies.

Following the feasibility demonstration during the CDR period, the CLIC study now moves forward based on a roadmap comprising two main phases before full construction can be initiated.

The first phase covers the period 2012 to 2016. For the accelerator study it is driven by technical and industrial objectives combined with studies towards optimal choices for the CLIC implementation, including the definition of the energy stages. This will result in a detailed project implementation plan at the end of this period, providing input for strategic particle physics decisions, expected by 2016–2017, on a next facility at the energy frontier. The work plan for this first phase has been defined in detail among all partners within the CLIC/CTF3 collaboration and is consistent with available resources. On the detector side, this phase is dedicated to a further exploration of the physics potential following LHC results, together with detector optimisation studies and hardware R&D. The hardware development focuses on technology demonstrators satisfying CLIC-specific detector requirements.

The above planning fits well into the current particle physics landscape. This first phase overlaps with further LHC data taking at 8 TeV and later 14 TeV, where the multi-TeV region will be much more accessible. It is anticipated that by 2016–2017 an informed choice can be made for a future collider at the energy frontier.

Such a choice would go together with a focused preparation programme to prepare for the construction of a future machine. The CLIC preparation phase starting in 2017 includes large-scale systems for performance optimisation and technical verification, detailed site studies and large-scale industrialisation efforts preparing for construction contracts. The drive beam and luminosity performances are foreseen to be addressed in larger systems and the plans include the construction of a significant initial drive beam facility with final hardware. For the detectors the same time-period will be dedicated to large-scale prototyping, detailed integration designs and industrialisation efforts in view of providing proof of construction readiness and cost optimisation.

The overall plan anticipates a start of the CLIC construction by 2023. With an estimated construction time of 7 years for a first CLIC energy stage of 400–500 GeV, the start of CLIC operation would coincide with the currently foreseen end of the LHC operation towards 2030. This would mark the start of a rich and long-term $e^+e^-$ physics programme providing precision data over an unprecedented energy span.

# Acknowledgements

The CLIC/CTF3 collaboration currently (August 2012) comprises 44 member institutes from 22 countries. The list of member institutes is provided in Appendix A. We acknowledge the support from these laboratories and from their technical staff for their sustained efforts. The CLIC physics and detector studies, described in this document, benefited from the ILC detector R&D efforts, which were supported by BMWF, Austria; MinObr, Belarus; FNRS and FWO, Belgium; NSERC, Canada; NSFC, China; MPO CR and VSC CR, Czech Republic; FP6 and FP7 European Commission, European Union; HIP, Finland; IN2P3-CNRS, CEA-DSM/IRFU, France; BMBF, DFG, HGF, MPG and AvH Foundation, Germany; DAE and DST, India; ISF, Israel; INFN, Italy; MEXT and JSPS, Japan; CRI(MST) and MOST/KOSEF, Korea; FOM, NWO, the Netherlands; NFR, Norway; MNSW, Poland; ANCS, Romania; MES of Russia and ROSATOM, Russian Federation; MON, Serbia and Montenegro; MSSR, Slovakia; MICINN and CPAN, Spain; SRC, Sweden; ETHZ and Uni-GE, Switzerland; STFC, United Kingdom; DOE and NSF, United States of America. We gratefully acknowledge the cooperation and exchange of knowledge with our colleagues engaged in ILC studies. Finally, we acknowledge the support from CERN as the host laboratory of this study.

# Appendices

# Appendix A

# Institutes participating in the CLIC/CTF3 collaboration

*Australian Collaboration for Accelerator Science, University of Melbourne, Melbourne, Australia*

*Joint Institute for Power and Nuclear Research - Sosny, Minsk, Belarus*

*Tsinghua University, Beijing, China*

*Institute of High Energy Physics, Beijing, China*

*Aarhus University, Aarhus, Denmark*

*Helsinki Institute of Physics, Helsinki, Finland*

*CEA, Irfu, Saclay, France*

*Laboratoire de l'Accélérateur Linéaire (LAL), Université de Paris-Sud XI, IN2P3/CNRS, Orsay, France*

*Laboratoire d'Annecy-le-Vieux de Physique des Particules (LAPP), Université de Savoie, IN2P3/CNRS, Annecy, France*

*Karlsruhe Institute of Technology, Karlsruhe, Germany*

*Democritus University of Thrace, Komotini, Greece*

*National Technical University of Athens, Athens, Greece*

*University of Patras, Rio Patras, Greece*

*Raja Ramanna Centre for Advanced Technology, DAE, Indore, India*

*INFN, Frascati, Italy*

*SISSA, Trieste, Italy*

*High Energy Accelerator Research Organisation, KEK, Tsukuba, Japan*

*Nikhef, Amsterdam, the Netherlands*

*University of Oslo, Oslo, Norway*

*National Centre for Physics, Islamabad, Pakistan*

*Budker Institute of Nuclear Physics, Akademgorodok, Russia*

*The Institute of Applied Physics of the Russian Academy of Sciences, Nizhny Novgorod, Russia*

*Joint Institute for Nuclear Research, Dubna, Russia*

*CIEMAT, Madrid, Spain*

*Instituto de Física Corpuscular, Valencia, Spain*

*Universitat Politècnica de Catalunya, BarcelonaTech, Barcelona, Spain*

*Universidade de Vigo, Vigo, Spain*

*Uppsala University, Uppsala, Sweden*
*CERN, Geneva, Switzerland*

*ETH Zurich, Zurich, Switzerland*

*Paul Scherrer Institute, Villigen, Switzerland*

*Ankara University, Ankara, Turkey*

*Gazi Universitesi Rektorlugu, Ankara, Turkey*

*Institute of Applied Physics, Sumy, Ukraine*

*The Cockcroft Institute, Daresbury, United Kingdom*

*The John Adams Institute for Accelerator Science, Oxford University, Oxford, United Kingdom*

*The John Adams Institute for Accelerator Science, Royal Holloway, University of London, Egham, United Kingdom*

*Argonne National Laboratory, Argonne, USA*

*Cornell University, Ithaca, USA*

*Fermi National Accelerator Laboratory, Batavia, USA*

*Jefferson Laboratory, Newport News, USA*

*Northwestern University, Illinois, USA*

*SLAC National Accelerator Laboratory, Menlo Park, USA*

*University of California, Santa Cruz, USA*

# Appendix B

# CLIC accelerator work-packages for the period 2012–2016

Table B.1: CLIC accelerator work-packages for the period 2012–2016 (part I).

| **Objectives and main activities** | **Work-package descriptions** |
|---|---|
| **Implementation studies** | |
| Define the scope, strategy and cost of the project implementation. Main input: the evolution of the physics findings at LHC and other relevant data, findings from the CDR and further studies, in particular concerning minimisation of the technical risks, cost, power as well as the site implementation.<br><br>A Governance Model as developed with partners | Civil engineering & services<br>Project implementation studies |
| **Parameters and design** | |
| Define and keep an up-to-date optimised overall baseline design that can achieve the scope within a reasonable schedule, budget and risk. Beyond beam line design, the energy and luminosity of the machine, key studies will address stability and alignment, timing and phasing, failure modes, stray fields and dynamic vacuum including collective effects. Other studies will address failure modes and operation issues. | Integrated baseline design and parameters<br>Integrated modelling and performance studies<br>Feedback design<br>Machine protection & operational scenarios<br>Background<br>Polarisation<br>Main beam electron source<br>Main beam positron source<br>Damping rings<br>Ring-to-main linac<br>Main linac – two-beam acceleration<br>Beam delivery system<br>Machine-detector interface (MDI) activities<br>Drive beam complex |

Table B.2: CLIC accelerator work-packages for the period 2012–2016 (part II).

| Objectives and main activities | Work-package descriptions |
|---|---|
| **Experimental verification** | |
| Identify and carry out system tests and programmes to address the key performance and operation goals and mitigate risks associated to the project implementation. The priorities are the measurements in: CTF3+, ATF and related to the CLIC0 injector addressing the issues of drive-beam stability, RF power generation and two-beam acceleration, as well as the beam delivery system. | CTF3 consolidation & upgrades |
| | Drive beam phase feed-forward and feed-backs |
| | TBL+, X-band high power RF production testing |
| | Two-beam module string, test with beam |
| | Drive beam source and injector system development |
| | Drive beam photo-injector |
| | Accelerator beam system tests (ATF, damping rings, FACET, . . . ) |
| Technical work-packages and studies addressing system performance parameters | Sources beam system tests |
| **Technological developments & X-band technology** | |
| Develop the technical design basis. i.e. move toward a technical design for crucial items of the machine and detectors, MDI, and the site. Priorities are the modulators/klystrons, module/structure development including testing facilities, alignment/stability and site studies. | Damping rings superconducting wiggler |
| | Survey & alignment |
| | Quadrupole stability |
| | Two-beam module development |
| | Warm magnet prototypes |
| | Beam instrumentation |
| | Collimation, mask and beam dumps |
| Technical work-packages providing input and interacting with all points above. | Controls |
| | RF systems (1 GHz klystrons & DB cavities, DR RF) |
| | Powering (modulators, magnet converters) |
| | Vacuum systems |
| | Magnetic stray fields measurements |
| | DR extraction system |
| | Creation of an "in house" technology center |
| | X-band RF structure design |
| | X-band RF structure production |
| | X-band RF structure high power testing |
| | Creation and operation of X-band high power testing facilities |
| | Basic high gradient R&D |

# Appendix C

# Acronyms

| | |
|---|---|
| **ATF** | Accelerator Test Facility |
| **BDS** | Beam Delivery System |
| **BDT** | Boosted Decision Tree |
| **BSM** | Beyond the Standard Model |
| **CDR** | Conceptual Design Report |
| **CLIC** | Compact Linear Collider |
| **CMOS** | Complementary Metal Oxide Semiconductor |
| **CTF3** | CLIC Test Facility 3 |
| **DBA** | Drive Beam Accelerator |
| **ECAL** | Electromagnetic Calorimeter |
| **FTE** | Full Time Equivalent |
| **GEM** | Gas Electron Multiplier |
| **HCAL** | Hadronic Calorimeter |
| **HVAC** | heating, ventilation and air conditioning |
| **ILC** | International Linear Collider |
| **ILD** | International Large Detector - one of the two validated ILC detector concepts |
| **ISR** | Initial State Radiation |
| **LEP** | Large Electron Positron Collider at CERN |
| **LHC** | Large Hadron Collider at CERN |
| **LoI** | Letter of Intent |
| **Micromegas** | Micro-MEsh Gaseous Structure |
| **MTP** | Medium Term Plan |
| **MPGD** | Micro-Pattern Gas Detector |
| **PETS** | Power Extraction and Transfer Structures |
| **PFO** | Particle Flow Object |
| **RF** | Radio Frequency |
| **SLS** | Swiss Light Source |
| **SLC** | Stanford Linear Collider at SLAC |

**SM** Standard Model

**SUSY** Supersymmetry

**SiD** Silicon Detector - one of the two validated ILC detector concepts

**SiPM** Silicon Photomultiplier

**TBL** test beam line

**TBM** tunnel-boring machines

**TBTS** two-beam test stand

**TDR** Technical Design Report